\documentclass{article}
\pdfoutput=1
\usepackage{jcappub}
\usepackage{graphicx}
\usepackage[latin1]{inputenc}
\usepackage{amsmath}
\usepackage{amssymb}
\usepackage{a4wide}
\usepackage{url}
\def\simlt{\ \raise -2.truept\hbox{\rlap{\hbox{$\sim$}}\raise5.truept   %
\hbox{$<$}\ }}
\def\simgt{\ \raise -2.truept\hbox{\rlap{\hbox{$\sim$}}\raise5.truept   %
\hbox{$>$}\ }}                                                          %

\def\be{\begin{equation}}
\def\ee{\end{equation}}
\def\newline{\hfil\break}

\def\la{\mathrel{\hbox{\rlap{\hbox{\lower4pt\hbox{$\sim$}}}\hbox{$<$}}}}
\def\ga{\mathrel{\hbox{\rlap{\hbox{\lower4pt\hbox{$\sim$}}}\hbox{$>$}}}}
\def\gs{{_>\atop^{\sim}}}

\newcommand{\pd}[3]{\frac{\partial^{#3} #1}{\partial {#2}^{#3}}} 
\newcommand{\td}[3]{\frac{d^{#3} #1}{d {#2}^{#3}}} 
\renewcommand{\v}[1]{\ensuremath{\mathbf{#1}}} 
\newcommand{\gv}[1]{\ensuremath{\mbox{\boldmath$ #1 $}}} 
\renewcommand{\bar}[1]{\ensuremath{\overline{#1}}}

\title{A Multi-frequency analysis of dark matter annihilation interpretations of recent anti-particle and $\gamma$-ray excesses in cosmic structures.} 
\author[a]{G. Beck}
\author[a,1]{S. Colafrancesco\note{Corresponding author.}}

\affiliation[a]{School of Physics, University of the Witwatersrand, Private Bag 3, WITS-2050, Johannesburg, South Africa}

\emailAdd{geoff.m.beck@gmail.com}
\emailAdd{sergio.colafrancesco@wits.ac.za}

\abstract{
The Fermi-LAT observation of a $\gamma$-ray excess from the galactic-centre, as well as the PAMELA, AMS, and AMS-2 anti-particle excesses,  and the recent indications of a Fermi-LAT $\gamma$-ray excess in the Reticulum II dwarf galaxy have all been variously put forward as possible indirect signatures of supersymmetric neutralino dark matter. 
These are of particular interest as the neutralino annihilation models which fit these observations must have observable consequences across the frequency spectrum, from radio to $\gamma$-ray emission. 
Moreover, since dark matter is expected to be a major constituent of cosmic structure, these multi-frequency consequences should be common to such structures across the mass spectrum, from dwarf galaxies to galaxy clusters. 
Thus, in this work we make predictions for the multi-frequency spectra of three well-known sources dominated by dark matter on cluster, galaxy and dwarf galaxy scales, e.g. the Coma cluster, the galaxy M81, and the Draco dwarf galaxy, using models favoured by dark matter interpretations of the aforementioned observations. 
We pay special attention to the consequences for these models when their cross-sections are renormalised to reproduce the recent $\gamma$-ray excess observed in the Reticulum II dwarf galaxy, as well as using cross-sections from the Fermi-LAT dwarf galaxy limits, which throw a dark matter interpretation of this excess into doubt. 
We find that the multi-frequency data of Coma and Draco are in conflict with the dark matter interpretation of the AMS, PAMELA and Fermi positron excess. Additionally, models derived from Fermi-LAT galactic centre observations, and AMS-2 re-analysis, present similar but less extensive conflicts.
Using the sensitivity projections for the Square Kilometre Array, the Cherenkov Telescope Array, as well as the ASTROGAM and ASTRO-H satellites, we determine the detection prospects for a subset of neutralino models that remain consistent with Planck cosmological constraints.
Although the SKA has the greatest sensitivity to dark matter models, we demonstrate that ASTRO-H is well positioned to probe the inverse-Compton scattering emissions from neutralino annihilation and identify characteristics of the spectra which contain information about the neutralino mass and annihilation channel. This means that, given environments with favourable X-ray backgrounds, multi-frequency observation with the next generation of experiments will allow for unprecedented sensitivity to the neutralino parameter space as well as offsetting the individual weaknesses of each observation mode. Finally we show that all of the studied models can be better tested with the SKA phase 1.}

\begin{document}
\maketitle
 

\section{Introduction}

The recent observed excesses of $\gamma$-ray emission from the galactic centre (GC) and anti-particle fluxes have been reported as possible signatures of dark matter (DM) annihilation~\cite{cholis2013,calore2014,hooper2011}.\\ 
In particular, the limits derived from both the Fermi-LAT~\cite{fermi-docs} data on the galactic centre $\gamma$-ray excess emission and the PAMELA~\cite{pamela-docs} anti-proton excess have been indicated to favour supersymmetric neutralino DM models with a particle mass of around $35$ GeV and a thermal annihilation cross-section of $\langle \sigma V\rangle \sim 3 \times 10^{-26}$ cm$^3$ s$^{-1}$~\cite{hooper2014,hooper2011}. However, this must be considered alongside the arguments in \cite{calore2014}, where it is indicated that background uncertainties for the GC imply that a far larger range of models, with masses between $10$ and $100$ GeV and annihilation cross-sections between $ 10^{-27}$ cm$^3$ s$^{-1}$ and $ 10^{-26}$ cm$^3$ s$^{-1}$, may be consistent with the observed GC $\gamma$-ray excess. 
A DM interpretation of these GC measurements has been, however, further disputed \cite{oleary2015,bartels2015b,lee2015,brandt2015,calore2015}, where these authors argue that unresolved populations of millisecond or young pulsars are sufficient to explain the observations.\\
In addition, the results from the Alpha Magnetic Spectrometer (AMS)~\cite{ams-doc} cosmic-ray detector have been used to claim that a dark matter mass of $\mathcal{O}(\mbox{TeV})$ with annihilation cross-section $\langle \sigma V\rangle \sim 3 \times 10^{-24}$ cm$^3$ s$^{-1}$ will consistently reproduce observed excesses via secondary positron production~\cite{cholis2013}, although the authors note that an unresolved population of young pulsars could equally account for the observations. 
It has been, nonetheless, demonstrated \cite{lopez2015} that the results of the Fermi-LAT dwarf galaxy observations~\cite{Fermidwarves2014} are largely incompatible with a DM explanation of the positron excess seen by AMS-2 for most annihilation channels, and masses below TeV scales. 
The aforementioned study \cite{Fermidwarves2014}, produced constraints ranging from $\langle \sigma V\rangle < 3 \times 10^{-26}$ cm$^3$ s$^{-1}$ for $m_{\chi} < 10$ GeV to $\langle \sigma V\rangle < 2 \times 10^{-23}$ cm$^3$ s$^{-1}$ for $m_{\chi} < 10^4$ GeV. These were further improved upon in the sub-TeV range \cite{geringer-sameth2014} with constraints $\langle \sigma V\rangle < 2.2 \times 10^{-26}$ cm$^3$ s$^{-1}$ for $m_{\chi} \leq 114$ GeV.\\
A recent re-analysis of the AMS-2 electron/positron data~\cite{mauro2015} also indicates that it is compatible with DM models with cross-sections $\sim (2 - 8) \times 10^{-26}$ cm$^{3}$ s$^{-1}$ and masses in the range $51 - 140$ GeV, depending on the annihilation channels studied. A very similar set of models are proposed to account for excess $\gamma$-ray emission observed by Fermi-LAT at the 2.3$\sigma$ confidence level, reported in the dwarf galaxy Reticulum II~\cite{geringer-sameth2015}. In particular, these authors argued for a DM interpretation with mass $\sim 40$ GeV and a cross-section $\langle \sigma V\rangle \sim 3 \times 10^{-26}$ cm$^3$ s$^{-1}$. We note that similar cross-sections are favoured based upon estimates of the astrophysical J-factor for Reticulum II~\cite{bonnivard2015}. However, a dark matter interpretation of this excess appears to be at odds with the recent Fermi-LAT dwarf galaxy analysis, which includes Ret. II~\cite{Fermidwarves2015}.

The release of the latest wave of Planck~\cite{planck-docs} cosmological results~\cite{planck2014} also marks the current status on the hunt for neutralino DM from the cosmological side, i.e. using the CMB anisotropy power spectrum analysis. Planck substantially curtailed the allowed regions of the mass vs. cross-section parameter space all but eliminating the models compatible with the reported AMS-2/Fermi/PAMELA (hereafter AFP) positron excess~\cite{cholis2013}, for cases where DM annihilation is efficient at depositing energy into the inter-galactic medium. Additionally, sub-TeV neutralinos with cross-section values above the relic density bound~\cite{jungman1996} were largely ruled out. The models favoured by the Fermi-LAT observations of the galactic centre (GC)~\cite{hooper2011,calore2014}, however, still remain largely unaffected by the Planck result. In cases where DM annihilation has a low energy deposition efficiency below $0.3$ the AFP region is also unaffected (as is true for the neutralino annihilation channels studied here). 

In this observational framework, it is worthwhile to examine the future prospects of the remaining models allowed by PAMELA/Fermi/AMS-2 data as well as those allowed by the Fermi-LAT GC data in terms of a possible explanation due to DM annihilation.\\
The purpose of this paper is, in fact, to explore the consequences of the possibly DM-consistent signals from AMS-2, PAMELA and those of the Ret. II dwarf galaxy observed with Fermi-LAT on the multi-frequency expectations for well known DM-dominated halos like galaxy clusters, galaxies and dwarf galaxies. Along this line of exploration, we define a multi-frequency observational strategy for neutralino hunting with coming experiments in the hard X-ray/soft $\gamma$-ray band (like ASTRO-H and ASTROGAM), in the very high-E $\gamma$-ray band (like  the Cherenkov Telescope Array, CTA), and in the very low-frequency radio range of the electromagnetic spectrum (like the Square Kilometre Array, SKA).
This examination will take the form of specific predictions of multi-frequency observation for a key reason, i.e. to confirm the possibility that these potential DM signatures are consistent with a larger set of observations or constraints and then to produce a consistent search for DM signals over the whole accessible frequency range of the electromagnetic spectrum.
%

For the aims of this paper we will consider representative models from each of the aforementioned regions of the parameter space and we will study their multi-frequency predictions in the light of the achievable sensitivities of the upcoming instruments at radio (e.g., SKA), hard X-ray (e.g., ASTRO-H), soft $\gamma$-rays (e.g., ASTROMEV and ASTROGAM) and high-E $\gamma$-rays (e.g., CTA). We will confine our analysis to a few well known target environments for which data and theoretical modelling are rich and available. These are the Coma cluster, the M81 galaxy, and the Draco dwarf spheroidal galaxy. This will allow us to make concrete predictions of the prospective ability of multi-frequency observations to explore the neutralino parameter space in these environments, as well as to compare the studied models to current observational data. We will use the spectral energy distributions (SED) of these sources to demonstrate the synergy between radio-frequency and high-energy observations, which will serve to increase the robustness of any purported indirect neutralino signatures as well as allow for the characterization of the neutralino through these observations.\\ 
In particular, we will show that the ASTRO-H space mission has an observational window on a portion of the DM-induced inverse-Compton scattering (ICS) spectrum which is sensitive to the neutralino mass and to the annihilation channel. In the environments we have considered here, ASTRO-H cannot provide better constraints than Fermi-LAT, due to the existence of strong X-ray backgrounds in these sources. Therefore, it remains a subject of future work to locate more favourable environments for the hard X-ray study of DM models.\\
In the case of $\gamma$-rays we find that both CTA and ASTROGAM will be able to make little impact in the study of the GC and AFP models. In the case of ASTROGAM this is because the instrument is insufficiently sensitive to detect the soft $\gamma$-ray spectrum produced by neutralino annihilation within these models. For CTA we find that it is sensitive to energies largely above the typical mass-dependent cut-off for the studied models, even in the case of the AFP with TeV scale masses. In galaxy cluster environments, like Coma, the CTA may be capable of marginal detection for  the $\tau^+\tau^-$ decay channel, which produces harder spectra, but this is complicated by the comparatively low sensitivity of CTA in this spectral range. Despite these issues, CTA may still have a role in determining whether observed hard $\gamma$-ray emission is not inconsistent with the aforementioned dark matter models, as the discovery of anomalous hard $\gamma$-ray excesses would pose difficulties for these models if DM were found to be the most likely explanation.\\
For the radio frequency search we find that SKA is well situated to study a large swathe of the dark matter parameter space, providing optimistic constraints up to 2 orders of magnitude below the current Planck limits in the studied environments. This despite accounting for the need to differentiate between sub-dominant DM emissions and astrophysical backgrounds. In addition the SKA will have access to a frequency range highly sensitive to neutralino annihilation channel and mass.

In our study, we also draw on the recently reported Reticulum II dwarf galaxy $\gamma$-ray excess~\cite{geringer-sameth2015}, and argue that its consequent radio and $\gamma$-ray emissions are incompatible with data available on the studied environments. Thus doubt is cast on the DM interpretation of the excess put forward in \cite{geringer-sameth2015}.

This paper is structured as follows: in Section~\ref{sec:models} we detail models of multi-frequency emission from DM annihilation and in Section~\ref{sec:neutralino} we discuss the neutralino parameters corresponding to our representative models. In Section~\ref{sec:data} we detail the relevant instrument parameters as well as the multi-frequency data used in this study. We provide multi-frequency predictions for the chosen environments in Section~\ref{sec:results}, as well as examining further detection prospects with SKA and ASTRO-H in Section~\ref{sec:constraint}, and discuss the results of our analysis in Section~\ref{sec:discussion} before drawing conclusions in Section~\ref{sec:conclusions}.

\begin{figure}[htbp]
\centering
\includegraphics[scale=0.8]{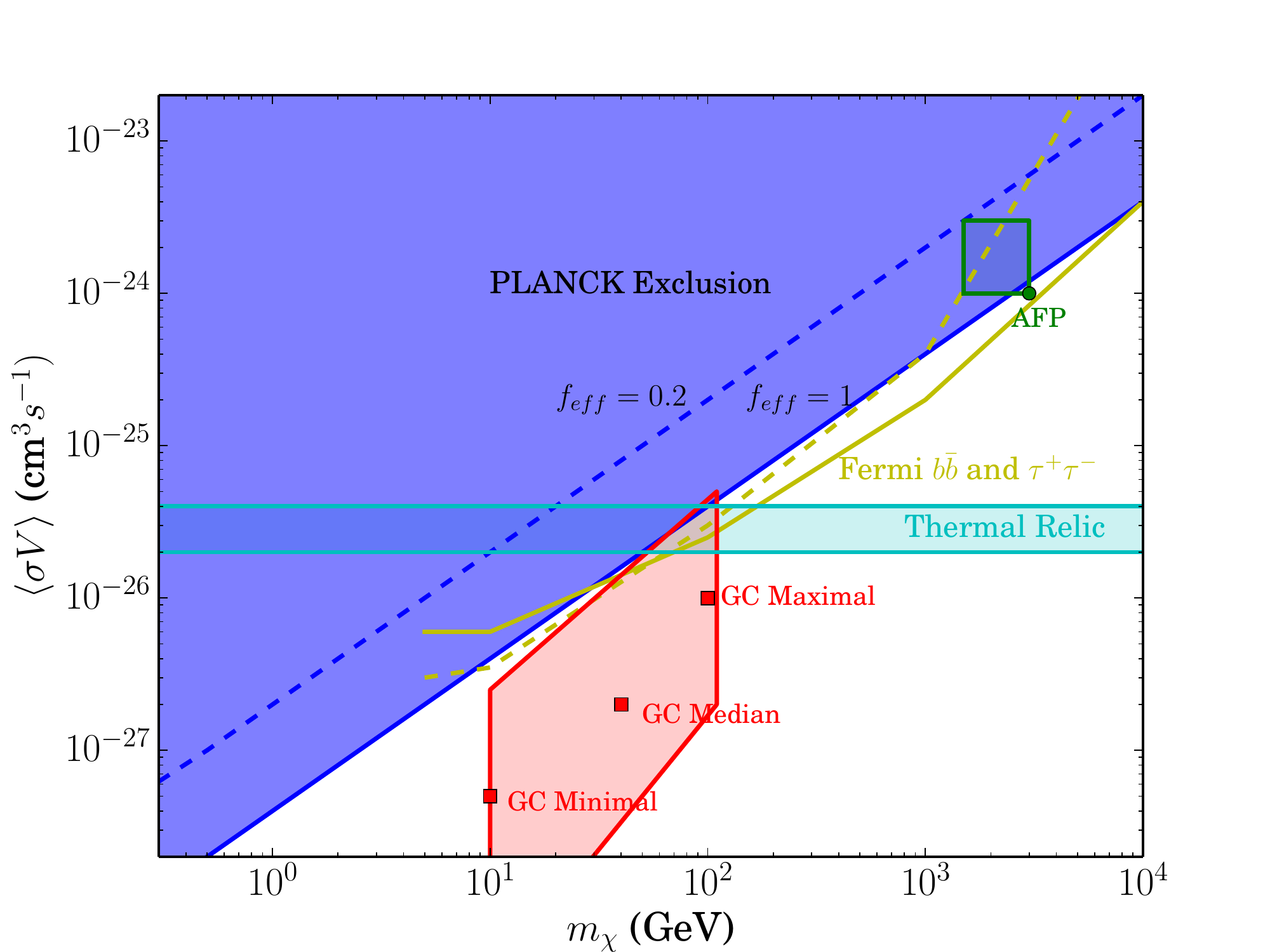}
\caption{Currently Excluded regions and DM signal scenarios  in the $\langle\sigma V\rangle $ vs. mass parameter space~\cite{planck2014}. $f_{eff}$ refers to the DM annihilation energy deposition efficiency factor (note that $b\bar{b}$ and $\tau^+\tau^-$ annihilation channels have $f_{eff} \sim 0.2 - 0.3$). The representative AMS-2/PAMELA/Fermi positron excess model is given by the AFP point, whereas the GC Maximal, Median, and Minimal are three representative models from the DM interpretation of the Fermi-LAT galactic centre observations. The Fermi-LAT dwarf-galaxy exclusion for $b\bar{b}$ and $\tau^+\tau^-$ (yellow, solid and dashed curves), as well as DM thermal relic abundance band on $\langle\sigma V\rangle $ are also shown for comparison.}
\label{fig:Planck}
\end{figure}

\section{Models of Dark Matter Halos and Multi-frequency Emission}
\label{sec:models}

In modelling the halos of our structures of interest we refer to both a cuspy DM Navarro-Frenk-White (NFW) density profile~\cite{nfw1996} and a cored Burkert profile ~\cite{Burkert1995}, that can bracket a larger range of possible phenomenological models.\\
The NFW profile is described by 
\begin{equation}
\rho(r)=\frac{\rho_s}{\frac{r}{r_s}\left(1+\frac{r}{r_s}\right)^{2}} \; ,\\
\label{eq:nfw}
\end{equation}
with $r_s$ being the scale radius of the profile, and $\rho_s$ is the halo characteristic density.\\
The Burkert profile is described by 
\begin{equation}
\rho(r)=\frac{\rho_s}{\left(1 + \frac{r}{r_s}\right)\left(1+\left(\frac{r}{r_s}\right)^2\right)}
\label{eq:burkert}
\end{equation}

We define the virial radius $R_{vir}$, of a halo with mass $M_{vir}$, as the radius within which the mean density of the halo is equal to the product of the collapse over-density $\Delta_{c}$ and the critical density $\rho_c$, where 
\begin{align}
\rho_c (z) & = \frac{3 H(z)^2}{8\pi G} \; , \\
M_{vir} & = \frac{4}{3}\pi \Delta_{c} \rho_c R_{vir}^3 \; ,
\end{align}
with $H(z)$ being the Hubble parameter. The density contrast parameter at collapse is given in a flat cosmology by the approximate expression~\cite{bryan1998}
\begin{equation}
\Delta_{c} \approx 18 \pi^2 - 82 x - 39 x^2 \; ,
\end{equation}
with $x = 1.0 - \Omega_m (z)$, where $\Omega_m (z)$ is the matter density parameter at redshift $z$ 
\begin{equation}
\Omega_m (z)  = \frac{1}{1 + \frac{\Omega_\Lambda (0)}{\Omega_m (0)}(1+z)^{-3}} \; .
\end{equation}
The concentration parameter for the halo defines the scale radius as follows
\begin{equation}
\begin{aligned}
r_{s} = \frac{R_{vir}}{c_{vir}} \; .
\end{aligned}
\end{equation}
where $c_{vir}$ is determined either for a particular environment or from the model for $c_{vir} (M_{vir})$ given in~\cite{munoz2010}. We note that this relation is true only for NFW and Einasto profiles, as it assumes that $r_s \equiv r_{-2}$, where $r_{-2}$ is the radius at which the effective logarithmic slope of $\rho$ is $-2$. In the case of a Burkert profile $r_s \approx \frac{R_{vir}}{1.52 c_{vir}}$.
The dimensionless characteristic density contrast $\frac{\rho_s}{\rho_c}$ is defined to ensure the normalisation
\begin{equation}
\int_0^{R_{vir}} dr \; 4\pi r^2 \rho (r) = M_{vir} \; . 
\end{equation}
In the case of the NFW halo this can written~\cite{ludlow2013} in terms of $c_{vir}$ as
\begin{equation}
\frac{\rho_s (c_{vir})}{\rho_{c}}=\frac{\Delta_{c}}{3}\frac{c_{vir}^3}{\ln(1+c_{vir})-\frac{c_{vir}}
{1+c_{vir}}} \; .
\end{equation}

In addition to their global structure (NFW and/or Burkert density profile), DM halos are generally thought to have sub-structure in the form of sub-halos that are denser than their parent halo~\cite{pieri2011,Colafrancesco2006} and can then boost the annihilation signals by a factor $b$, which has a radial dependence based on the assumption that sub-halo distribution follows a similar pattern to the DM density of the parent halo, with a longer scale radius~\cite{pieri2011,Colafrancesco2006}. 
In order to derive the boost factor $b$ we follow the prescription presented in \cite{prada2013}, as this conforms to recent analysis performed on Fermi-LAT data. This sub-structure boost factor is defined as a luminosity increase caused by integrating over sub-halo luminosities determined by the virial mass and by halo concentration parameters found numerically according to the method discussed in~\cite{prada2012}; we note that a similar method is provided in~\cite{ng2014}. These methods place the boosting factor of cluster sized halos around $b \sim 30$, with the large dwarf-like galaxy Draco having $b \sim 3$. As noted in \cite{prada2013}, these values of $b$ are substantially smaller than those of many popular models quoted in the literature, and are based upon the $c_{vir}$ - $M_{vir}$ relations that agree well with N-body simulations~\cite{prada2012}. We note, however, that this form of the boost factor has no explicit radial dependence $b(r) \equiv b$ as it represents a total contribution from all sub-halos.
In addition to these considerations, a recent study \cite{bartels2015} indicates that tidal-stripping of sub-halos has the potential to enhance the sub-halo luminosity boost by a factor of $2 - 3$. This indicates that there may be additional dynamical considerations that increase boost factors, strengthening the certainty of the conclusions reached here with more conservative boosts that do not account for dynamical effects.

An additional consideration in studying DM emissions is diffusion of the electrons/positrons which are responsible for synchrotron  and inverse-Compton scattering emissions. In~\cite{gsp2015,Colafrancesco2007}, it was argued that energy-loss dominates particle diffusion in large-scale structures like large galaxies and galaxy clusters, while diffusion is significant in small-size galaxies like dwarf galaxies \cite{Colafrancesco2007} particularly when observation is confined to a small angular segment of the target. For this reason we will consider the impact of spatial diffusion in dwarf galaxy environments according to the following method.

The equation for the equilibrium electron spectrum is found as follows:
\begin{equation}
\begin{aligned}
\pd{}{t}{}\td{n_e}{E}{} = & \; \gv{\nabla} \left( D(E,\v{r})\gv{\nabla}\td{n_e}{E}{}\right) + \pd{}{E}{}\left( b(E,\v{r}) \td{n_e}{E}{}\right) + Q_e(E,\v{r}) \; ,
\end{aligned}
\end{equation}
where $\td{n_e}{E}{}$ is the electron equilibrium spectrum, $D(E,\v{r})$ and $b(E,\v{r})$ are the spatial diffusion and energy-loss functions respectively (see~\cite{gsp2015,Colafrancesco2006} and below), and $Q_e(E,\v{r})$ is the electron source function. A detailed analysis of the solution to this equation in the case of electron production via neutralino annihilation can be found in \cite{Colafrancesco2006}. 

In order to take into account the effects of the magnetic field and thermal plasma on electron diffusion we take average values for the field strength and thermal plasma density, being $\overline{B} \equiv \sqrt{\langle B(r)^2 \rangle}$ and $\overline{n} \equiv \langle n(r) \rangle$, respectively. We use $\langle \; \rangle$ to denote a spatial average over the target structure out to the virial radius.
We then define the spatial diffusion coefficient as~\cite{Colafrancesco1998}
\begin{equation}
D(E) = \frac{1}{3}c r_L (E) \frac{\overline{B}^2}{\int^{\infty}_{k_L} dk P(k)} \; ,
\end{equation}
where $r_L$ is the Larmour radius of a relativistic particle with energy $E$ and charge $e$ and $k_L = \frac{1}{r_L}$, and require that
\begin{equation}
\int^{\infty}_{k_0} dk P(k) = \overline{B}^2 \; .
\end{equation}
This leads us to the result that
\begin{equation}
D(E) = D_0 d_0^{\frac{2}{3}} \left(\frac{\overline{B}}{1 \mu\mbox{G}}\right)^{-\frac{1}{3}} \left(\frac{E}{1 \mbox{GeV}}\right)^{\frac{1}{3}}  \; , \label{eq:diff}
\end{equation}
where $D_0 = 3.1\times 10^{28}$ cm$^2$ s$^{-1}$. It is worth noting that the diffusion coefficient is assumed to be lacking radial dependence. While it is possible to implement diffusion without this simplification, we present results here under the assumption we can substitute the averaged value of the magnetic field in the diffusion coefficient as it is evident that the weak radial dependence of the magnetic fields and the weak dependence of the diffusion coefficient on the field strength imply that our approximation is not unwarranted. For this analysis we will assume the minimum scale on which a dwarf galaxy magnetic field is homogeneous is given by $d_0 \sim 100$ pc.

Finally, the energy-loss function takes the form,
\begin{equation}
\begin{aligned}
b(E) = & b_{IC} E^2 (1+z)^4 + b_{sync} E^2 \overline{B}^2 \\&\; + b_{Coul} \overline{n} (1+z)^3 \left(1 + \frac{1}{75}\log\left(\frac{\gamma}{\overline{n} (1+z)^3}\right)\right) \\&\; + b_{brem} \overline{n} (1+z)^3 \left( \log\left(\frac{\gamma}{\overline{n} (1+z)^3 }\right) + 0.36 \right)
\end{aligned}
\label{eq:loss}
\end{equation}
where $\overline{n}$ is given in cm$^{-3}$ and $b_{IC}$, $b_{synch}$, $b_{col}$, and $b_{brem}$ are the Inverse Compton, synchrotron, Coulomb and Bremsstrahlung energy loss factors, taken to be $0.25$, $0.0254$, $6.13$, and $1.51$ respectively in units of $10^{-16}$ GeV s$^{-1}$. Here $E$ is the energy in GeV and the B-field is in $\mu$G.


Three particular dark matter halos with very different mass will be of interest in our study: that of the Coma cluster, the M81 galaxy, and the Draco dwarf galaxy. We also considered the case of the Reticulum II dwarf galaxy for its recent interest as a possible source of $\gamma$-ray emission.\\ 
For the Coma cluster DM halo we consider the model described in~\cite{Colafrancesco2006}. 
The virial mass of this cluster is taken to be $M_{vir} = 1.33 \times 10^{15}$ M$_{\odot}$, with virial concentration  $c_{vir} = 10$, at the redshift $z = 0.0231$.
The thermal electron density of the ICM in Coma $n(r)$ is given by 
\begin{equation}
n_e(r) = n_0 \left(1 + \left[\frac{r}{r_s}\right]^2\right)^{-q_e} \; ,
\end{equation}
with $r_s$ being a characteristic radius (taken equal to the halo scale radius), $n_0 = 3.44 \times 10^{-3}$ cm$^{-3}$ and $q_e = 1.125$~\cite{briel1992}.
The magnetic field in Coma is assumed to follow the one derived by~\cite{bonafede2010} having a radial profile given by
\begin{equation}
B(r) = B_0 \left(\frac{n_e(r)}{n_0}\right)^{q_b} \; ,
\end{equation}
where $r$ is the distance from the cluster centre, $B_0 = 4.7$ $\mu$G, and $q_b = 0.5$. 
Additionally, this magnetic field has a Kolmogorov turbulence power spectrum with a minimal coherence length of $\approx2$ kpc.\\
In M81 we use of the following magnetic field model:.
\begin{equation}
B(r) = B_0 \left(1 + \left(\frac{r}{r_b}\right)^2\right)^{-q_b q_e} \; ,
\end{equation} 
here, $r_b = 13$ kpc, $q_b = 0.5$, $q_e = 1.125$, and $B_0= 7.5$ $\mu$G~\cite{beck1985}. In the case of the thermal electron density we use a central value of $n_0 = 0.03$ cm$^{-3}$~\cite{beck1985} with a similar radial profile to the one used for Coma but with scale radius of $3$ kpc~\cite{beck1985}, and the DM halo of this galaxy is taken to have a virial mass $M_{vir} = 1.4 \times 10^{11}$ M$_{\odot}$ at a distance of $3.6$ Mpc~\cite{kostov2006-ngc3198-m81}.\\
For the case of the Draco dwarf galaxy we take the virial mass to be $M_{vir} = 7 \times 10^7$ M$_{\odot}$ at a distance of $\sim 80$ kpc~\cite{lokas2005-draco}, with a constant magnetic field model with magnitude $B(r) = B_0 = 1$ $\mu$G, and a thermal electron density $n_e(r) = n_0 = 10^{-6}$ cm$^{-3}$, both in accordance with \cite{Colafrancesco2007}.\\ 
Finally, we make use of a conservative model for the Reticulum II dwarf galaxy, using a distance of $\sim 30$ kpc~\cite{geringer-sameth2015}, and assuming a J-factor from~\cite{bonnivard2015} with a constant magnetic field model with magnitude $B(r) = B_0 = 1$ 
$\mu$G, and a thermal electron density $n_e(r) = n_0 = 10^{-6}$ cm$^{-3}$.

The structure parameters on the three target environments are reported in Table~\ref{tab:phys}.
\begin{table}[htbp]
\centering
\begin{tabular}{|l|l|l|l|}
\hline
Quantity & Draco & M81 & Coma \\
\hline
$M_{vir}$ (M$_{\odot}$) & $7\times 10^7$ & $1.4 \times 10^{11}$ & $1.33 \times 10^{15}$\\
boost-factor & 3.43 & 10.1 & 35.2 \\
$B_0$ ($\mu$G) & $1.0$ & $7.5$ & $4.7$ \\
$\langle B \rangle$ ($\mu$G) & $1.0$ & $1.9$ & $1.2$ \\ 
$n_0$ (cm$^{-3}$)& $10^{-6}$ & $0.03$ & $3.44 \times 10^{-3}$ \\
$\langle n_0 \rangle$ (cm$^{-3}$)& $10^{-6}$ & $8.8 \times 10^{-4}$ & $4.8 \times 10^{-4}$\\
\hline 
\end{tabular}
\label{tab:phys}
\caption{Summary of the relevant halo parameters.}
\end{table}

As a supplementary summary of pertinent halo parameters, we calculate the astrophysical ``J-factor" for each halo:
\begin{equation}
J = \int_{\Delta\Omega} \int_{l} \rho_{DM}^2(\v{r}) d l^{\prime}d\Omega^{\prime} \; ,
\end{equation}
where the integral is performed over the line of sight $l$ and the solid angle $\Delta \Omega$. For reference we present the J-factor calculated when integrating over the solid angle of the entire halo virial radius given in Table~\ref{tab:jfac}, along with the Ret. II J from~\cite{bonnivard2015}.
\begin{table}[htbp]
\centering
\begin{tabular}{|l|l|l|}
\hline
Halo & NFW & Burkert \\
\hline
Draco & $1.1 \times 10^{17}$  & $3.1 \times 10^{16}$ \\
Coma & $1.0 \times 10^{18}$ & $2.8 \times 10^{17}$ \\
M81 & $3.0\times 10^{16}$  & $1.0 \times 10^{16}$ \\
Ret. II & $2.0 \times 10^{19}$ & $2.0 \times 10^{19}$ \\
\hline
\end{tabular}
\label{tab:jfac}
\caption{J-factors for each studied environment and for Reticulum II. These include sub-structure boosting factors (where appropriate), and are given in units of GeV$^2$ cm$^{-5}$.}
\end{table}

For the general description  of DM halos and synchrotron emissions we follow the approach described in~\cite{gsp2015} and in the references contained therein, while for the high-energy emission properties of DM annihilation we follow the approach of~\cite{Colafrancesco2006}. Here in the following we report the basic formulae we will use for the multi-frequency spectral energy distribution produced by DM annihilation.

The average power of the synchrotron radiation at observed frequency $\nu$ emitted by an electron with energy $E$ in a magnetic field with amplitude $B$ is given by~\cite{longair1994}
\begin{equation}
P_{synch} (\nu,E,r,z) = \int_0^\pi d\theta \, \frac{\sin{\theta}^2}{2}2\pi \sqrt{3} r_e m_e c \nu_g F_{synch}\left(\frac{\kappa}{\sin{\theta}}\right) \; ,
\label{eq:power}
\end{equation}
where $m_e$ is the electron mass, $\nu_g = \frac{e B}{2\pi m_e c}$ is the non-relativistic gyro-frequency, $r_e = \frac{e^2}{m_e c^2}$ is the classical electron radius, and the quantities $\kappa$ and $F_{synch}$ are defined
\begin{equation}
\kappa = \frac{2\nu (1+z)}{3\nu_g \gamma^2}\left[1 +\left(\frac{\gamma \nu_p}{\nu (1+z)}\right)^2\right]^{\frac{3}{2}} \; ,
\end{equation}
with $\nu_p \propto \sqrt{n_e}$, and
\begin{equation}
F_{synch}(x) = x \int_x^{\infty} dy \, K_{5/3}(y) \simeq 1.25 x^{\frac{1}{3}} \mbox{e}^{-x} \left(648 + x^2\right)^{\frac{1}{12}} \; .
\end{equation}
and the average power of inverse-Compton Scattering (ICS) is given by
\begin{equation}
P_{IC} (\nu,E,z) = c E_{\gamma}(z) \int d\epsilon \; n(\epsilon) \sigma(E,\epsilon,E_{\gamma}(z)) \; ,
\label{eq:ics_power}
\end{equation}
where $E_{\gamma}(z) = h \nu (1+z)$ is the emitted photon energy, $n(\epsilon)$ is the black-body spectrum of the CMB photons, and $E$ is the electron energy. Here we consider mainly the ICS on CMB photons because this is the largest radiation background available in the universe.\\
Additionally,
\begin{equation}
\sigma(E,\epsilon,E_{\gamma}) = \frac{3\sigma_T}{4\epsilon\gamma^2}G(q,\Gamma_e) \; ,
\end{equation}
where $\sigma_T$ is the Thompson cross-section, $\gamma$ is the electron Lorentz factor, and
\begin{equation}
G(q,\Gamma_e) = 2 q \ln{q} + (1+2 q)(1-q) + \frac{(\Gamma_e q)^2(1-q)}{2(1+\Gamma_e q)} \; ,
\end{equation}
with
\begin{equation}
\begin{aligned}
q & = \frac{E_{\gamma}}{\Gamma_e(\gamma m_e c^2 + E_{\gamma})} \; , \\
\Gamma_e & = \frac{4\epsilon\gamma}{m_e c^2}
\end{aligned}
\end{equation}

Bremstrahlung emission of DM-produced secondary electrons from the intra-cluster medium (ICM) and from the inter-stellar medium (ISM) has an average power
\begin{equation}
P_B (E_{\gamma},E,r) = c E_{\gamma}(z)\sum\limits_{j} n_j(r) \sigma_B (E_{\gamma},E) \; ,
\end{equation}
where $n_j(r)$ is the density of intra-cluster species $j$, and
\begin{equation}
\sigma_B (E_{\gamma},E) = \frac{3\alpha \sigma_T}{8\pi E_{\gamma}}\left[ \left(1+\left(1-\frac{E_{\gamma}}{E}\right)^2\right)\phi_1 - \frac{2}{3}\left(1-\frac{E_{\gamma}}{E}\right)\phi_2 \right] \; ,
\end{equation}
with $\phi_1$ and $\phi_2$ being energy dependent factors determined by the species $j$(see \cite{longair1994}).

For the DM-induced $\gamma$-ray production through $\pi^0 \to \gamma \gamma$ decay the flux calculation is somewhat simplified
\begin{equation}
S_{\gamma} (\nu,z) = \int_0^r d^3r^{\prime} \, \frac{Q_{\gamma}(\nu,z,r)}{4\pi D_L^2} \; ,
\end{equation} 
with $Q_{\gamma}(\nu,z,r)$ being the source function for neutral pion decay within the given DM halo.

The local emissivity for the $i-th$ emission mechanism  (Synchrotron, ICS, Bremstrahlung) can then be found as a function of the electron and positron equilibrium distributions as well as the associated power
\begin{equation}
j_{i} (\nu,r,z) = \int_{m_e}^{M_\chi} dE \, \left(\td{n_{e^-}}{E}{} + \td{n_{e^+}}{E}{}\right) P_{i} (\nu,E,r,z) \; .
\label{eq:emm}
\end{equation}
The flux density spectrum within a radius $r$ is then written as 
\begin{equation}
S_{i} (\nu,z) = \int_0^r d^3r^{\prime} \, \frac{j_{i}(\nu,r^{\prime},z)}{4 \pi D_L^2} \; ,
\label{eq:flux} 
\end{equation}
where $D_L$ is the luminosity distance to the halo.

\section{Neutralino Models}
\label{sec:neutralino}

In this paper our neutralino DM particle is drawn from the minimal supersymmetric extension to the standard model, following the DarkSUSY package~\cite{darkSUSY}. The source function for the production of a stable particle $i$, produced promptly by neutralino annihilation or ancillary processes is given by
\begin{equation}
Q_i (r,E) = \langle \sigma V\rangle \sum\limits_{f}^{} \td{N^f_i}{E}{} B_f \mathcal{N}_{\chi} (r) \; ,
\end{equation}
where $\langle \sigma V\rangle$ is the non-relativistic velocity-averaged neutralino annihilation cross-section at $0$ K, the index $f$ labels kinematically allowed annihilation final states with branching ratios $B_f$ and spectra $\td{N^f_i}{E}{}$,  and $\mathcal{N}_{\chi} (r)$ is the neutralino pair density at a given halo radius $r$. In keeping with standard procedure in indirect detection studies we will focus on one annihilation channel at a time and assume a branching ratio of $1$ for the channel of interest. We will examine the $b\bar{b}$ and $\tau^+\tau^-$ channels. The factor $\td{N^f_i}{E}{}$ is determined using the PYTHIA Monte-Carlo routines in DarkSUSY~\cite{pythia}.

Our study examines four neutralino mass models, each of which is then further differentiated by three cross-section values: a best-fit cross-section, the one derived from Reticulum II $\gamma$-ray excess (as detailed below), and the one derived from Fermi-LAT dwarf studies~\cite{Fermidwarves2015}. 
The first model is taken to represent the neutralino model interpretation of the PAMELA/AMS-2/Fermi (AFP) positron excess, which is still accommodated by the Planck results regardless of DM annihilation energy deposition efficiency factor, and has $M_{\chi} \sim 3$ TeV and best-fit cross-section $\langle \sigma V \rangle \sim 10^{-24}$ cm$^3$ s$^{-1}$ (see Fig.\ref{fig:Planck}). The other three models are representative of the minimal, median, and maximal cases of the neutralino model interpretation of the Fermi-LAT galactic centre (GC) observations: these have values $M_{\chi} \sim 10$ GeV, best-fit $\langle \sigma V \rangle \sim 10^{-28}$ cm$^3$ s$^{-1}$, $M_{\chi} \sim 40$ GeV, best-fit $\langle \sigma V \rangle \sim 10^{-27}$ cm$^3$ s$^{-1}$, and $M_{\chi} \sim 100$ GeV, best-fit $\langle \sigma V \rangle \sim 10^{-26}$ cm$^3$ s$^{-1}$, respectively (see Fig.\ref{fig:Planck}). The choice of the GC models is predicated on covering the range of the parameter space favoured by the analysis of the Fermi-LAT data~\cite{hooper2011,calore2014}.

Each of these GC models will be tested for both $b\bar{b}$ and $\tau^+\tau^-$ channels with the same cross-section used in both cases. However, we note that the best-fit cross-sections for these two channels will typically differ for any given neutralino mass. In the case of GC excess models~\cite{calore2014} the $b\bar{b}$ best-fit model is $\langle \sigma V\rangle \sim 10^{-26}$ cm$^3$ s${-1}$ for $m_{\chi} \sim 50$ GeV. Whereas for $\tau^+\tau^-$ it is $\langle \sigma V\rangle \sim 3 \times 10^{-27}$ cm$^3$ s${-1}$ for $m_{\chi} \sim 10$ GeV. Thus GC region, and its representative points, displayed in Fig.~\ref{fig:Planck} attempts to encompass the whole range of models favoured by the GC excess for $b\bar{b}$ and $\tau^+\tau^-$ channels.

In order to determine the  annihilation cross-section required to match DM emissions to the Reticulum II $\gamma$-ray signal we use the reported $2.3 \sigma$ Fermi-LAT excess and calculate the relative value of the annihilation cross-section $\langle \sigma V \rangle$ by normalising the maximum $\gamma$-ray flux, for a given neutralino mass and from the Reticulum II halo model, to match the $2.3 \sigma$ Fermi-LAT excess at the appropriate observed energy (these values are determined separately for $b\bar{b}$ and $\tau^+\tau^-$ channels).

\section{Multi-frequency Data and Instruments}
\label{sec:data}

Our choice of particular DM dominated cosmic structures is justified by two considerations, the first being the availability of data/limits for optimal comparison to predictions. The second is that emissions of any structure expected to host DM should be compatible with models derived from DM interpretations of observed excesses in other environments. Moreover, we choose environments that are not ideal detection test-beds in order to strengthen our conclusions, as if a neutralino model fails to accommodate existing data from non-ideal detection environments then this serves as stronger evidence against it.

For the Coma cluster we use the diffuse radio data set from~\cite{coma-radio2003}, total X-ray flux data~\cite{coma-xray2004} from the ASCA experiment~\cite{asca-docs} in the 2-10 keV band, as well as from INTEGRAL~\cite{integral} in the 20-50 keV band~\cite{coma-xray2005}. These X-ray sources were selected from the NED SED builder~\cite{ned}, under the requirement they were broad-band measurements yielding a total flux integrated over a map of the target. In addition to this, we use direct Fermi-LAT limits on Coma $\gamma$-ray emission~\cite{fermicoma2015}, and those derived from stacked cluster analysis of Fermi-LAT data~\cite{Fermiclusters2014}. These are compared to the neutralino-induced emission over the entire virial radius, due to the extended nature of these limits. The neutralino predictions can be tested against this data, as if they exceed either the measured points or limits then the model is unviable within the Coma environment.

As the radio data for Coma prove highly important in this work, we must note that there are difficulties in the precise measurement of a diffuse radio flux from an extended target like Coma. This is because such diffuse emission can only be found by subtracting out identified point sources and known extended radio sources (like e.g. radio-galaxies). Thus the accuracy of such a flux determination is dependent on the ability of the instrument to resolve point and extended astrophysical sources within the cluster environment.


For the M81 galaxy we make use of an SED composed of data points, from broad-band measurements only, where the flux is integrated over a map of M81~\cite{m81sed1,m81sed2,m81sed3,m81sed4,m81sed5,m81sed6,m81sed7,m81sed8,m81sed9,m81sed10,m81sed11,m81sed12,m81sed13,m81sed14,m81sed15,m81sed16}. These points were selected by hand using the ASDC and NED SED builders~\cite{asdc,ned}, on the criterion that they are integrated over a map and not limited to an aperture area. The chosen points span the radio and far infra-red spectrum, with a slight incursion into soft X-rays~\cite{young2007,smith1985} from Chandra~\cite{chandra-docs1,chandra-docs2} and EXOSAT~\cite{exosat-docs}, respectively.

For the Draco dwarf we make use of the VLA radio limit~\cite{vladraco}, integrated over a 4$^{\prime}$ $\times$ 4$^{\prime}$ area around the centre of the galaxy (comparing it to DM synchrotron emission within the same region). In addition we use Fermi-LAT upper limits from dwarf observations on Draco~\cite{Fermidwarves2014} compared to the flux integrated over the virial radius of Draco. Given the appropriate area of flux integration, it is clear that neutralino emission predictions cannot exceed these upper limits in Draco.

We note that the ASTRO-H telescope has a 34$^{\prime}$ $\times$ 34$^{\prime}$ field of view in the soft X-ray band while in harder X-rays it is limited to 9$^{\prime}$ $\times$ 9$^{\prime}$~\cite{astroH}. We will take this into account by limiting the area of flux integration appropriately when comparing the ASTRO-H sensitivities to the model predictions. The size of the soft X-ray FOV, when compared to DM-induced surface brightness profiles, means we will use the virial radius in this spectral region (as the FOV captures upwards of 95\% of the flux). This consideration is unnecessary for SKA which has a field of view of at least one square degree. 

In all of the following comparisons we will show point-source sensitivities for the considered instruments (we note that extended source sensitivities are not officially available for the SKA or ASTRO-H, so we use the point-source information as a benchmark), typically at 1000 hours of observation, but for Fermi-LAT we use the 10 year sensitivity for P8R2\_SOURCE\_V6~\cite{Fermidetails}. For comparisons with SKA and ASTRO-H in M81 and Coma we will be accounting for observed backgrounds, thus we are comparing SKA sensitivity to DM-induced fluxes against these backgrounds. This approach is in accordance with the fact that NFW halos are point sources within 1$^{\prime}$ (resolution of ASTRO-H is around 1.5$^{\prime}$~\cite{astroH}), for Burkert halos there will be some extension of the source but ASTRO-H/SKA extended sensitivities are unavailable as yet and CTA is largely affected only in the upper regions of its observation window~\cite{silverwood2014} (which are not reached with the studied models). 

\section{Multi-frequency analysis}
\label{sec:results}

In this section we present the results of our multi-frequency analysis, discussing separately each one of the cosmic environments we consider.

\subsection{Coma Cluster}
\label{sec:coma}

\begin{figure}[htbp]
\centering
\includegraphics[scale=0.6]{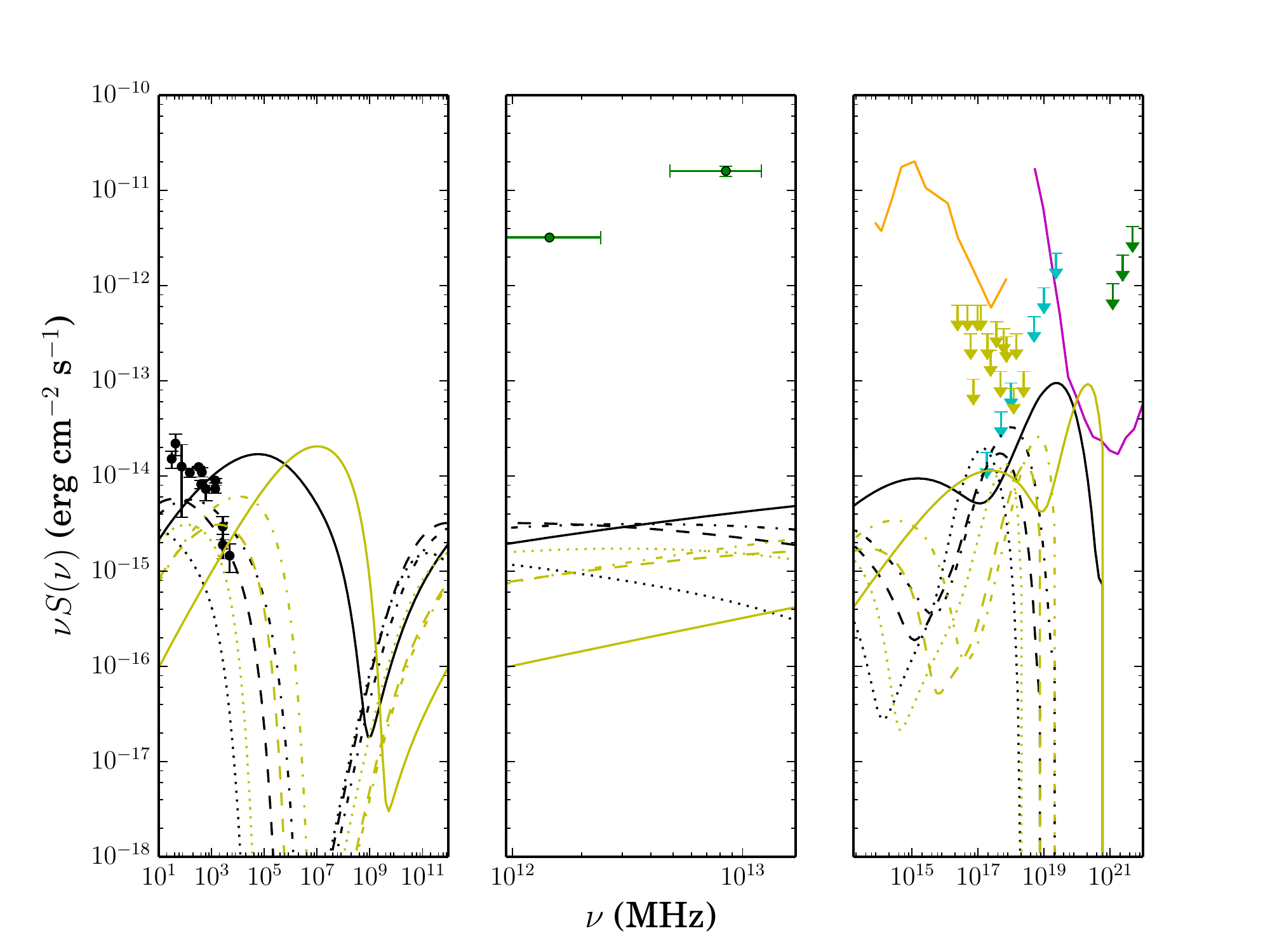}
\includegraphics[scale=0.6]{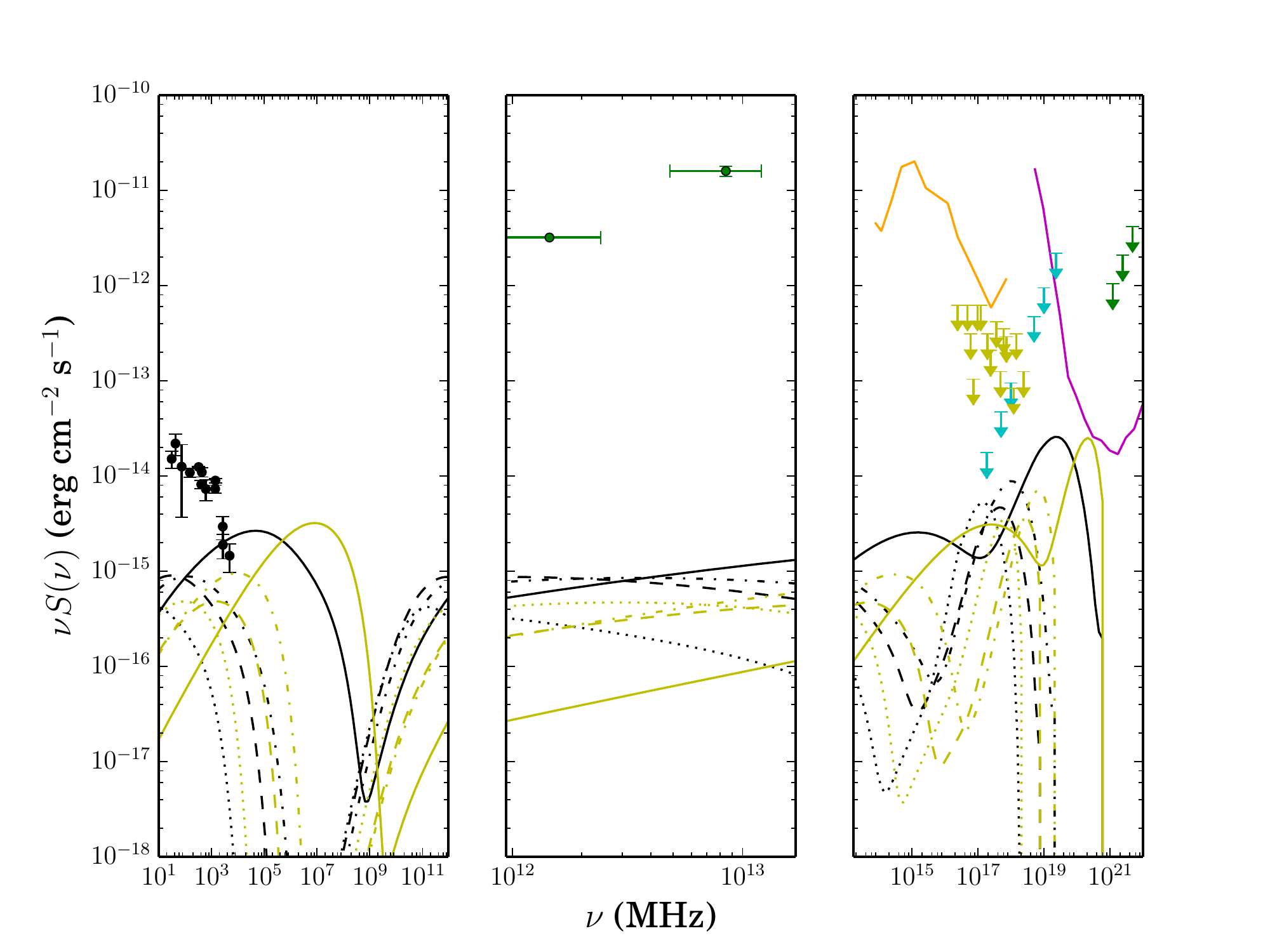}
\caption{Dark matter annihilation spectra for the Coma cluster with best-fit cross-sections from Section~\ref{sec:neutralino}. Black lines indicate predicted spectra for $b\bar{b}$, while yellow correspond to $\tau^+\tau^-$, with the solid curve corresponding to the AFP model, the dash-dotted, dashed, and dotted curves correspond to maximal, median, and minimal GC models respectively. The solid pink curve corresponds to the 1000 hours sensitivity of the CTA~\cite{funk-cta2013}. The black points correspond to the coma radio data~\cite{coma-radio2003}, green points are X-ray data from \cite{coma-xray2004,coma-xray2005}, the the cyan arrows to the Fermi-LAT stacked cluster limit~\cite{Fermiclusters2014}, yellow are Fermi-LAT Coma limits~\cite{fermicoma2015}, while green arrows are the HESS Coma limit~\cite{hesscoma}. The solid orange curve is the ASTROGAM 1-year sensitivity~\cite{astrogam}. Upper panel: halos use NFW profile. Lower panel: halos use Burkert profile. All fluxes are integrated over the virial radius. The central frequency windows cover the ASTRO-H frequency range.}
\label{fig:coma}
\end{figure}

For the Coma cluster we begin by presenting the upper panel of Figure~\ref{fig:coma}, which shows the multi-frequency spectra for the considered neutralino models. It is clear that the shape of the observed synchrotron spectrum in Coma is incompatible with the predictions of the AFP model because these exceed the Coma radio halo flux and also show a spectral flattening for $\nu > 1$ GHz which is not observed. 

The maximal and median GC models ($\tau^+\tau^-$ only for median GC) also predict unobserved flux excesses as well as spectral flattening for $\nu > 1$ GHz. The aforementioned models also have issues with their amplitude exceeding the observed spectrum. However, we find that the synchrotron flux amplitude and slope are not an issue for the minimal GC model. Only the GC minimal model with $b\bar{b}$ is in tension with the Fermi-LAT stacked cluster limit.

In the lower panel of Fig.~\ref{fig:coma} we see that only the AFP $b\bar{b}$ model conflicts with the radio data when a Burkert profile is used. However, this halo profile does lead to sub-dominant DM emissions. In general the Burkert profile will be seen to reduce the flux by more than an order of magnitude at all frequencies.

\begin{figure}[htbp]
\centering
\includegraphics[scale=0.6]{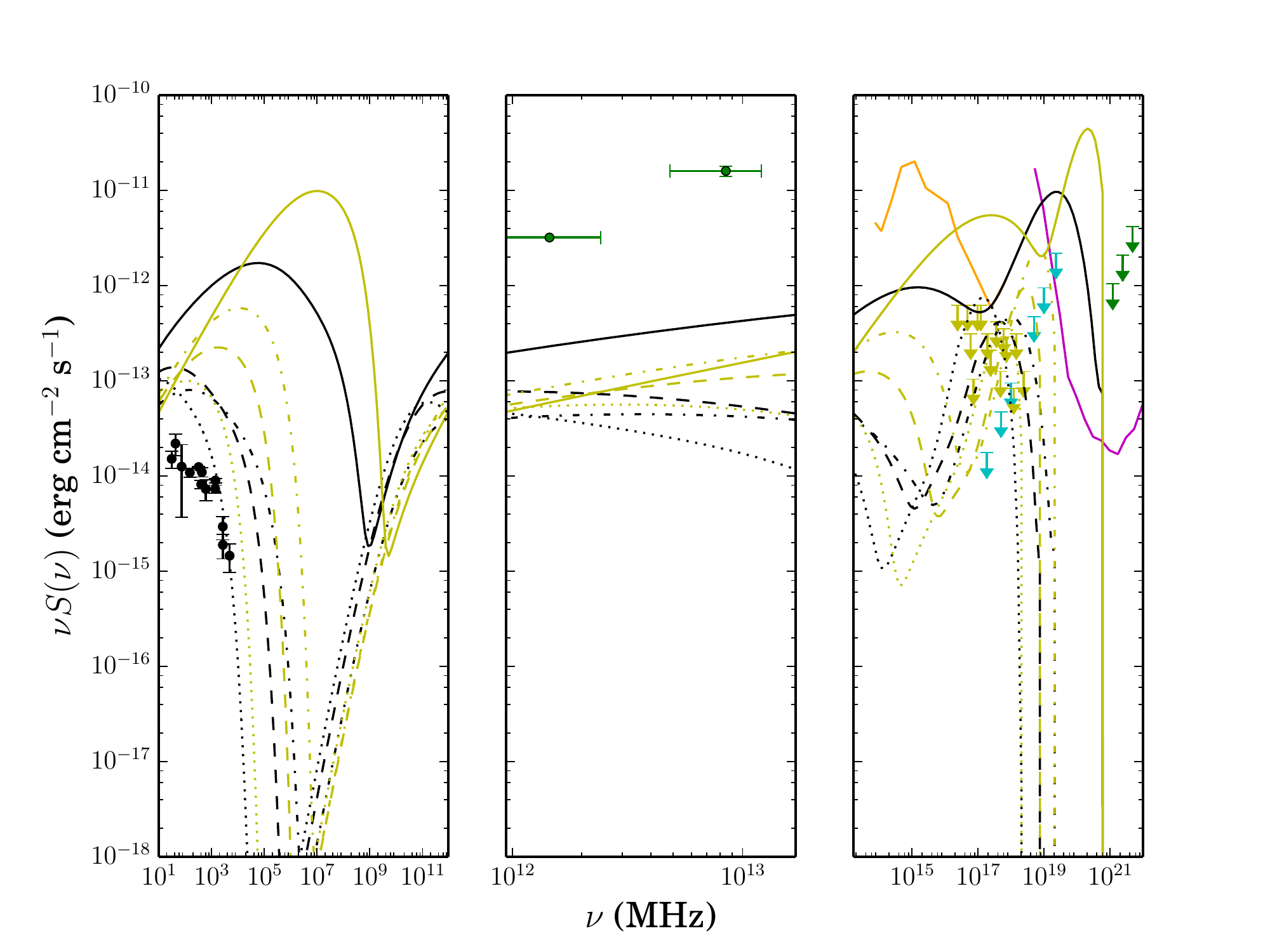}
\includegraphics[scale=0.6]{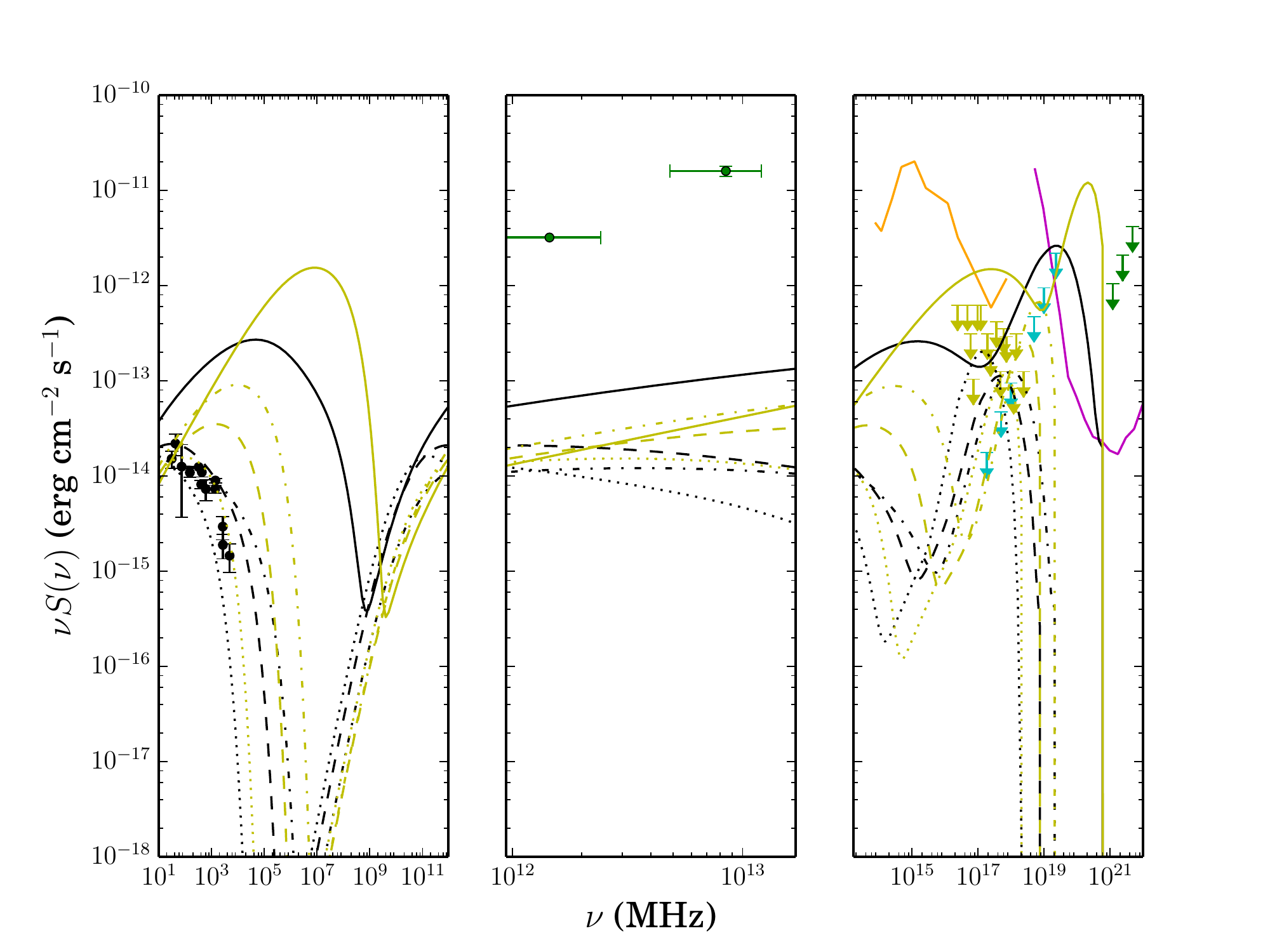}
\caption{Dark matter $b\bar{b}$ annihilation spectra for the Coma cluster with cross-sections determined from Reticulum II excess as detailed in text. Solid curves correspond to the 3 TeV model, the dash-dotted, dashed, and dotted curves correspond to 10, 40, and 100 GeV models respectively. The solid pink curve corresponds to the 1000 hours sensitivity of the CTA~\cite{funk-cta2013}. The black points correspond to the coma radio data~\cite{coma-radio2003}, green points are X-ray data from \cite{coma-xray2004,coma-xray2005}, the cyan arrows to the Fermi-LAT stacked cluster limit~\cite{Fermiclusters2014}, while yellow are Fermi-LAT Coma limits~\cite{fermicoma2015}. Green arrows are the HESS Coma limit~\cite{hesscoma}. Upper panel: halos use NFW profile. Lower panel: halos use Burkert profile. All fluxes are integrated over the virial radius. The central frequency windows cover the ASTRO-H frequency range.}
\label{fig:coma-2}
\end{figure}

\begin{figure}[htbp]
\centering
\includegraphics[scale=0.6]{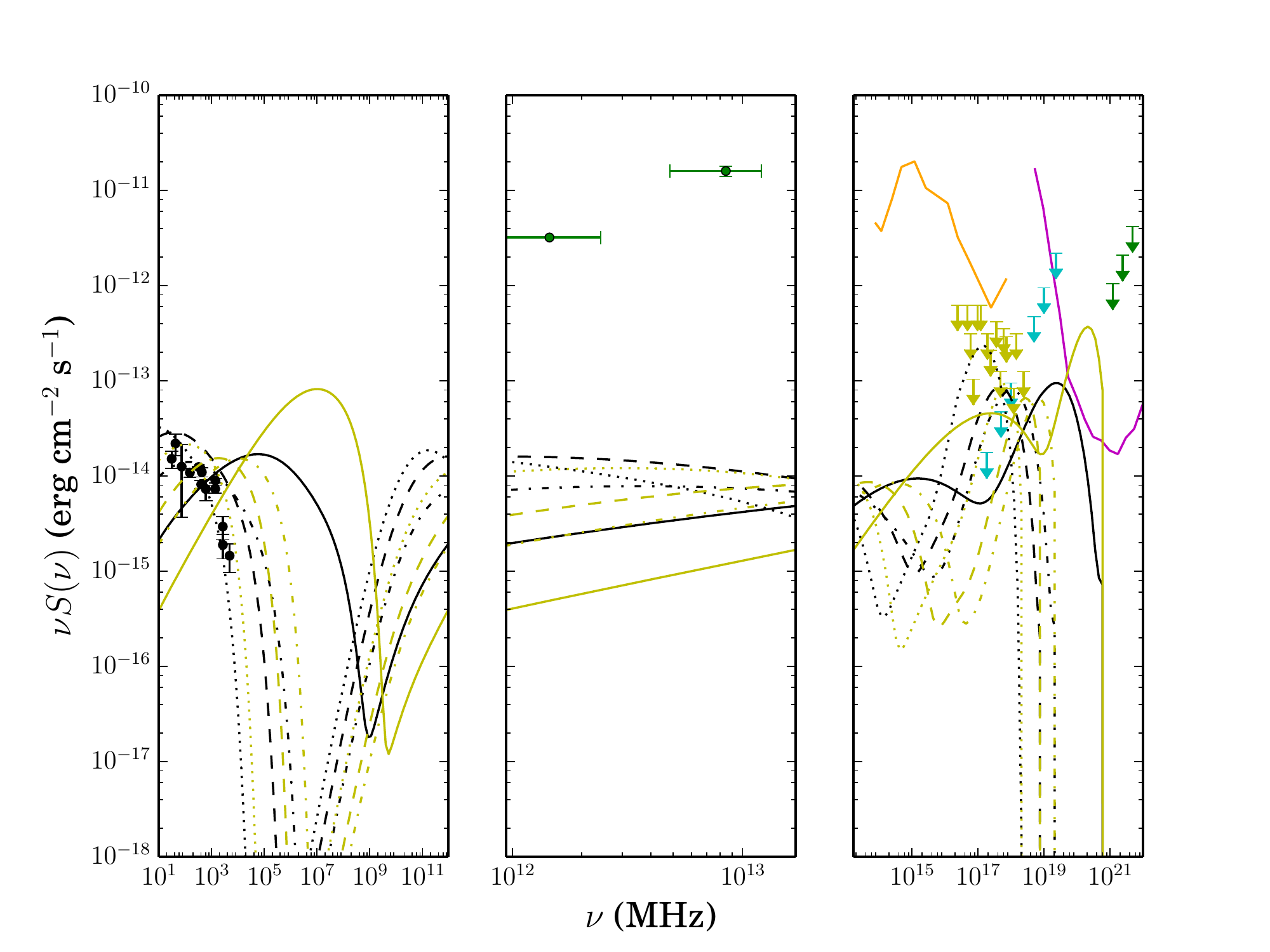}
\includegraphics[scale=0.6]{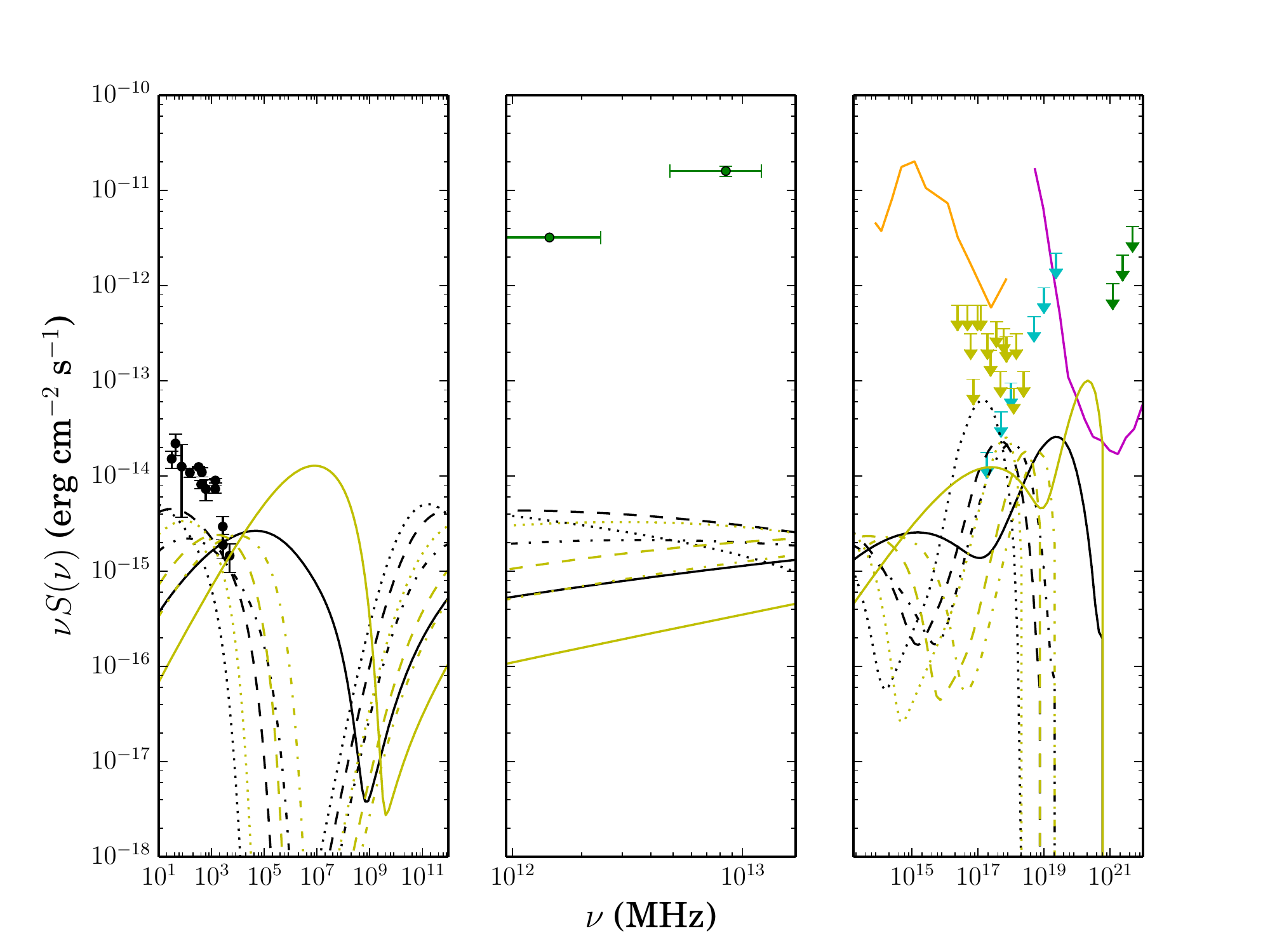}
\caption{Dark matter $b\bar{b}$ annihilation spectra for the Coma cluster with cross-sections from Fermi-LAT dwarf limits. Solid curves correspond to the 3 TeV model, the dash-dotted, dashed, and dotted curves correspond to 10, 40, and 100 GeV models respectively. The solid pink curve corresponds to the 1000 hours sensitivity of the CTA~\cite{funk-cta2013}. The black points correspond to the coma radio data~\cite{coma-radio2003}, green points are X-ray data from \cite{coma-xray2004,coma-xray2005}, the cyan arrows to the Fermi-LAT stacked cluster limit~\cite{Fermiclusters2014}, while yellow are Fermi-LAT Coma limits~\cite{fermicoma2015}. Green arrows are the HESS Coma limit~\cite{hesscoma}. Upper panel: halos use NFW profile. Lower panel: halos use Burkert profile. All fluxes are integrated over the virial radius. The central frequency windows cover the ASTRO-H frequency range.}
\label{fig:coma-3}
\end{figure}

\begin{figure}[htbp]
\centering
\includegraphics[scale=0.8]{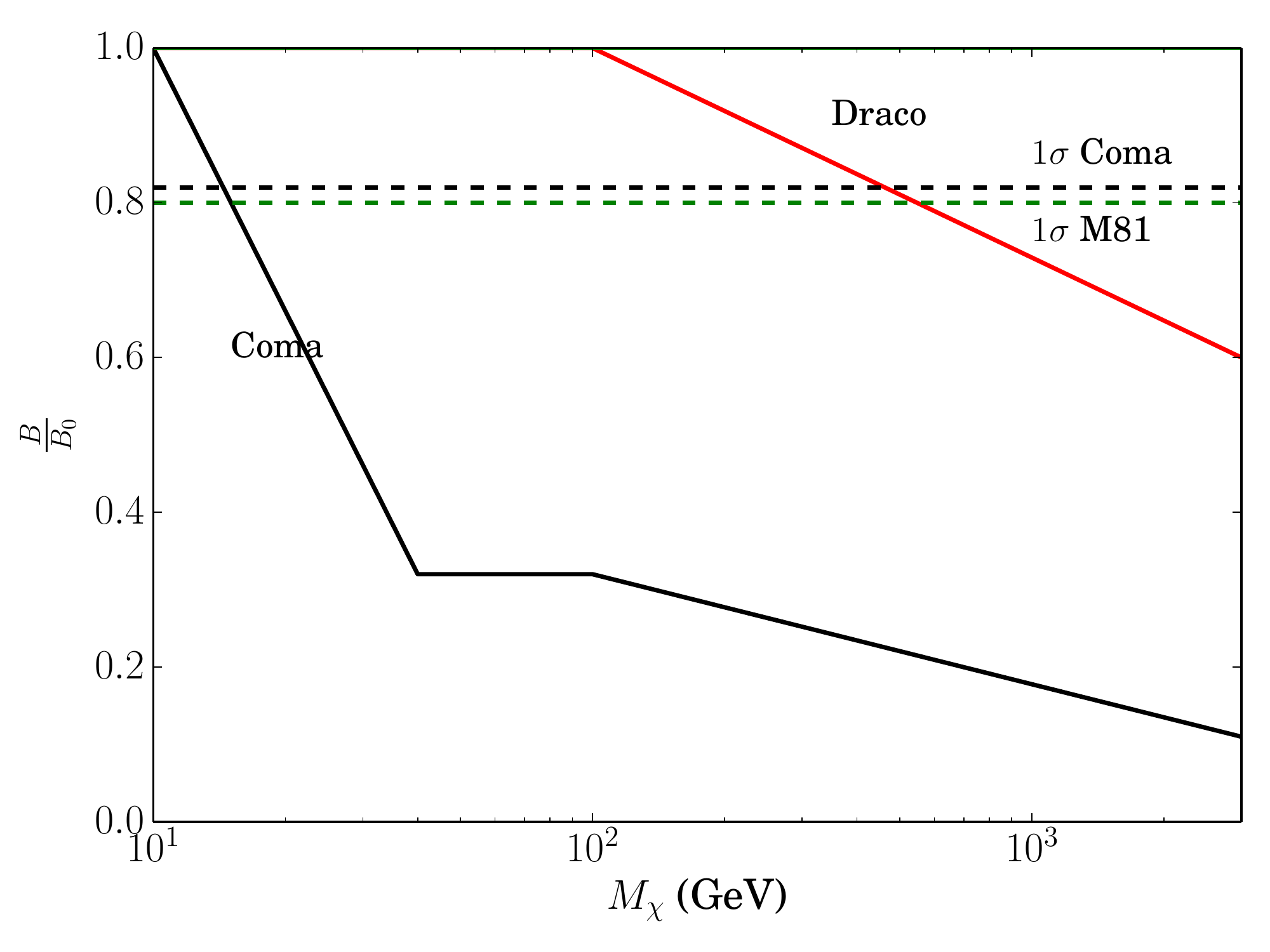}
\caption{Magnetic field strength reduction factor required to bring spectra in Figs.~\ref{fig:coma-2}, \ref{fig:m81-2}, and \ref{fig:draco-2} into consistency with available radio data. Coma is plotted in black, M81 in green, and Draco in red. The dashed lines represent the $1\sigma$ error range for the magnetic field values quoted in Section~\ref{sec:models}. The data is plotted only for points corresponding to the neutralino masses studied here.}
\label{fig:bfields}
\end{figure}

\begin{table}[htbp]
\label{tab:bfields}
\centering
\begin{tabular}{|l|l|l|l|l|}
\hline
Halo & 10 GeV & 40 GeV & 100 GeV & 3 TeV \\
\hline
Coma & - & $\sim 3.5\sigma$ & $\sim 3.5 \sigma$ & $\sim 4\sigma$ \\
M81 & - & - & - & - \\
\hline
\end{tabular}
\caption{Magnetic field deviations needed to keep neutralino $b\bar{b}$ models with given masses, and Fermi-LAT dwarf cross-sections, consistent with available synchrotron data.}
\end{table}

In terms of differentiating between neutralino spectra, substantial differences are apparent between the spectral slopes of different annihilation channels in both the radio (SKA) and X-ray (ASTRO-H) observation windows for all of the models. Notably the $\tau^+\tau^-$ spectra are of lower amplitude at low frequencies (with respect to form of emission) and cross over the $b\bar{b}$ spectrum at higher frequencies, resulting in a harder spectrum. This is true for all the models and all forms of emission, and this kind of spectral crossing also extends to the effect of variations in the mass of the neutralino. At low frequencies the spectra of heavy neutralino models fall below their lighter counterparts, which is in complement to the hardening of the spectrum for heavy neutralinos. It is notable that such regions of spectral difference between the GC models fall within the observational window of ASTRO-H (middle figure panels), whether this is observable will be discussed in Section~\ref{sec:constraint}. It seems, therefore, that ASTRO-H will be well positioned to discriminate between the various neutralino compositions. 

We note that the ASTRO-H observation window also encompasses the region of the ICS spectrum that shows significant differences between various choices of neutralino mass and annihilation channel within the Coma cluster. 
Importantly, CTA seems only to be able to observe heavier DM models (like AFP) in this environment, with the GC masses falling below the region of CTA sensitivity.
In the case of ASTROGAM it is evident that it is insufficiently sensitive to observe emissions from the studied DM models, even in the case of a large DM halo like the Coma cluster.

In Figure~\ref{fig:coma-2} we display the predictions of assuming the annihilation cross-section necessary to reproduce the Reticulum II $\gamma$-ray excess~\cite{geringer-sameth2015}. This prediction was derived by assuming a J-factor for Ret. II from~\cite{bonnivard2015} (see Table~\ref{tab:jfac}). and using this to determine a $\gamma$-ray SED. Then we took a $2.3\sigma$ Fermi-LAT excess reported by~\cite{geringer-sameth2015} and normalised our model of Reticulum II so that it is the maximal $\gamma$-ray flux matching the Fermi-LAT excess. The cross-sections thus derived are listed in order of model mass as: $1.9 \times 10^{-26}$ cm$^{3}$ s$^{-1}$, $4.8 \times 10^{-26}$ cm$^{3}$ s$^{-1}$, $1.4 \times 10^{-25}$ cm$^{3}$ s$^{-1}$, and $1.0 \times 10^{-22}$ cm$^{3}$ s$^{-1}$.
When the model is applied to other DM halos, we see that in the radio and $\gamma$-ray frequency ranges the predicted fluxes for this models greatly exceed the known measurements/limits for the Coma cluster. It is evident then that the consequences of a DM interpretation of the excess $\gamma$-rays in the Reticulum II dwarf galaxy are unacceptable for the considered neutralino mass range (10-3000 GeV) in the case of Coma. We note that the use of the Burkert profile in the lower panel only removes conflicts between the 10 GeV mass and radio data, while $\gamma$-ray incompatibility remains for all models.
In conclusion, the available SED of Coma discards the DM interpretation of the Reticulum II $\gamma$-ray excess in the case that the same DM model is responsible for the formation of the halo of dwarf galaxies and galaxy clusters.

Since the previous results have been obtained with a NFW profile  and with the relative boost factor (as described in Sect. 2), we show in Figure~\ref{fig:coma-3} a conservative version of Fig.~\ref{fig:coma-2}, which considers the same set of neutralino masses but with the cross-sections derived by the Fermi-LAT collaboration from dwarf galaxy observations, including Reticulum II~\cite{Fermidwarves2015}. We can see that many of the features we highlight for Fig.~\ref{fig:coma-2} remain in evidence, and particularly the predictions being in excess in the synchrotron spectrum and in violating both Fermi-LAT Coma and stacked cluster limits in $\gamma$-rays. The lower panel shows that the Burkert profile has only the 100 GeV $\tau^+\tau^-$ and 3 TeV $b\bar{b}$ in conflict with the radio data. However, 10 GeV $b\bar{b}$ is in tension with the Fermi-LAT stacked cluster limit in this case.

We also attempted this exercise by assuming a virial mass of $10^6$ M$_{\odot}$ and a distance of 30 kpc for Ret. II, in this case the required cross-sections are far larger and the resulting excesses in Coma are thus greater.
 
Given the sensitivity of synchrotron radiation to magnetic field strength we show in Fig.~\ref{fig:bfields} the factor by which the magnetic field strength must be multiplied in order to return the predictions in Fig.~\ref{fig:coma-3} (for the $b\bar{b}$ cases) to consistency with available radio data; this is also summarised in Table~\ref{tab:bfields}. 
We find that the $4\sigma$ deviation that would be required for the synchrotron spectrum to be accommodated by the data, for all but the $10$ GeV neutralino mass model, demonstrate that the inconsistencies we previously highlight cannot be solely blamed upon magnetic field values. 
In fact, a value as low as $\approx 1$ $\mu$G for Coma is in sharp contrast with the results of Bonafede et al.~\cite{bonafede2010} indicating $B \approx 5$ $\mu$G.
Therefore, we conclude that even the cross-sections derived by the Fermi-LAT Collaboration for dwarf galaxies cannot support an interpretation of a $\gamma$-ray excess being the result of neutralino DM annihilation.\\ 
We note that the Fermi-LAT cross-sections are similar to the value reported in \cite{geringer-sameth2015} for neutralinos around $\approx 40$ GeV, this increases the robustness of our results as the highlighted conflicts with available data will not be exorcised by such a sub-order-of-magnitude cross-section reduction.\\
Finally, we note that recent Fermi-LAT analysis for Coma~\cite{ackermann2015} produces more stringent limits upon the $\gamma$-ray flux and thus will strengthen the results here, as well as exclude any currently marginal cases.

\subsection{M81 Galaxy}
\label{sec:m81}

\begin{figure}[htbp]
\centering
\includegraphics[scale=0.6]{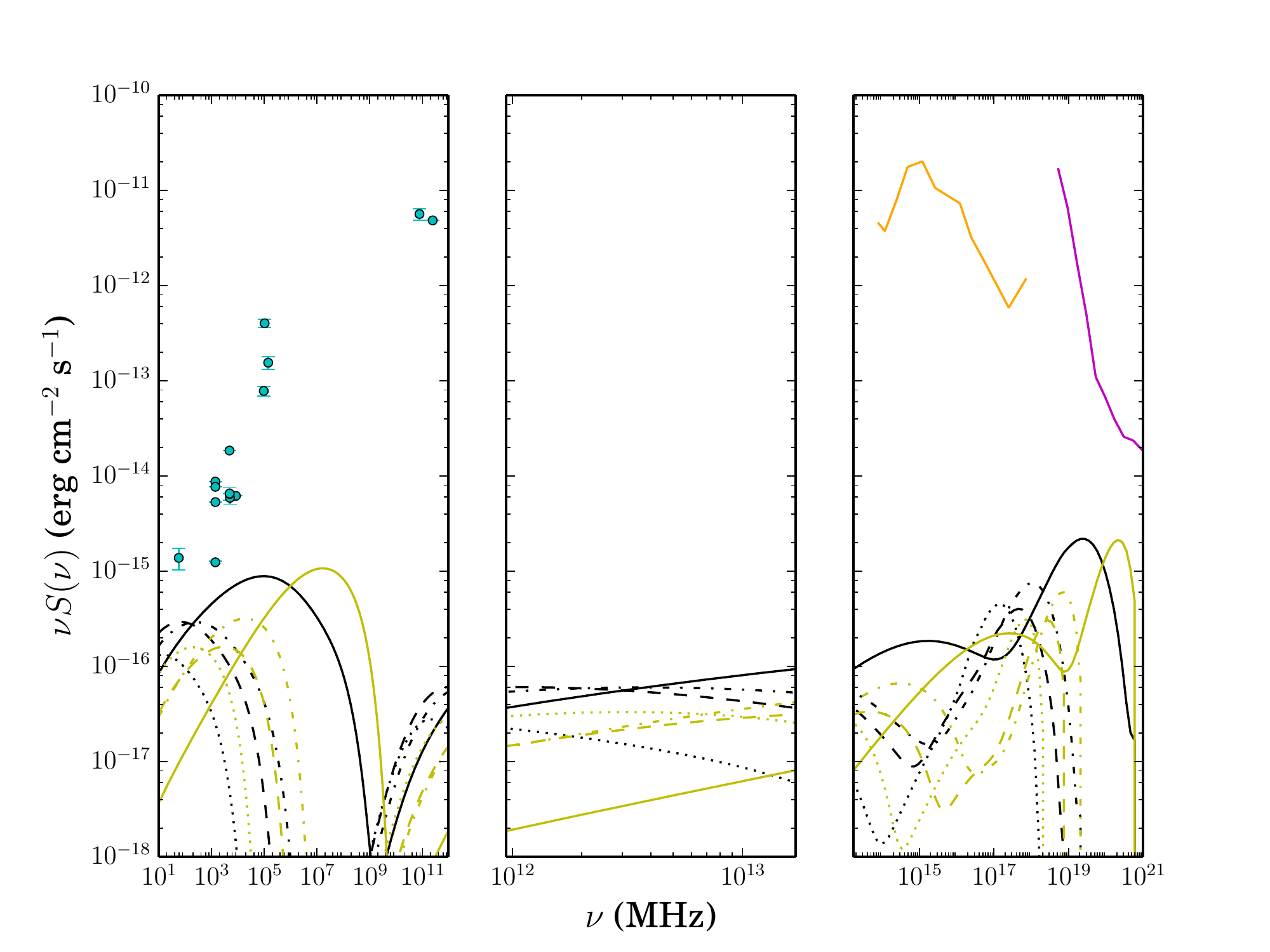}
\includegraphics[scale=0.6]{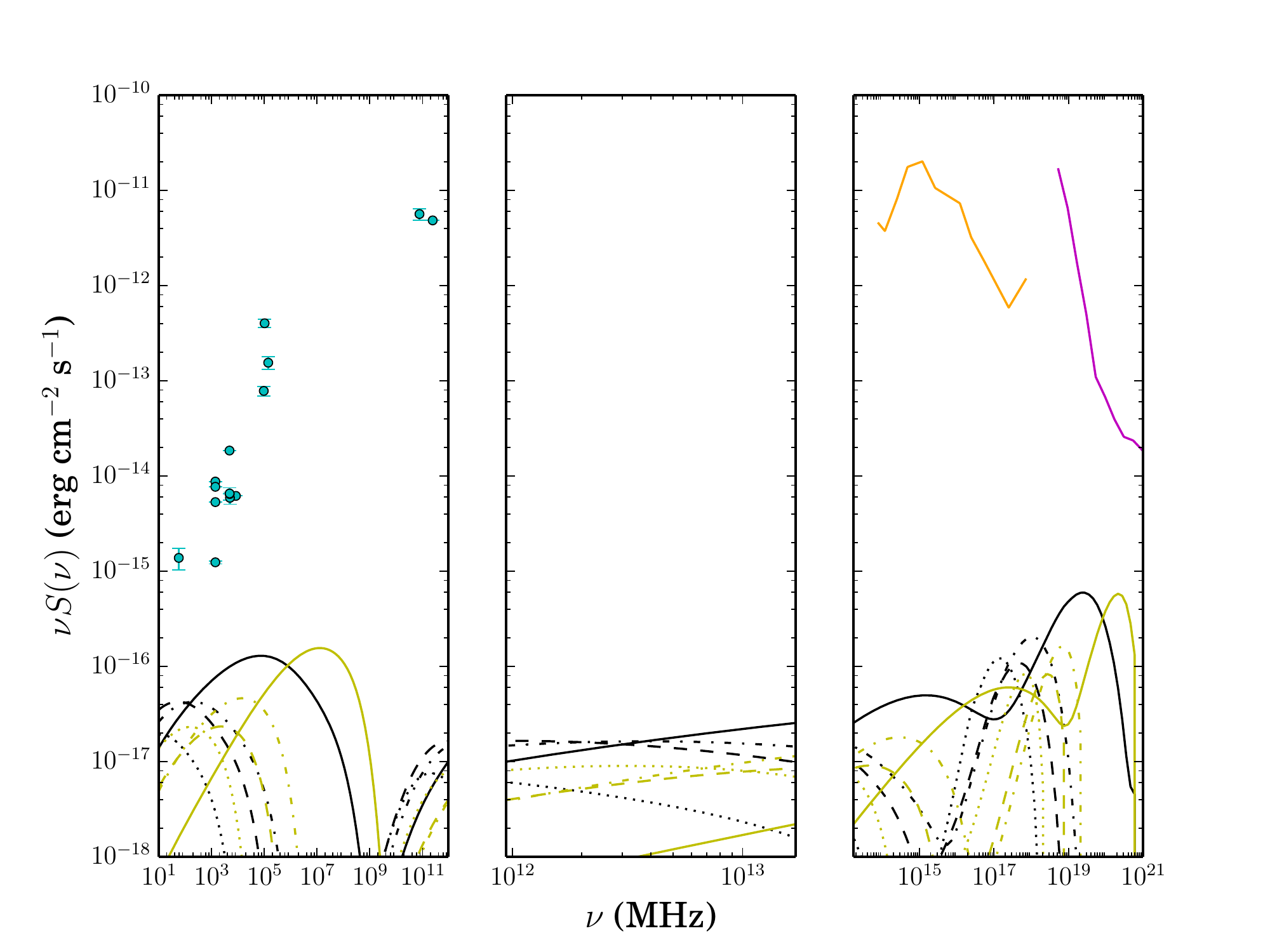}
\caption{Dark matter annihilation spectra for the M81 galaxy with best-fit cross-sections from Section~\ref{sec:neutralino}. Black lines indicate predicted spectra for $b\bar{b}$, while yellow correspond to $\tau^+\tau^-$, with the solid curve corresponding to the AFP model, the dash-dotted, dashed, and dotted curves correspond to maximal, median, and minimal GC models respectively. The solid pink curve corresponds to the 1000 hours sensitivity of the CTA~\cite{funk-cta2013}. Green points correspond to the M81 SED~\cite{m81sed1,m81sed2,m81sed3,m81sed4,m81sed5,m81sed6,m81sed7,m81sed8,m81sed9,m81sed10,m81sed11,m81sed12,m81sed13,m81sed14,m81sed15,m81sed16}. The solid red and blue curves are the 1000 hours SKA-1 and ASTRO-H sensitivities~\cite{ska2012,astroH}. The solid orange curve is the ASTROGAM 1 year sensitivity~\cite{astrogam}. Upper panel: halos use NFW profile. Lower panel: halos use Burkert profile. All fluxes are integrated over the virial radius. The central frequency windows cover the ASTRO-H frequency range.}
\label{fig:m81}
\end{figure}

\begin{figure}[htbp]
\centering
\includegraphics[scale=0.6]{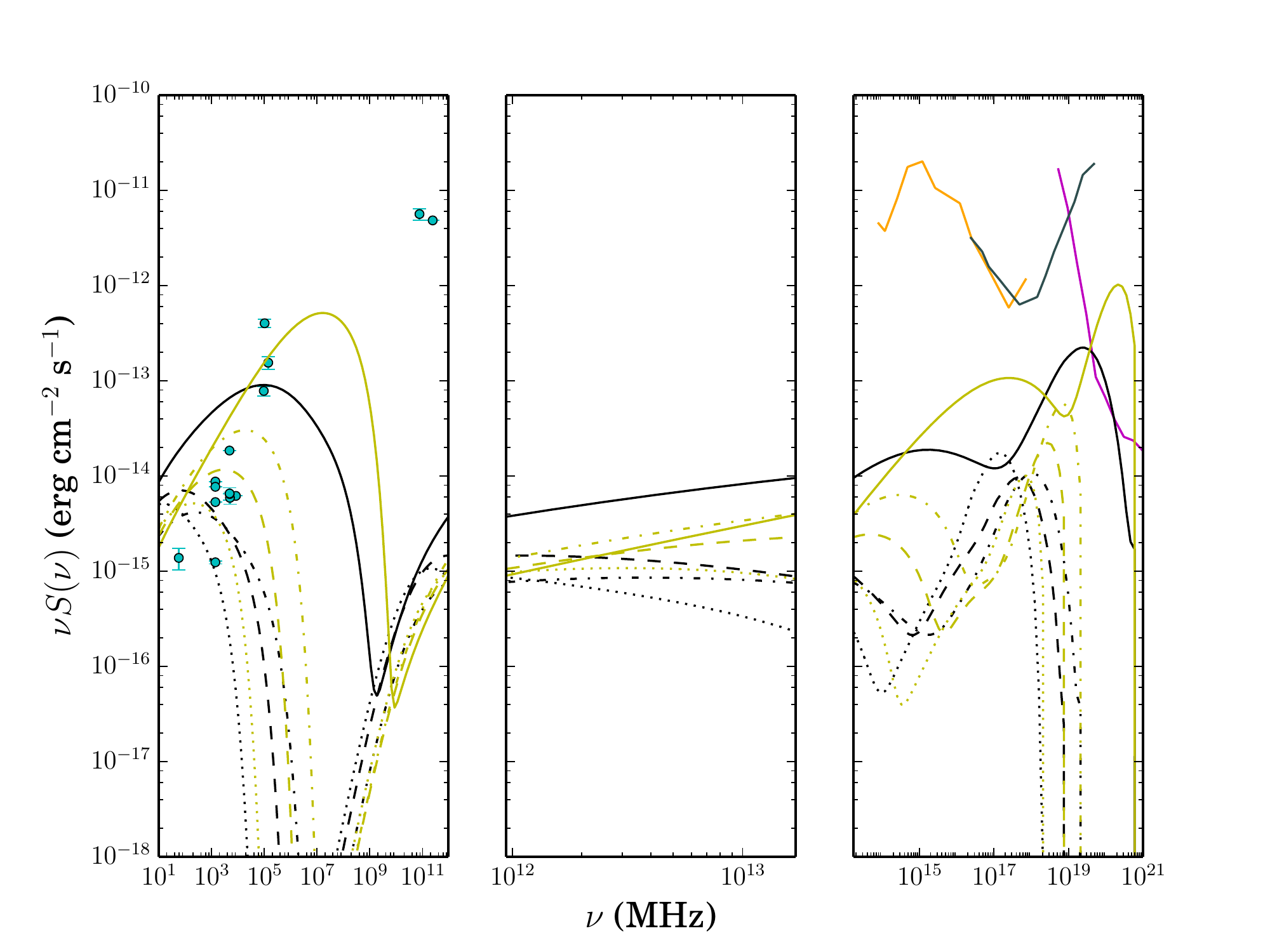}
\includegraphics[scale=0.6]{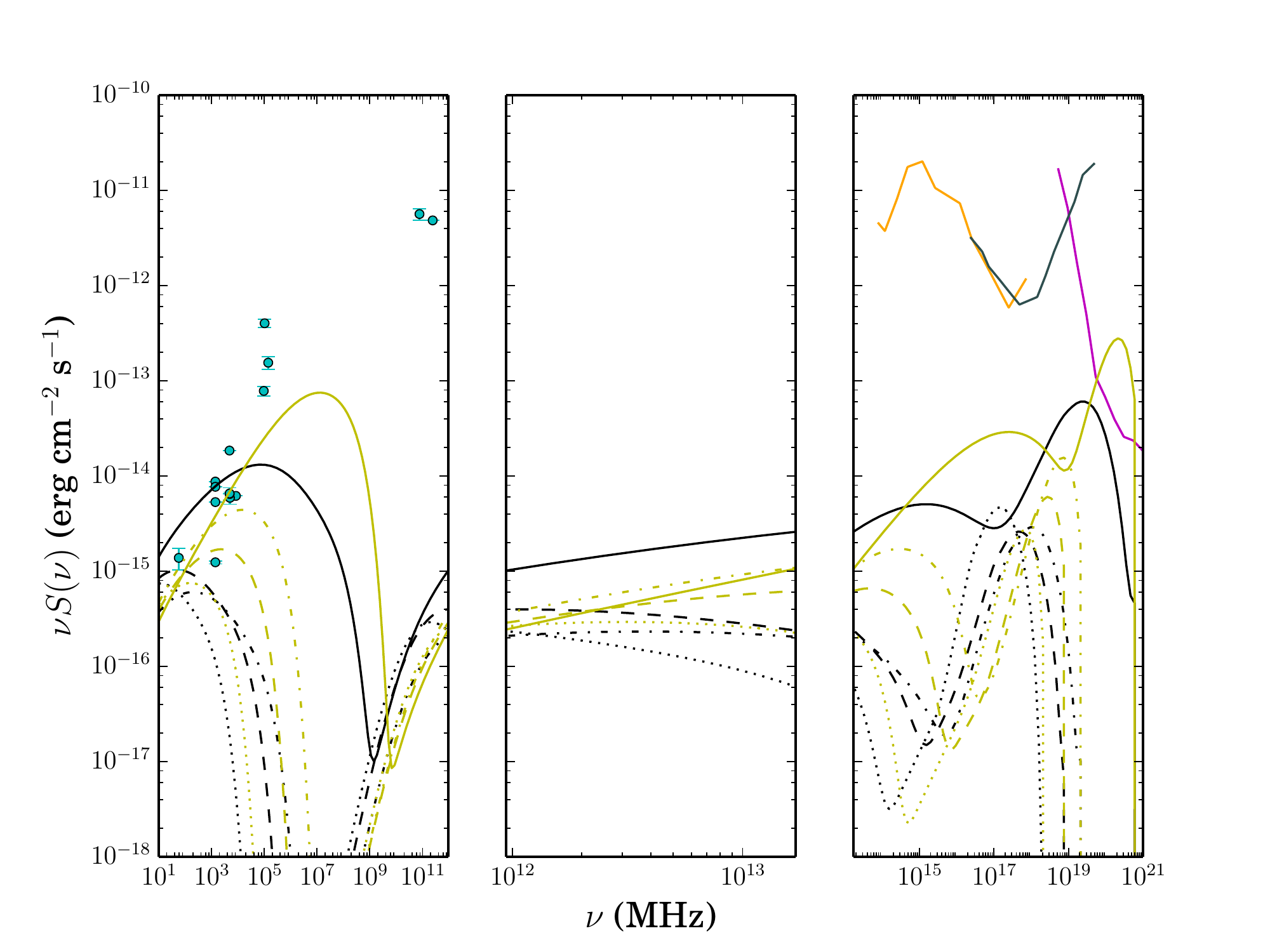}
\caption{Dark matter annihilation spectra for the M81 galaxy with cross-sections determined from Reticulum II excess as detailed in text. Black lines indicate predicted spectra for $b\bar{b}$, while yellow correspond to $\tau^+\tau^-$, with the solid curves corresponding to the 3 TeV model, the dash-dotted, dashed, and dotted curves correspond to 10, 40, and 100 GeV models respectively. The solid pink curve corresponds to the 1000 hours sensitivity of the CTA~\cite{funk-cta2013}, the solid grey curve shows the Fermi-LAT 10 year point sensitivity~\cite{Fermidetails}. Green points correspond to the M81 SED~\cite{m81sed1,m81sed2,m81sed3,m81sed4,m81sed5,m81sed6,m81sed7,m81sed8,m81sed9,m81sed10,m81sed11,m81sed12,m81sed13,m81sed14,m81sed15,m81sed16}. Upper panel: halos use NFW profile. Lower panel: halos use Burkert profile. The left and right panels integrate flux over $R_{vir}$, while the centre does so over a $4.5^{\prime}$ radius.}
\label{fig:m81-2}
\end{figure}

\begin{figure}[htbp]
\centering
\includegraphics[scale=0.6]{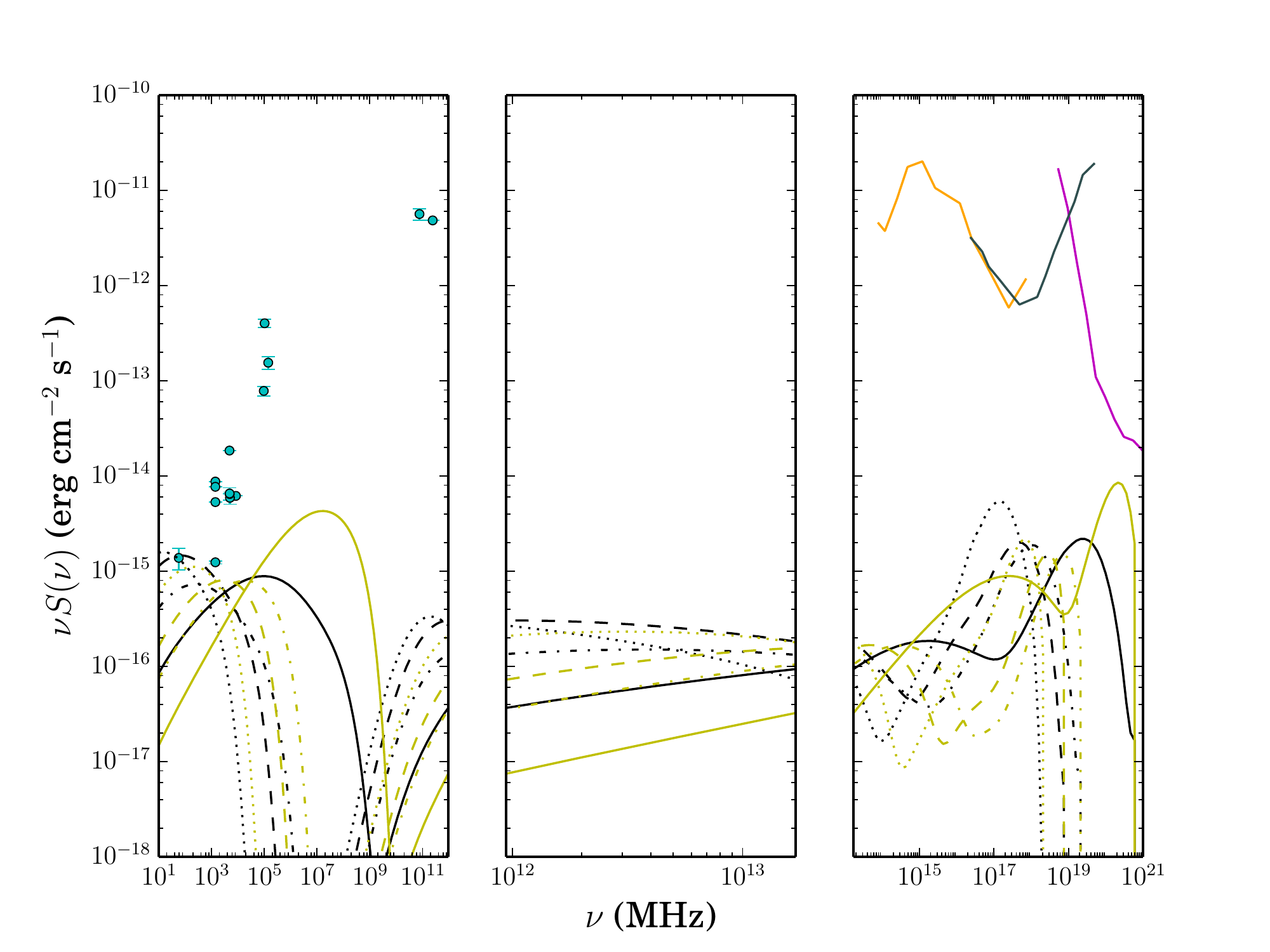}
\includegraphics[scale=0.6]{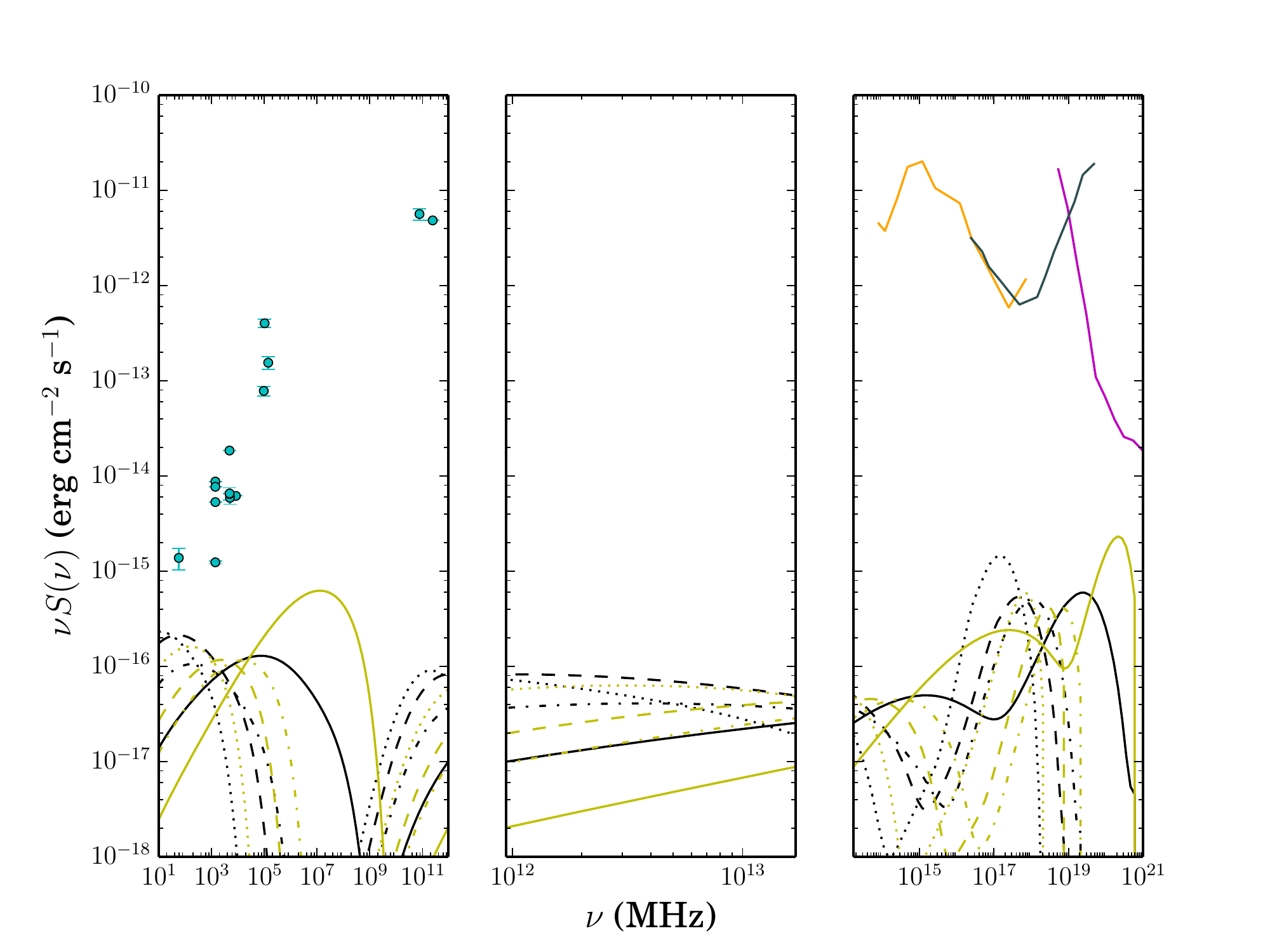}
\caption{Dark matter $b\bar{b}$ annihilation spectra for the M81 galaxy with cross-sections from Fermi-LAT dwarf limits. Solid curves correspond to the 3 TeV model, the dash-dotted, dashed, and dotted curves correspond to 10, 40, and 100 GeV models respectively. The solid pink curve corresponds to the 1000 hours sensitivity of the CTA~\cite{funk-cta2013}, the solid grey curve shows the Fermi-LAT 10 year point sensitivity~\cite{Fermidetails}. Green points correspond to the M81 SED~\cite{m81sed1,m81sed2,m81sed3,m81sed4,m81sed5,m81sed6,m81sed7,m81sed8,m81sed9,m81sed10,m81sed11,m81sed12,m81sed13,m81sed14,m81sed15,m81sed16}. Upper panel: halos use NFW profile. Lower panel: halos use Burkert profile. All fluxes are integrated over the virial radius. The central panel covers the ASTRO-H frequency range.}
\label{fig:m81-3}
\end{figure}

Figure~\ref{fig:m81} shows the spectrum expected in the given DM models for the M81 galaxy environment. CTA observation of the predicted DM models are less likely from this source, even with the super-TeV masses, and its study is complicated by the extreme dominance of its active nucleus over a broad range of frequencies. As in the Coma environment, the ASTRO-H window covers several ICS spectral crossings between differing neutralino masses and annihilation channels. In the radio frequency range it is notable that, regardless of halo profile, there are no conflicts between any of the models and the data. 

Figure~\ref{fig:m81-2} shows the consequences of assuming the Reticulum II DM annihilation cross-section. All of the neutralino masses are incompatible with the available radio measurements.
The DM models would be observable by Fermi-LAT for $\sim 10$ year observations at $E > 0.04$ GeV and the 3 TeV mass neutralino should produce a sufficiently high flux for CTA observation with $\sim 1000$ hours exposure.
Despite the large flux produced by the active nucleus of M81, the Reticulum II excess annihilation cross-section proves to be incompatible with existing measurements of M81. When the Burkert profile is used in Figure~\ref{fig:m81-2} we see that 3 TeV models as well as $\tau^+\tau^-$ with 100 and 40 GeV masses remain in tension with the data. The case of particular relevance is the best-fit Reticulum II model of $m_\chi \sim 40$ GeV and $\langle \sigma V\rangle \sim 3 \times 10^{-26}$ cm$^3$ s$^{-1}$, which is excluded by the data for an NFW profile and is marginal with $\tau^+\tau^-$ annihilation and a Burkert halo.

In Figure~\ref{fig:m81-3} we show the consequences of the cross-sections derived from Fermi-LAT dwarf studies including Reticulum II. In the case there are once again no conflicts between the data and the models.

\subsection{Draco dwarf galaxy}

\begin{figure}[htbp]
\centering
\includegraphics[scale=0.6]{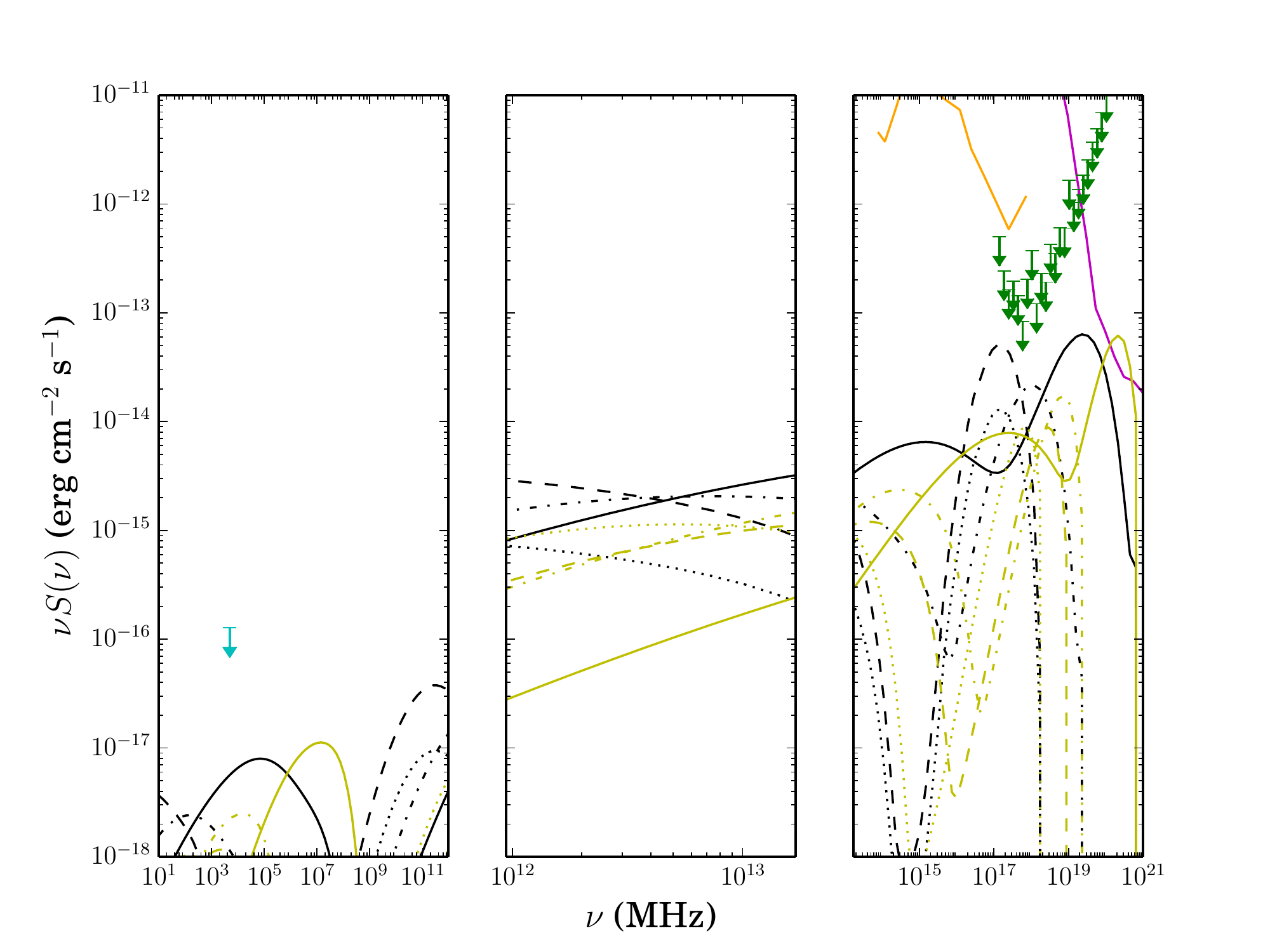}
\includegraphics[scale=0.6]{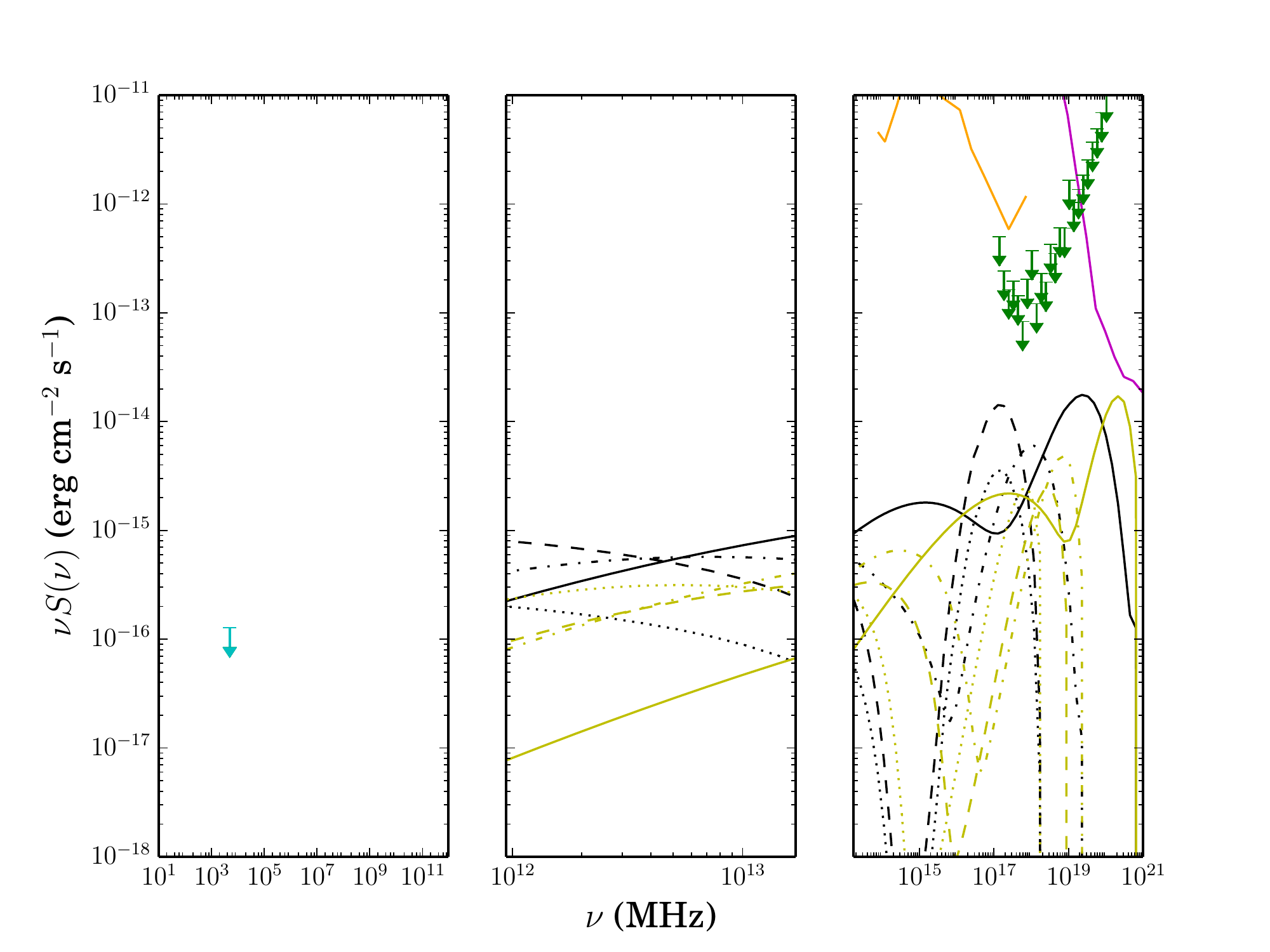}
\caption{Dark matter annihilation spectra for the draco dwarf galaxy with best-fit cross-sections from Section~\ref{sec:neutralino}. Black lines indicate predicted spectra for $b\bar{b}$, while yellow correspond to $\tau^+\tau^-$, with the solid curve corresponding to the AFP model, the dash-dotted, dashed, and dotted curves correspond to maximal, median, and minimal GC models respectively. The solid pink curve corresponds to the 1000 hours sensitivity of the CTA~\cite{funk-cta2013}. Green arrows indicate the upper limits set by the Fermi-LAT observations~\cite{Fermidwarves2014}, while the cyan arrow corresponds to the VLA limit~\cite{vladraco}. The solid red and blue curves are the 1000 hours SKA-1 and ASTRO-H sensitivities~\cite{ska2012,astroH}. The solid orange curve is the ASTROGAM 1 year sensitivity~\cite{astrogam}. Upper panel: halos use NFW profile. Lower panel: halos use Burkert profile. The left panel integrates flux over a $2^{\prime}$ radius, other panels have fluxes integrated over the virial radius. The central panel covers the ASTRO-H frequency range.}
\label{fig:draco}
\end{figure}

\begin{figure}[htbp]
\centering
\includegraphics[scale=0.6]{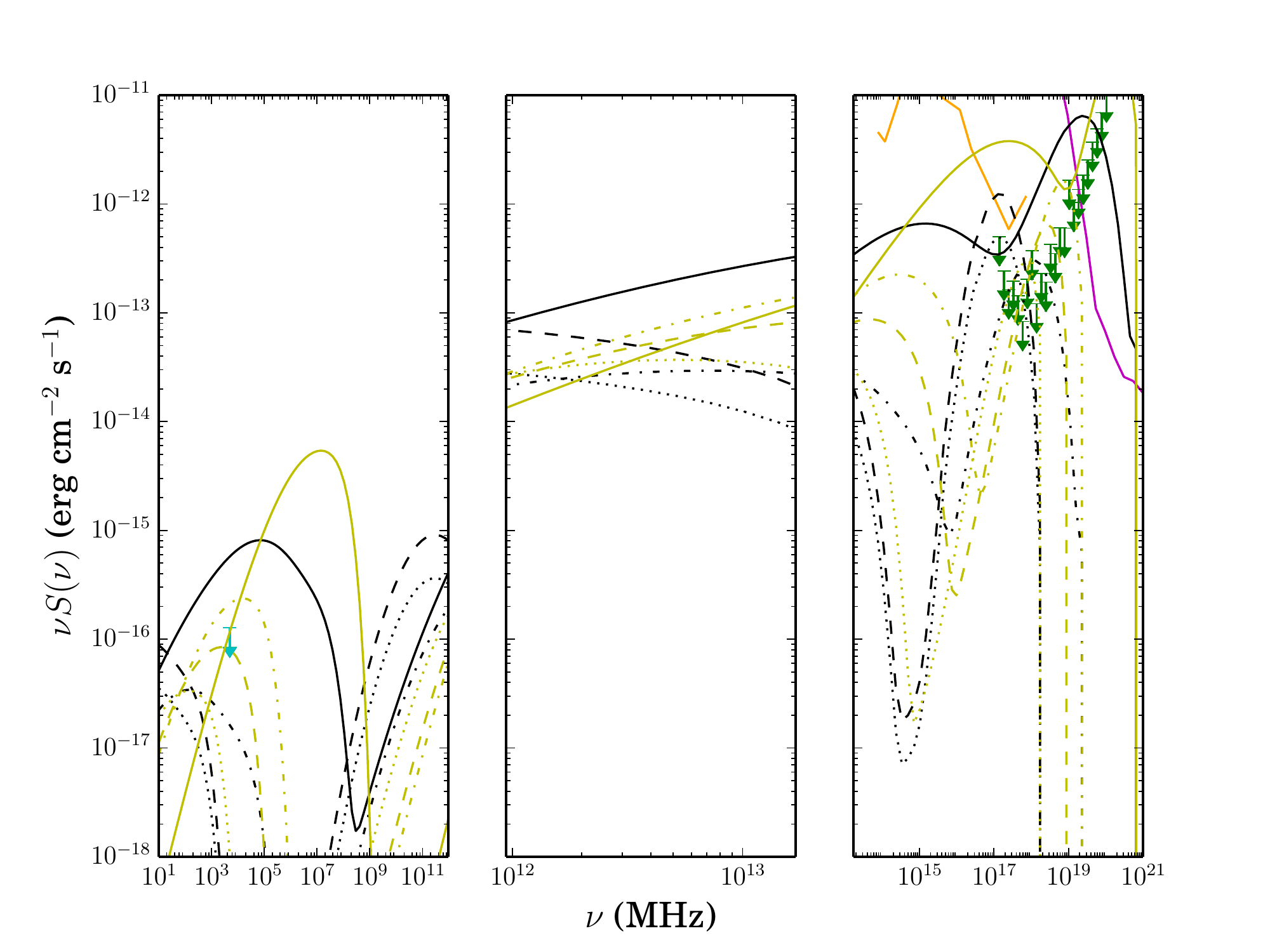}
\includegraphics[scale=0.6]{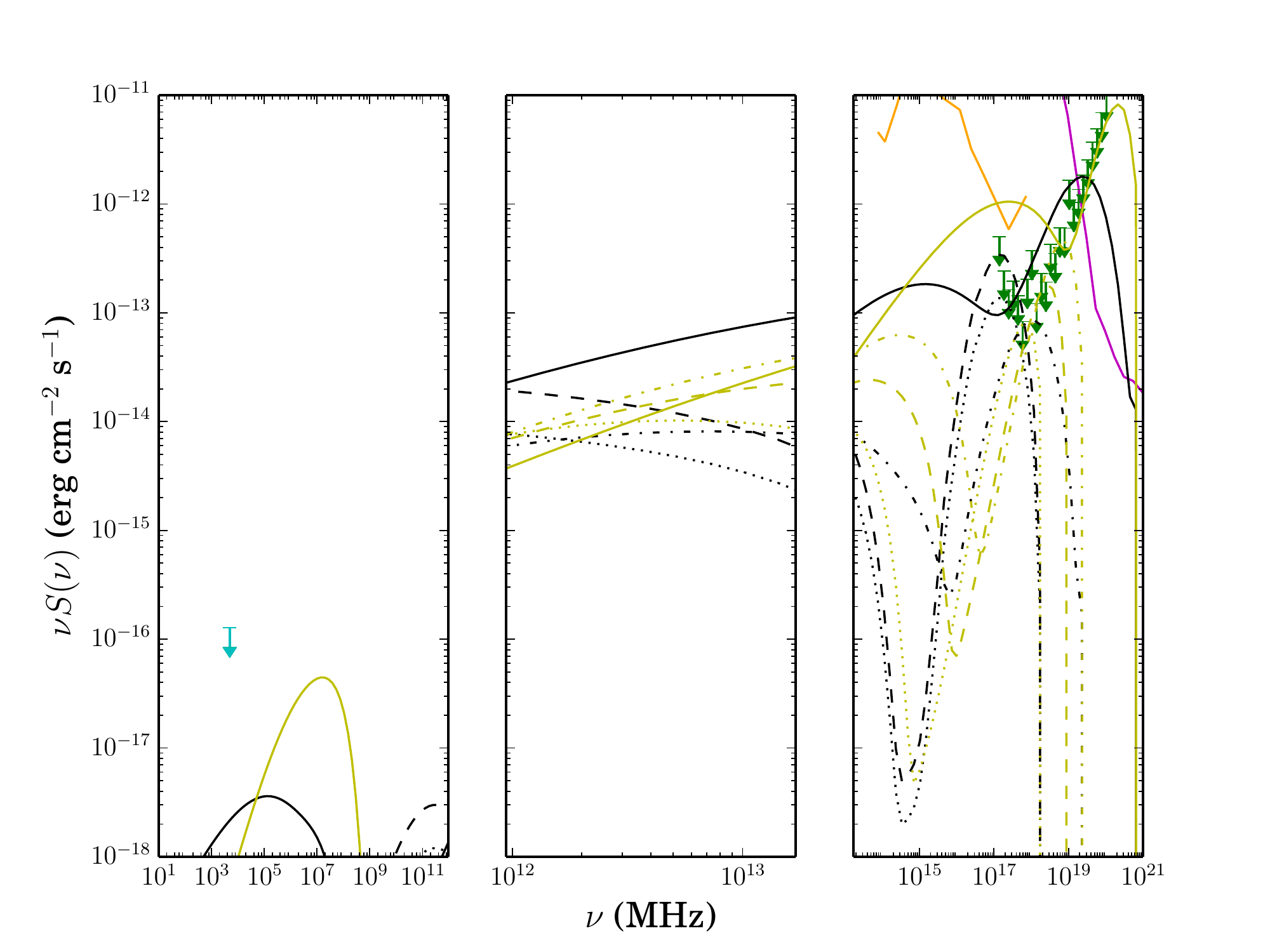}
\caption{Dark matter annihilation spectra for the draco dwarf galaxy with cross-sections determined from Reticulum II excess as detailed in text. Black lines indicate predicted spectra for $b\bar{b}$, while yellow correspond to $\tau^+\tau^-$, with the solid curve corresponding to the 3 TeV model, the dash-dotted, dashed, and dotted curves correspond to 10, 40, and 100 GeV models respectively. The solid pink curve corresponds to the 1000 hours sensitivity of the CTA~\cite{funk-cta2013}. Green arrows indicate the upper limits set by the Fermi-LAT observations~\cite{Fermidwarves2014}, while the cyan arrow corresponds to the VLA limit~\cite{vladraco}. Upper panel: halos use NFW profile. Lower panel: halos use Burkert profile. The left panel integrates flux over a $2^{\prime}$ radius, other panels have fluxes integrated over the virial radius. The central panel covers the ASTRO-H frequency range.}
\label{fig:draco-2}
\end{figure}

\begin{figure}[htbp]
\centering
\includegraphics[scale=0.6]{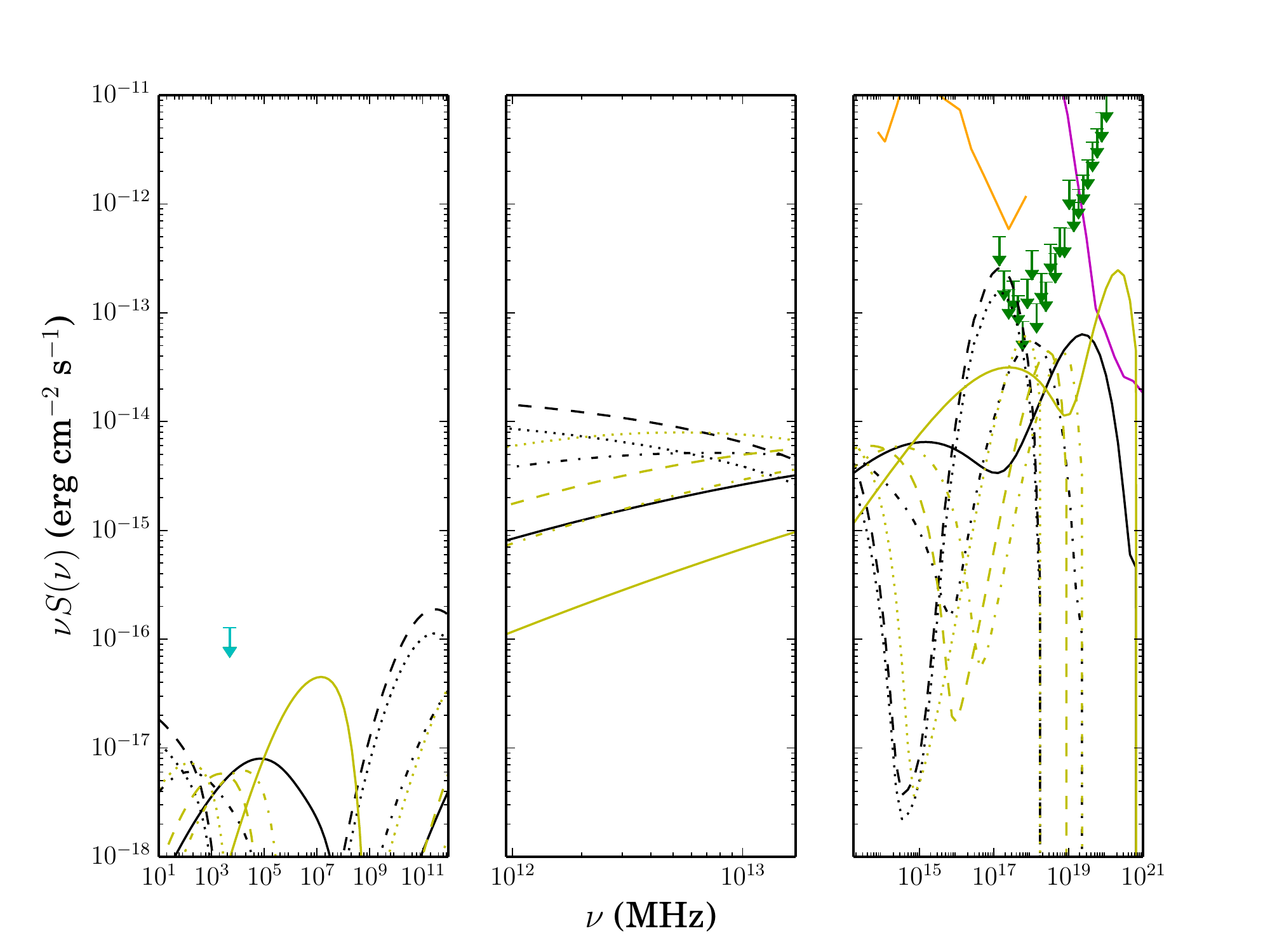}
\includegraphics[scale=0.6]{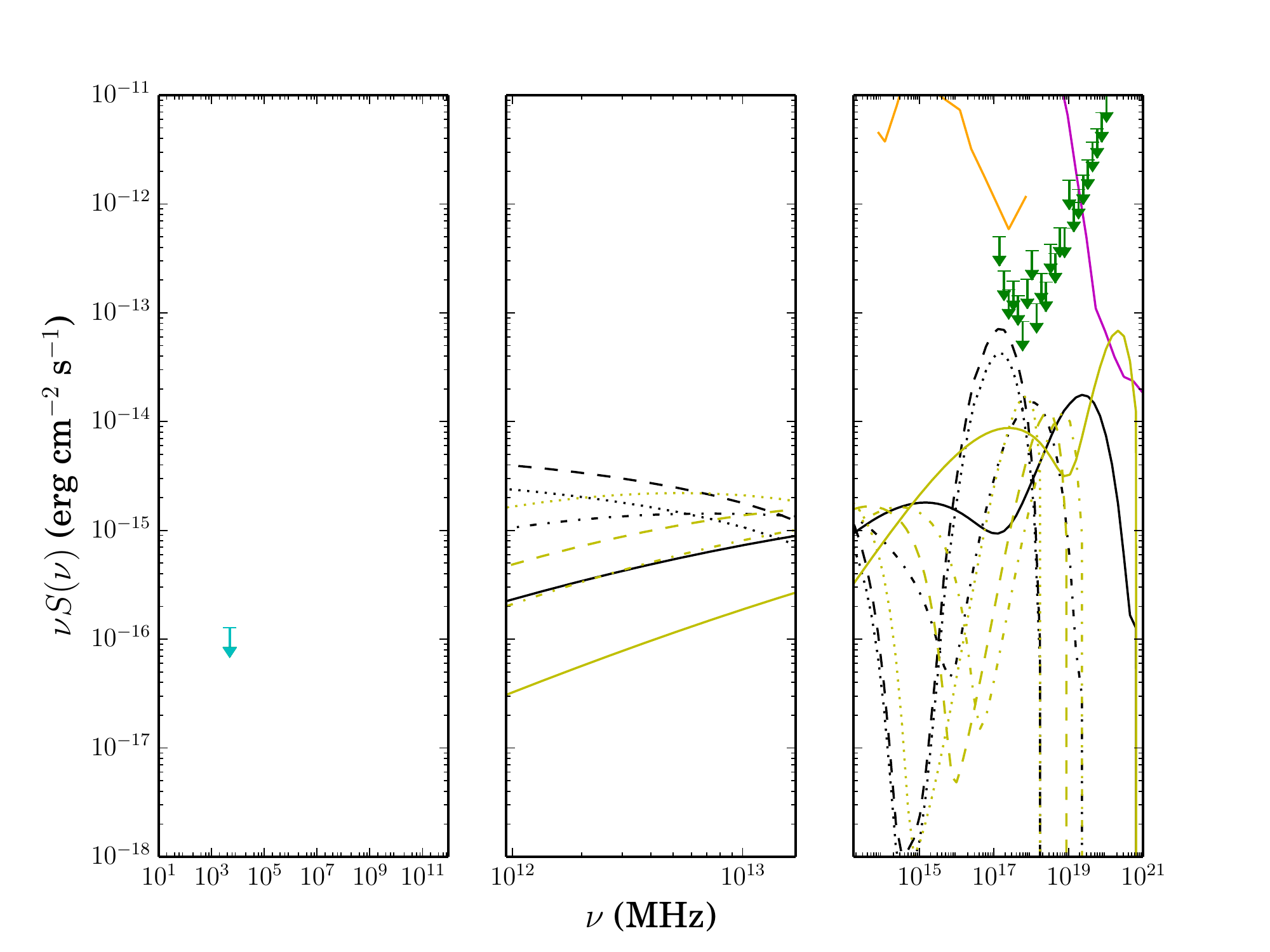}
\caption{Dark matter $b\bar{b}$ annihilation spectra for the draco dwarf galaxy with Fermi-LAT dwarf limits. Solid curves correspond to the 3 TeV model, the dash-dotted, dashed, and dotted curves correspond to 10, 40, and 100 GeV models respectively. The solid pink curve corresponds to the 1000 hours sensitivity of the CTA~\cite{funk-cta2013}. Green arrows indicate the upper limits set by the Fermi-LAT observations~\cite{Fermidwarves2014}, while the cyan arrow corresponds to the VLA limit~\cite{vladraco}. Upper panel: halos use NFW profile. Lower panel: halos use Burkert profile. The left panel integrates flux over a $2^{\prime}$ radius, other panels have fluxes integrated over the virial radius. The central panel covers the ASTRO-H frequency range.}
\label{fig:draco-3}
\end{figure}

Figure~\ref{fig:draco} shows the spectrum of the considered DM models in the Draco dwarf galaxy environment. In the radio range we find that there is no tension with the VLA limit on Draco~\cite{vladraco} (while integrating the flux over an appropriate 4$^{\prime}$ $\times$ 4$^{\prime}$ area at the centre of Draco) but stronger magnetic field/diffusion characterisation would be needed to robustly support this conclusion. Additionally there are no conflicts with the Fermi-LAT upper limits. These results holds with both NFW and Burkert halo profiles. We note that this work does not account for possible diffuse foreground emissions~\cite{regis2015}, which may substantially impact upon Draco observations. 

Once again, for each form of emission, at low frequencies the $\tau^+\tau^-$ spectrum lies below the $b\bar{b}$ one but then it crosses over at high frequencies. The exact frequency at which the cross-over occurs is dependent on the neutralino mass and is trivially red-shifted as discussed in~\cite{gsp2015} and shows a mild sensitivity to the conditions of the halo in the form of being shifted to higher frequencies for larger magnetic fields (synchrotron only) and ICM densities (as can be seen from comparison of Figs.~\ref{fig:coma},\ref{fig:m81}, and \ref{fig:draco}). 
The fact that this cross-over behaviour appears in each region of the spectrum, although the ICS cross-over can be hidden by the $\gamma$-ray spectrum, suggests that such features should be attributed to the underlying differences in the particle distributions produced by these neutralino annihilation channels, and thus the shape of the emission spectrum constitutes a signature of the dominant channel. All of the models are compatible with the Fermi-LAT dwarf upper bounds, which conforms to the slope and shape of the high-energy spectrum. 
Although more sensitive measurements will be necessary to constrain the $\gamma$-ray spectrum more effectively. 

The ICS spectra are significantly different within the energy window of ASTRO-H, similar to the environment of Coma, with the crossing of spectra due to differing annihilation channels being present for the GC models. Where the spectra for different models/channels have similar amplitudes, they differ substantially in slope over the observational region of ASTRO-H, greatly increasing the possibility of identifying the neutralino mass and annihilation channel from the nature of an observed signal.

Figure~\ref{fig:draco-2} shows the consequences of assuming the Reticulum II annihilation cross-section. 
All of the models explored here are incompatible with the Fermi-LAT $\gamma$-ray limits on Draco, while all but the 10 GeV cases and the 40 GeV $b\bar{b}$ models are incompatible with the VLA limit on Draco. This is slightly affected by the use of the Burkert profile in the lower panels, with only no models in tension with the VLA limit, but $\gamma$-ray limits still exclude all models. Thus, the Reticulum II DM interpretation seems untenable for all the studied masses and this is not subject to the halo profile uncertainty between NFW and Burkert.

Figure~\ref{fig:draco-3} displays the effects of assuming the Fermi-LAT annihilation cross-section for dwarf spheroidal galaxies. Once again the effects of diffusion ensure that no models are in conflict with the VLA limit. In the case without spatial diffusion, the 3 TeV and $\tau^+\tau^-$ 100 GeV models are in tension with the VLA radio limit with an NFW profile (the conflict is removed by the Burkert profile in this case). Additionally, there is no tension with the Fermi-LAT limits as one should expect. The use of the Burkert profile (see lower panel of this figure) does not affect this. 
For neutralino masses above $100$ GeV without spatial diffusion, a magnetic field reduction of at most $\sim 40$\% would be necessary for consistency of the featured models with the VLA data, as can be seen in Fig.~\ref{fig:bfields} for an NFW profile. 
Thus uncertainty is provided by both the diffusion/magnetic-field and the DM halo profile, the latter is particularly important in Draco given the compatibility of many dwarf halos with cored distributions~\cite{Walker2009,Adams2014}. Therefore we take the Draco results to provide some support for the conclusion that only cross-sections below the Fermi-LAT limit for dwarf galaxies could justify a DM interpretation of the Reticulum II $\gamma$-ray excess. The radio frequency uncertainty emanates mainly from the magnetic field and diffusion characteristics.

\subsection{Dark matter constraints}
\label{sec:constraints}

It is worth noting that the synchrotron portions of the spectra displayed in Figs.~\ref{fig:coma},\ref{fig:m81}, and \ref{fig:draco} are sensitive to the assumed magnetic field strength, as seen in Eq.~(\ref{eq:power}). This is significant because it will be necessary to obtain accurate estimates of the magnetic fields within cosmic structures in order to properly constrain the synchrotron spectra resulting from DM annihilation. In this regard, the SKA is expected to play a prominent role, as discussed in~\cite{Colafrancesco2015}. This is because, for arcminute resolution at flux levels of $\sim 1$ $\mu$Jy, it has been shown~\cite{Stiletal2014} that polarisation stacking calculations indicate an expected polarised source density of the order of $\approx 1300$ sources per square degree, with the analysis of~\cite{Govonietal2014} indicating that this involves an overall uncertainty of $\approx 50\%$. This means that already the SKA-1 will be able to derive stringent constraints for cosmic magnetic fields on the required scale with a sufficiently large source count for spatial profiling. 

\begin{figure}[htbp]
\centering
\includegraphics[scale=0.6]{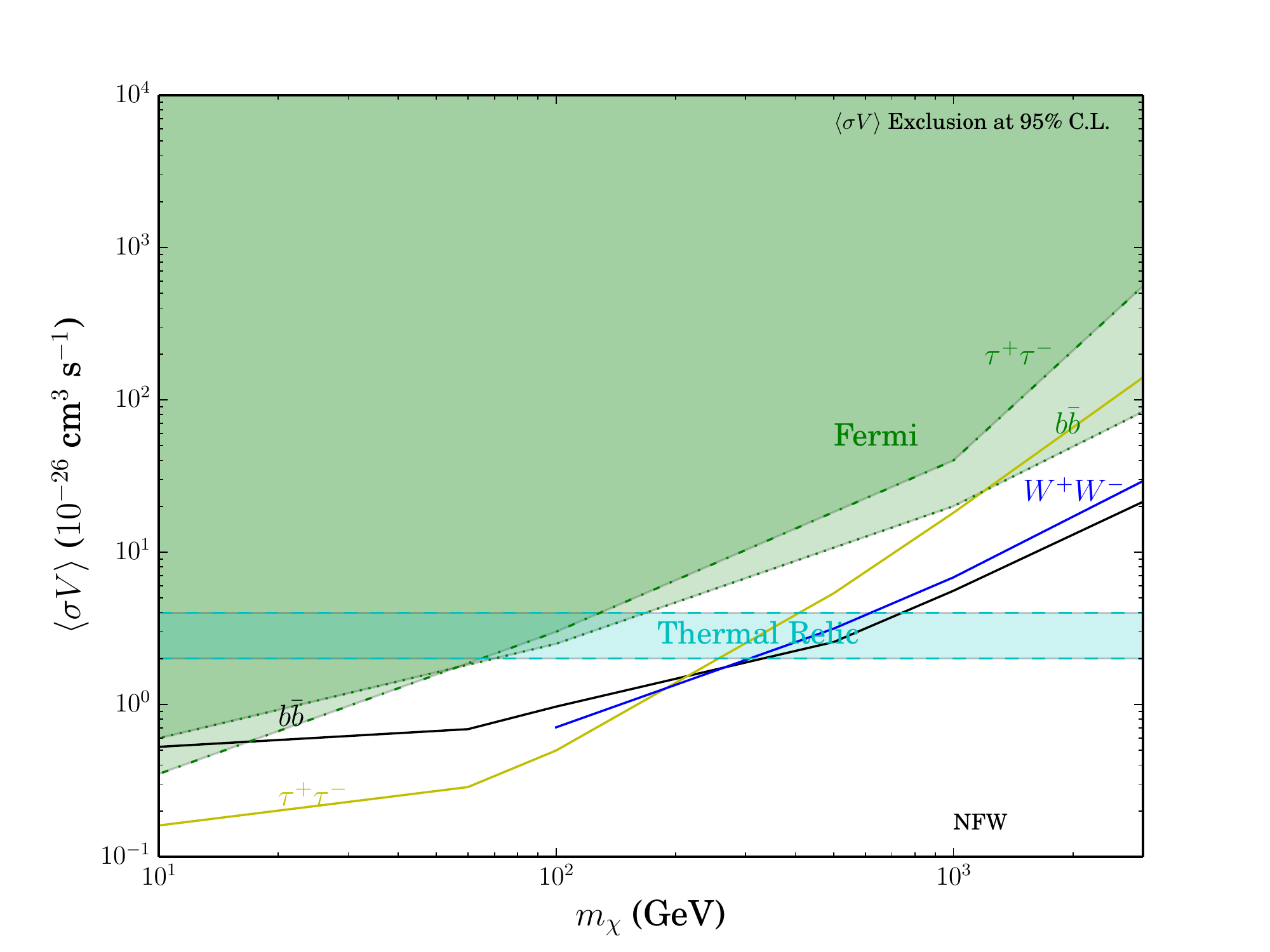}
\includegraphics[scale=0.6]{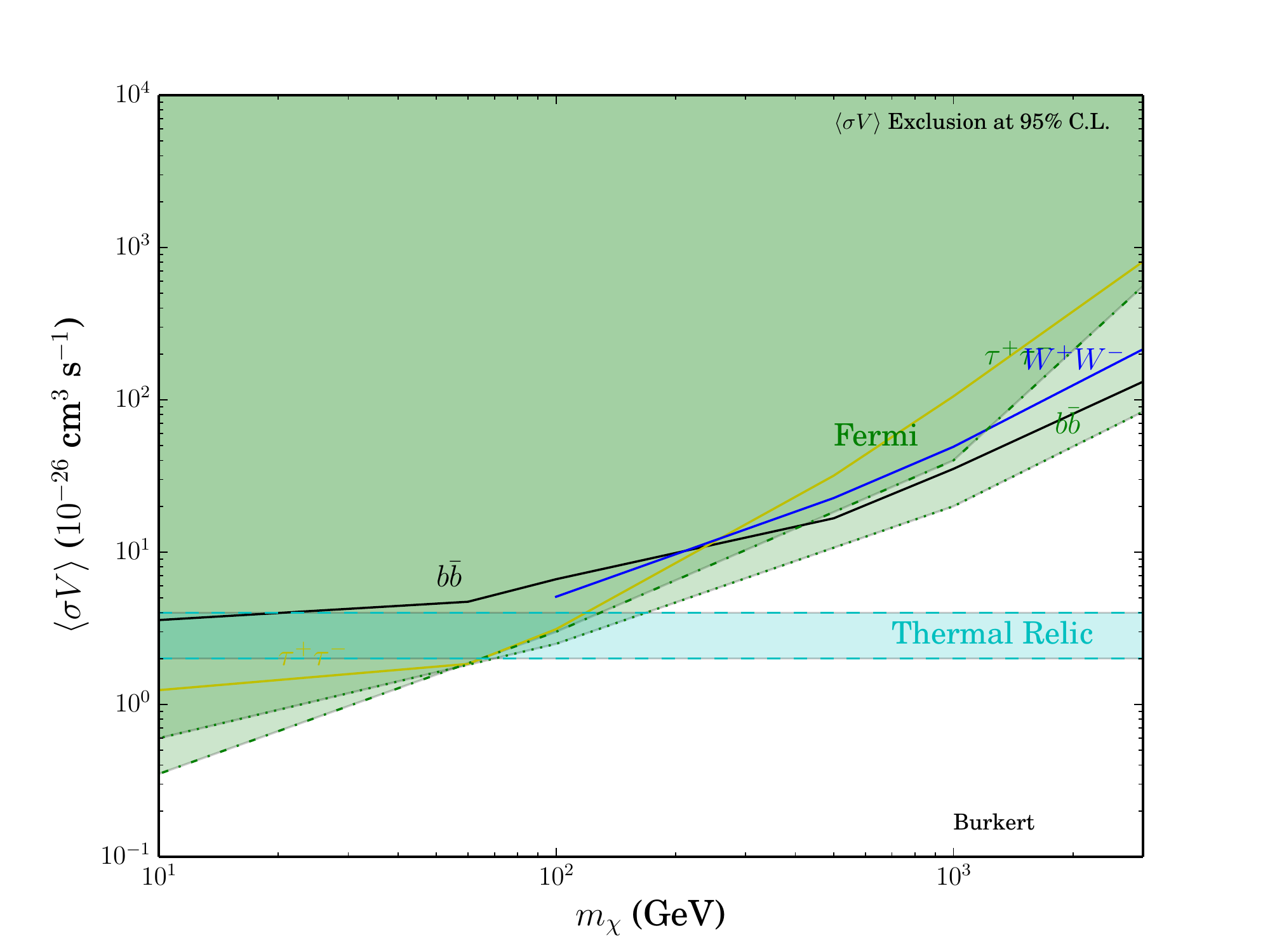}
\caption{The $3\sigma$ cross-section limit derived from Coma is shown as a function of neutralino mass. The black curve corresponds to $b\bar{b}$, yellow to $\tau^+\tau^-$, and blue to $W^+W^-$. The green region is for the Fermi-LAT exclusion derived via J-factor estimation in dwarf galaxies which were assumed to be point-sources. The cyan region shows the thermal relic region. Upper panel: NFW halo profile. Lower panel: Burkert halo profile.}
\label{fig:sigv_coma}
\end{figure}

\begin{figure}[htbp]
\centering
\includegraphics[scale=0.6]{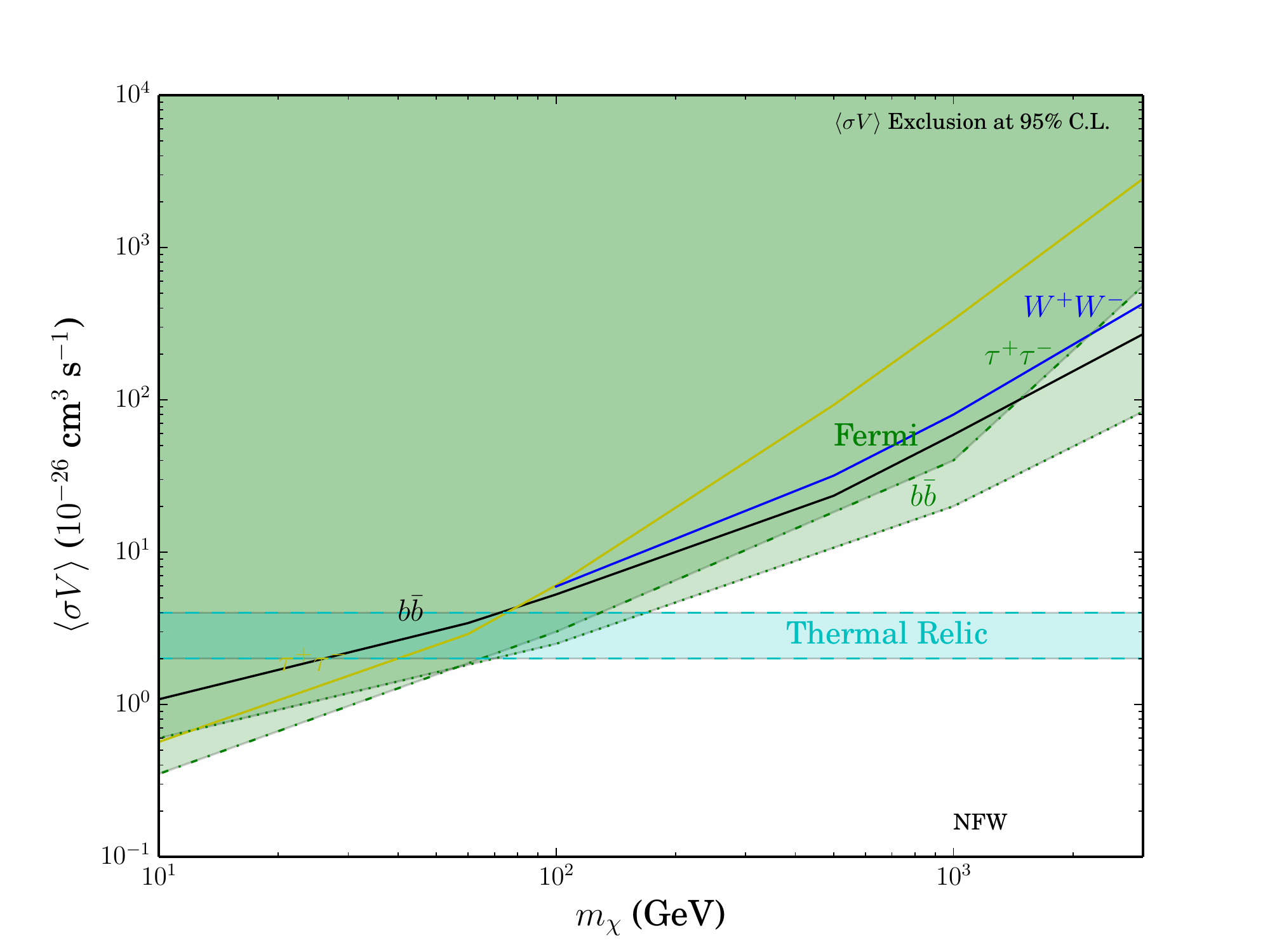}
\includegraphics[scale=0.6]{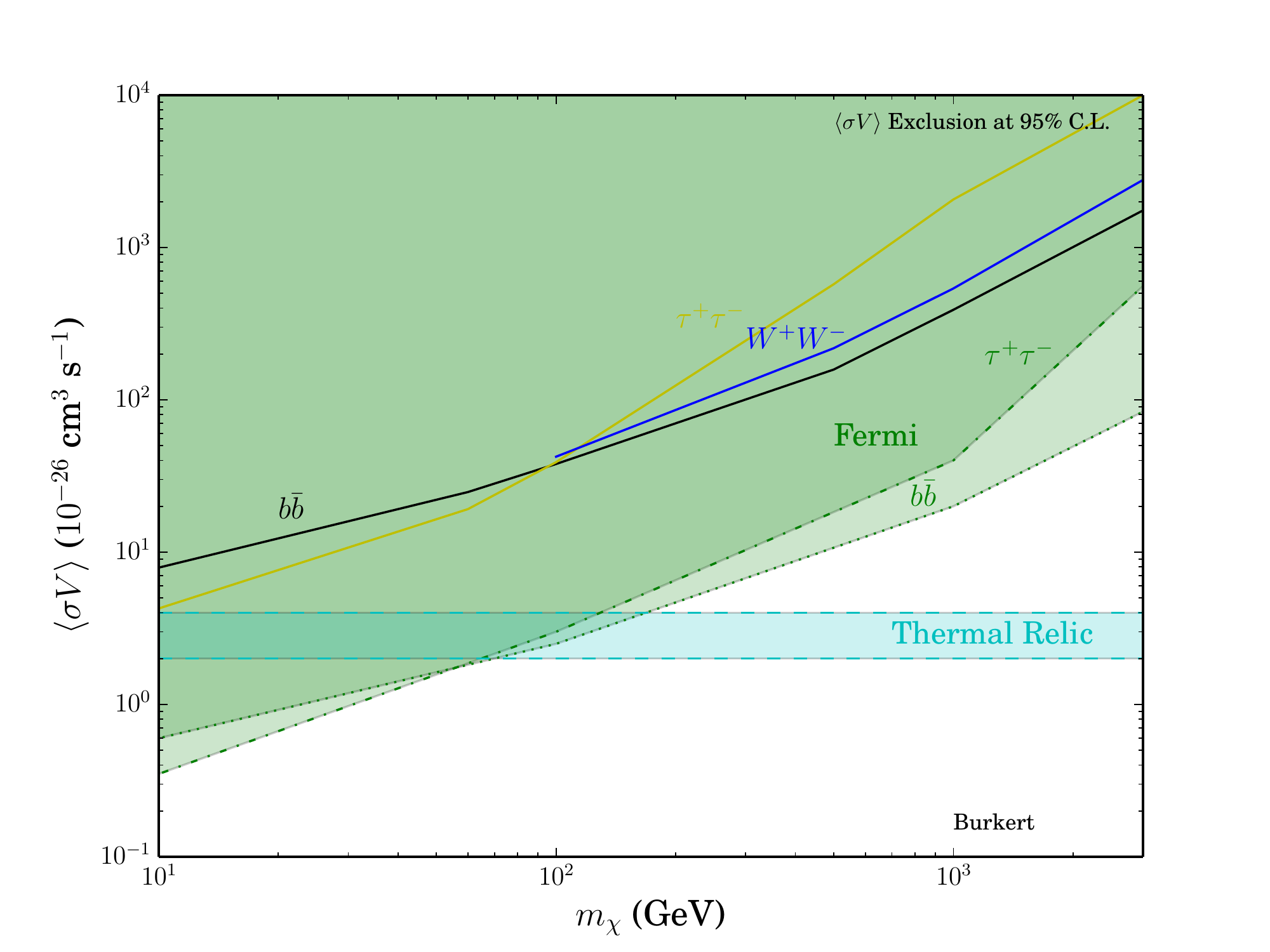}
\caption{The $3\sigma$ cross-section limit derived from M81 data is shown as a function of neutralino mass. The black curve corresponds to $b\bar{b}$, yellow to $\tau^+\tau^-$, and blue to $W^+W^-$. The green region is for the Fermi-LAT exclusion derived via J-factor estimation in dwarf galaxies which were assumed to be point-sources. The cyan region shows the thermal relic region. Upper: NFW halo profile. Lower: Burkert halo profile.}
\label{fig:sigv_m81}
\end{figure}

In Figures~\ref{fig:sigv_coma} and \ref{fig:sigv_m81}, we derive the cross-section limits that can be placed on the parameter space using the Coma radio data, as well as the M81 spectrum. In the Coma cluster case (Fig.~\ref{fig:sigv_coma}), the data provides $3\sigma$ constraints that are about an order of magnitude stronger than the Fermi-LAT dwarf limits at all masses above 10 GeV, where Fermi-LAT is similar. We note that the strength of these limits is significantly affected by the use of the Burkert profile in the lower panel, weakening constraints by roughly an order of magnitude at all masses. 

It is instructive to compare these derived limits to previous works such as \cite{Bertone2008,Crocker2010}, which derive limits from radio observation of the galactic centre. We see that our Coma limits produce similar constraints to \cite{Bertone2008} around 100 GeV ($\langle \sigma V \rangle \lesssim 10^{-26}$ cm$^3$ s$^{-1}$ for $b\bar{b}$ with NFW profile) but improve over \cite{Bertone2008} substantially towards 1 TeV ($\langle \sigma V \rangle \lesssim 10^{-25}$ cm$^3$ s$^{-1}$ for $b\bar{b}$ with NFW profile). In the case of \cite{Crocker2010} we find that their best-case magnetic field results for 10 GeV, $\langle \sigma V \rangle \lesssim 10^{-26}$ cm$^3$ s$^{-1}$ for $b\bar{b}$ with NFW profile, are somewhat weaker than those derived here, in addition to this they scale more severely with $m_{\chi}$, reaching $\langle \sigma V \rangle \lesssim 10^{-23}$ cm$^3$ s$^{-1}$ for $b\bar{b}$ with NFW profile by masses of 1 TeV.

\begin{figure}[htbp]
	\centering
	\includegraphics[scale=0.6]{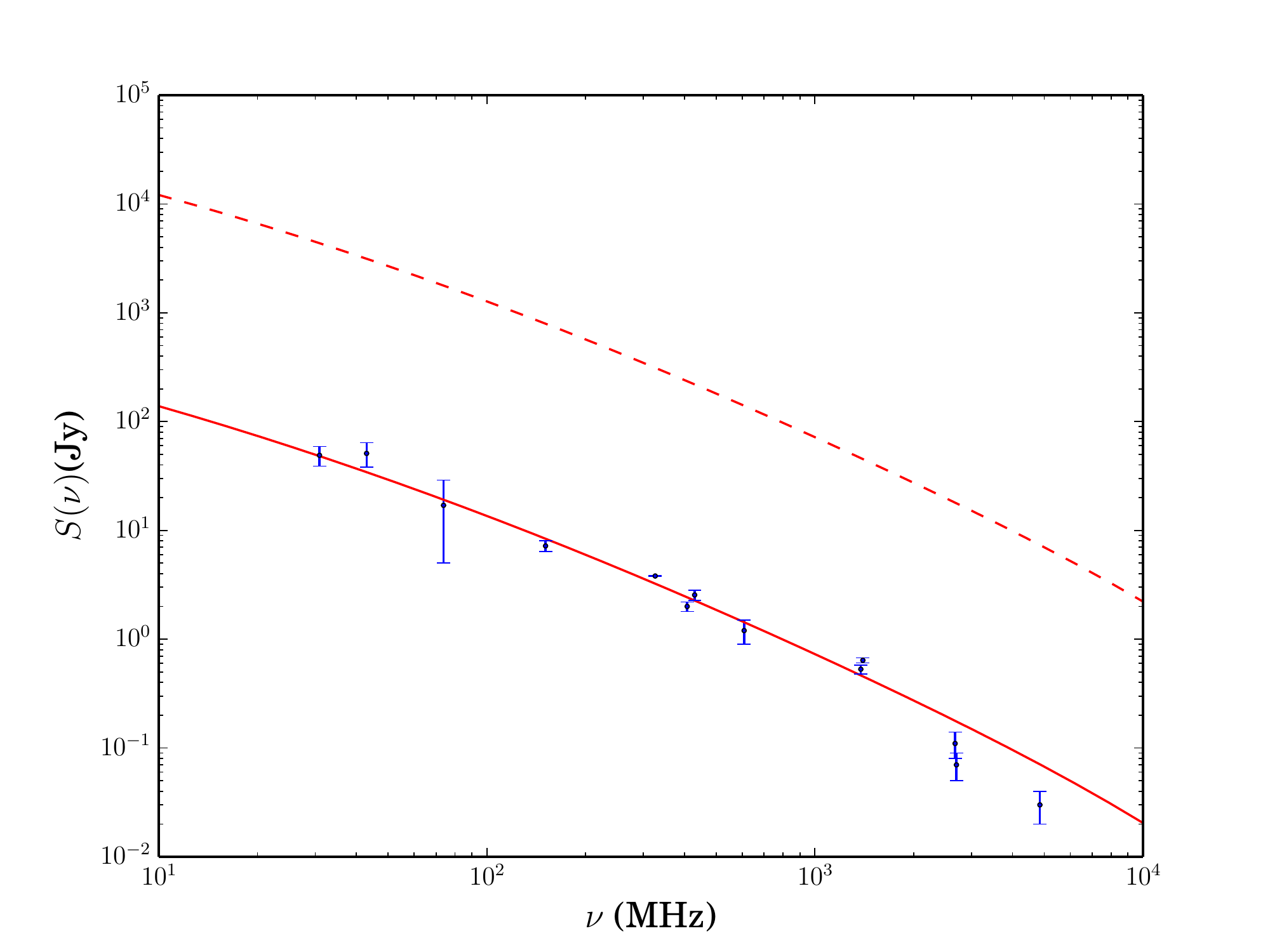}
	\caption{Comparison of spectra for $m_{\chi} = 40$ GeV in the $b\bar{b}$ channel with the best-fit cross-section from~\cite{Colafrancesco2006}: $\langle \sigma V \rangle = 4.7 \times 10^{-25}$ cm$^3$ s$^{-1}$. The solid curve is that using the parameters for \cite{Colafrancesco2006} (see text) and the dashed curve uses those of this work.}
	\label{fig:coma_cpu2}
\end{figure}

\begin{figure}[htbp]
	\centering
	\includegraphics[scale=0.6]{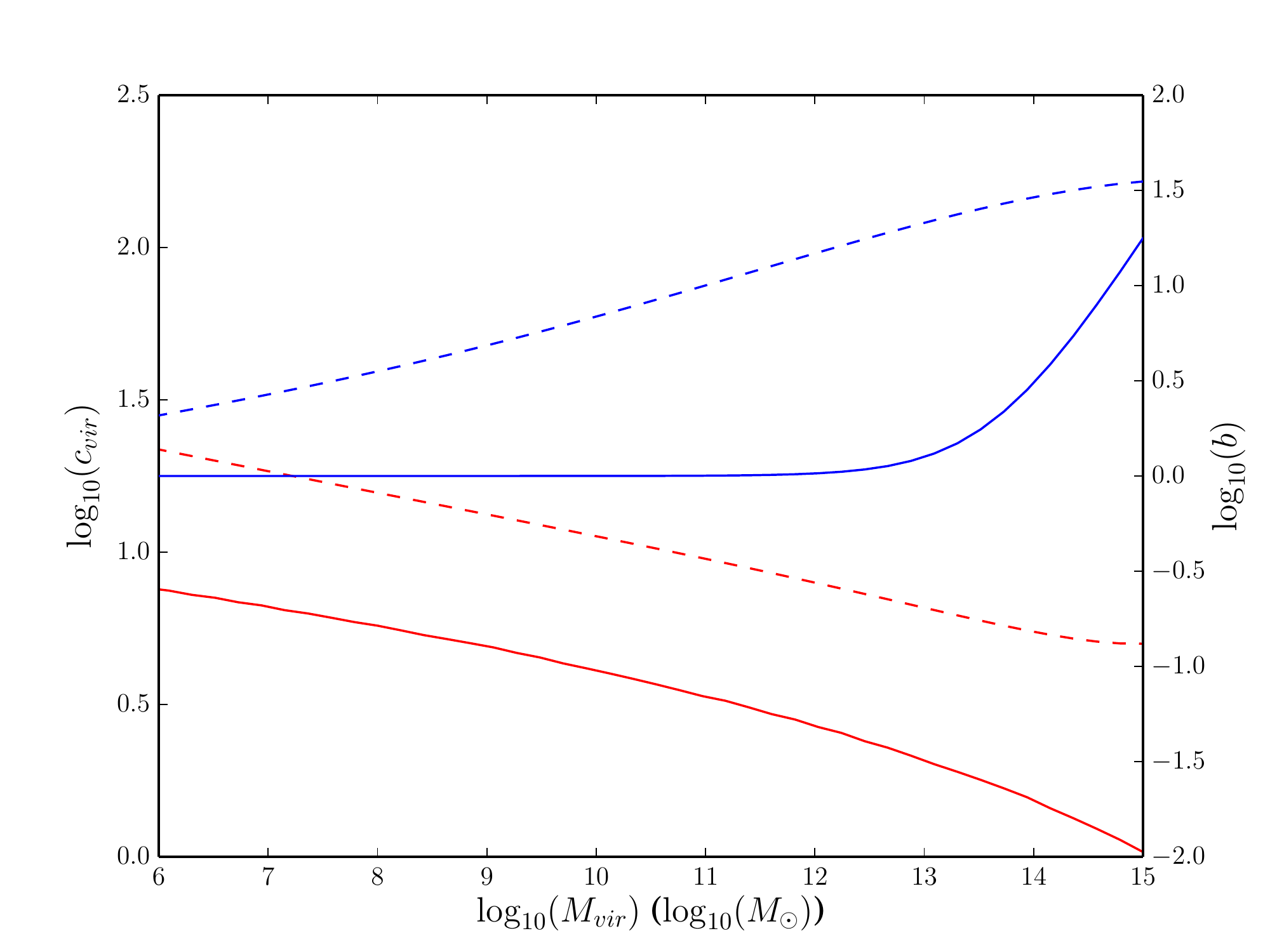}
	\includegraphics[scale=0.6]{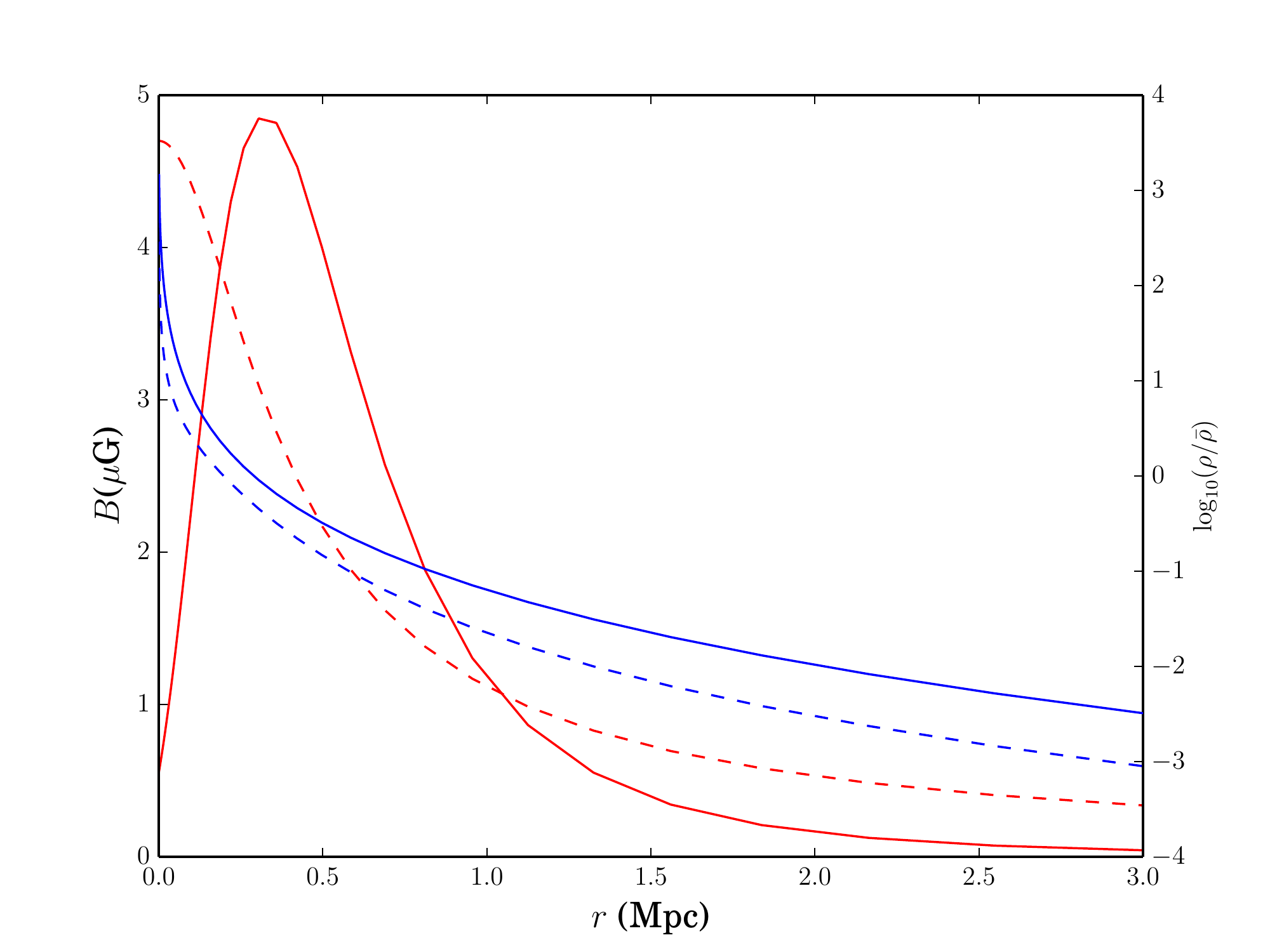}
	\caption{Comparison plots between this work and \cite{Colafrancesco2006} for parameters relevant to Coma. Solid lines are for the \cite{Colafrancesco2006} case while dashed lines display the parameters used in this work. Upper panel: $c_{vir}$ and $b$ as functions of the halo virial mass. Blue curves show the sub-structure boost factor $b$, while red show the parameter $c_{vir}$. Lower panel: $\rho_{DM}$ (blue) and $B$ (red) as functions of radius $r$.}
	\label{fig:coma_cpu1}
\end{figure}

We note that the Coma radio flux we obtained here differs from previous derivations of DM-induced radio emission in this cluster, such as in \cite{Colafrancesco2006}. In order to understand this difference, we show in Figure~\ref{fig:coma_cpu2} the two calculations, and we see a two order-of-magnitude difference in flux between our model and that performed using the model from the aforementioned work \cite{Colafrancesco2006}. In this figure we use the best-fit cross-section derived in \cite{Colafrancesco2006} to illustrate that using our approach we can reproduce the results of \cite{Colafrancesco2006} when we employ the same halo model. Due to the magnitude of the difference between the fluxes obtained in these two cases, the differences in the underlying models of Coma deserve to be remarked upon.\\
In the upper-panel of Figure~\ref{fig:coma_cpu1} we show a comparison of $c_{vir}$ and sub-structure boost factor values between our work and the aforementioned earlier study of Coma. As previously stated, we take our $c_{vir}$ value for Coma from the fits done in \cite{Colafrancesco2006}. Therefore, in this regard the two calculations do not differ; however, this $c_{vir}$ calculational method is of interest for its contribution to the sub-structure boosting factor (as it will be used to find $c_{vir}$ for sub-halos). 
It is important to note that when we determine the boosting factor for a halo with Coma-like mass using the model from~\cite{prada2013} we obtain a boost factor that is twice as large as that derived from the sub-structure calculations used in~\cite{Colafrancesco2006} (see Table~\ref{tab:boost}). However, the differences in these parameters are clearly insufficient to explain the difference between the radio fluxes shown in Fig.~\ref{fig:coma_cpu2}. 
In the lower panel of Fig.~\ref{fig:coma_cpu1} we also note that the DM halo density profiles are similar: we use an NFW profile while \cite{Colafrancesco2006} use an NFW-like Einasto profile with $\alpha = 0.17$. 
However, we point out that the magnetic field model from \cite{Colafrancesco2006} peaks well outside the scale-radius of Coma ($\sim 0.29$ Mpc), while the one we employ in this paper following a later analysis from \cite{bonafede2010} peaks at the cluster centre. It is clear from Figure~\ref{fig:coma_cpu1} that DM density has dropped by two orders of magnitude before we reach the peak of the magnetic field model used in \cite{Colafrancesco2006}. This effect is able to reduce substantially the synchrotron flux generated by annihilations (which is proportional to $\rho^2$) in the dense central region of the cluster, in comparison to our model, and accounts for the remaining difference between the radio flux density curves displayed in Fig.~\ref{fig:coma_cpu2}.\\
Therefore, we conclude that the main difference between our radio flux calculations in Coma and those of \cite{Colafrancesco2006} is due to the spatial profile of the magnetic field within the inner parts of the halo, with the boosting factor only accounting for a factor of 2 of the difference.

We also note that the boosting factor we employ, following \cite{prada2013}, depends only the DM sub-halo mass distribution within the parent halo and it is therefore a DM pair-annihilation boost; as such it does not account for the spatial variation of the magnetic field which would affect how much synchrotron radio flux is produced by sub-halos. In other words, by applying this method one assumes that all types of DM-induced e.m. emission will benefit similarly from sub-structure boosting. 
In the specific case of DM-induced radio synchrotron emission, since sub-halos are distributed radially within their parent halo, they will not all encounter the same value of the magnetic field strength and thus the secondary electrons produced in DM annihilations within sub-halos at different radii will provide differing boosts to the total synchrotron flux. Taking this into account would then have the effect of reducing the total synchrotron flux produced by sub-halos~\cite{storm2013}. 
However, it was also shown in \cite{Colafrancesco2006} that, if the sub-halos distributions follow a similar profile to the DM density profile of their parent halo, most of the substructure boosting occurs near the cusp/core region, which will mitigate the aforementioned synchrotron reduction effect given a magnetic field model that peaks centrally within the halo, as in our case. 

For the case of M81 (Fig.~\ref{fig:sigv_m81}), we note that the use of the Burkert profile has a significant effect, reducing the limits from slightly weaker than Fermi-LAT to significantly weaker than this benchmark.

\begin{table}[htbp]
\centering
\begin{tabular}{|l|l|l|}
\hline
Halo & Boost A & Boost B \\
\hline
Coma & 35.2 & 17.7 \\
M81 & 10.1  & 1.00 \\
Draco & 3.43 & 1.00 \\
\hline
\end{tabular}
\caption{Substructure boosting factors for two cases. A: from \cite{prada2012} as used in this work. B: from \cite{Colafrancesco2006} used to study Coma.}
\label{tab:boost}
\end{table}

\section{SKA and ASTRO-H Constraints}
\label{sec:constraint}

In order to determine how far into the $m_{\chi}-\langle \sigma V\rangle$ parameter future experiments could probe in the studied environments, we will determine the smallest cross-section to which they are sensitive. We do this first by locating the smallest cross-section observable by the instrument (assuming 100\% of emissions results from DM annihilation) and then by determining the minimal cross-section for which the DM-induced emissions can be disentangled from the dominant foreground emission with a power-law spectrum. This is important as DM-induced radio emissions will likely be sub-dominant in all the environments for cross-sections below those shown in Fig.~\ref{fig:sigv_coma}; for X-rays the sub-dominance in the studied sources will evidently begin at much larger cross-sections (see Figs.~\ref{fig:coma} and \ref{fig:m81}). 

This disentangling process is performed by assuming a power-law spectrum $S_n \nu^{-\eta}$ with either $\eta = 0.85$ or $\eta=0.7$ as appropriate (resulting from an electron distribution with power-law index 2.7 or 2.4 respectively) with $S_n$ chosen to closely match the flux level of available data (same points as displayed in previous plots). We then assume that the total flux from the source will be the sum of the DM and power-law fluxes characterized to within a 1\% error. The DM flux will then be found subtracting off the power-law, characterized to within a 2\% error, and averaged over many random realizations of this ``simulated measurement''. The error assigned to the resulting ``measurement'' of the DM-induced flux at each frequency is taken to be the variance of the set of simulated measurements. The given error limits are chosen to reasonably match the capabilities of the SKA and ASTRO-H experiments. This analysis is performed separately for both the ICS and synchrotron spectra.

In the case of Draco and ASTRO-H, we chose to normalize $S_n$ using the WISE~\cite{wise} 3-Band limit  for Draco, lacking for any hard X-ray data.

The minimal cross-section to which the instrument is sensitive is then taken to be the smallest that can be resolved from the power-law spectrum (such that it is not dominated by the $1\sigma$ uncertainties), provided this is larger than the 100\% DM analysis result. We also require that the case of power-law plus DM has a $\chi^2$ value $10^3$ times smaller than power-law alone (when comparing these cases to the total flux) in order to ensure that it is the strongly preferred hypothesis.

\begin{figure}[htbp]
\centering
\includegraphics[scale=0.5]{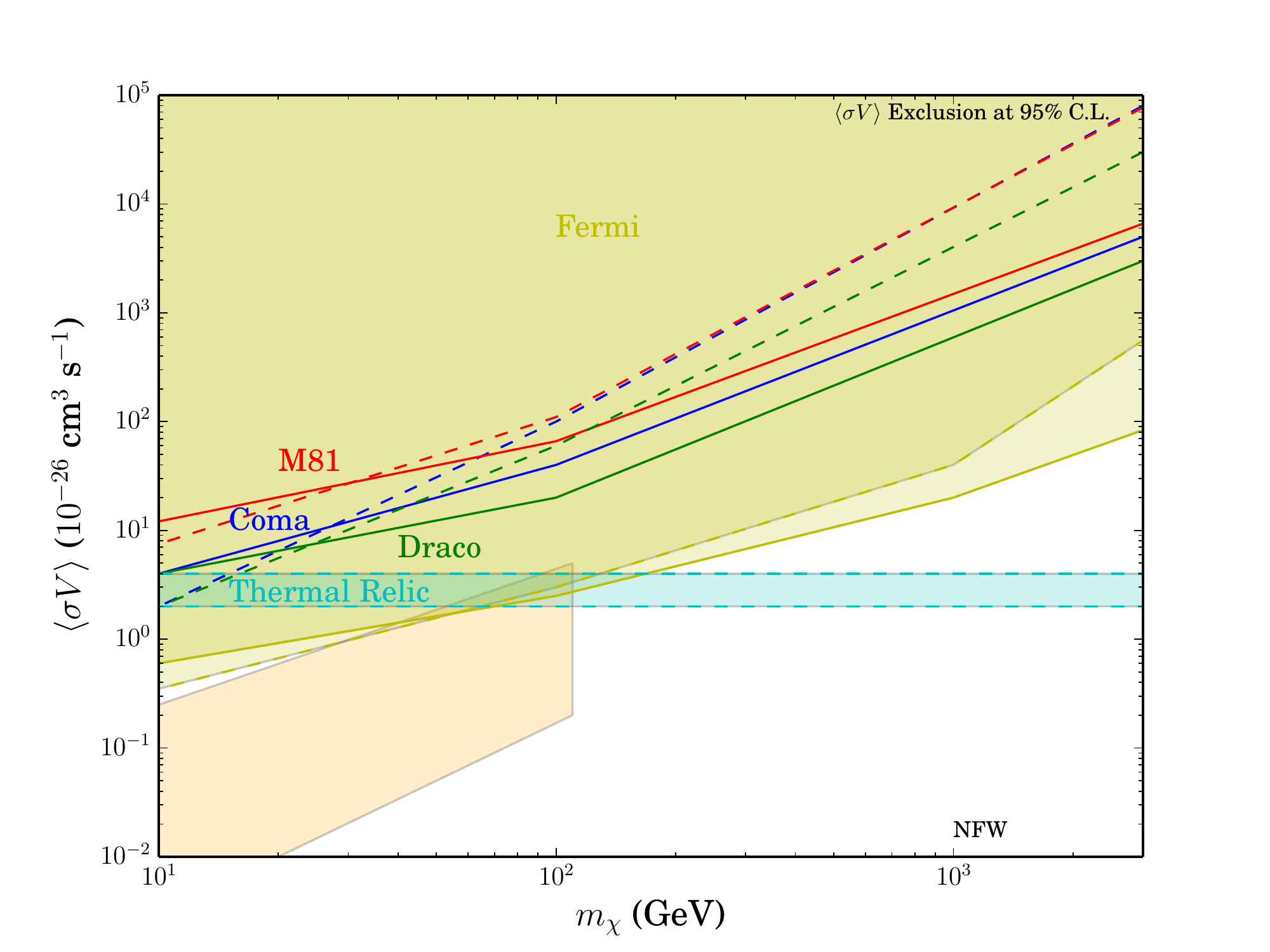}
\includegraphics[scale=0.5]{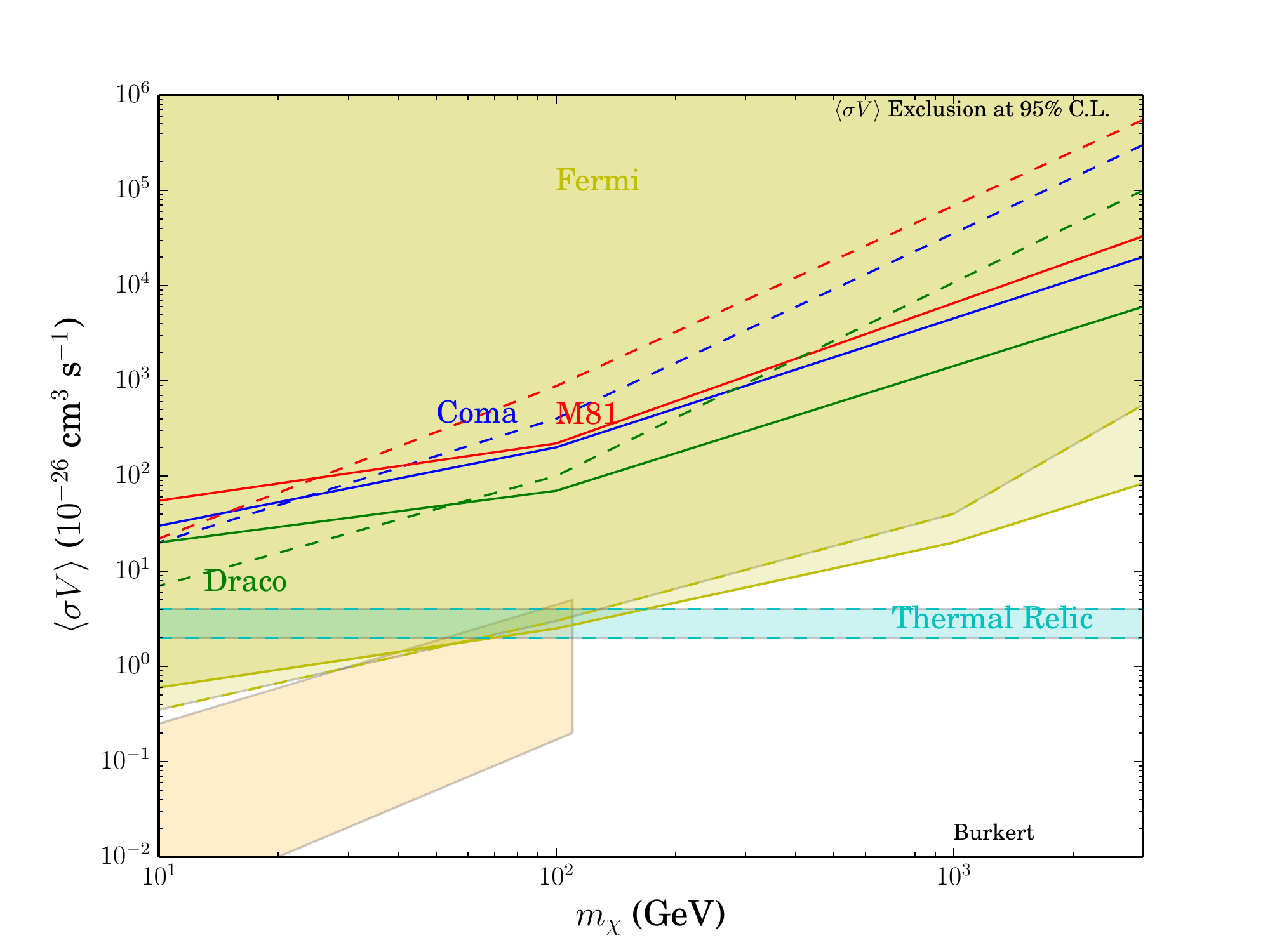}
\caption{ASTRO-H 1000 hrs sensitivity cross-section as a function of neutralino mass. The red curve corresponds to the M81 galaxy, the green curve to the Draco dwarf and the blue to the Coma cluster. The yellow region is for the Fermi-LAT exclusion derived via J-factor estimation in dwarf galaxies which were assumed to be point-sources~\cite{Fermidwarves2015}, while the cyan shaded area is the thermal wimp region, and the orange region covers the Fermi-LAT GC models. Solid curves are $b\bar{b}$ and dashed are $\tau^+\tau^-$. Upper panel: NFW halo profile. Lower panel: Burkert halo profile.}
\label{fig:sigv}
\end{figure}

\begin{figure}[htbp]
\centering
\includegraphics[scale=0.5]{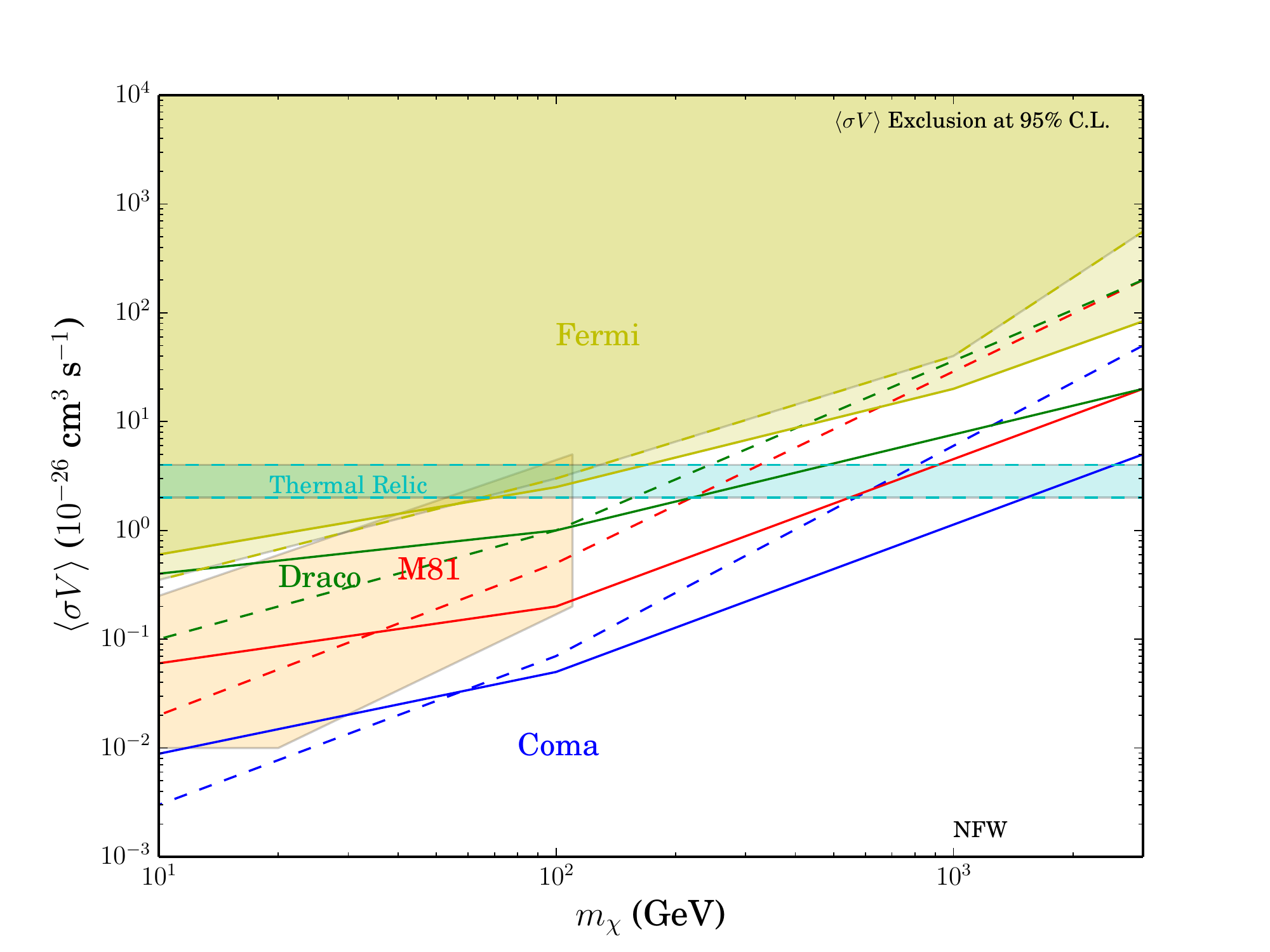}
\includegraphics[scale=0.5]{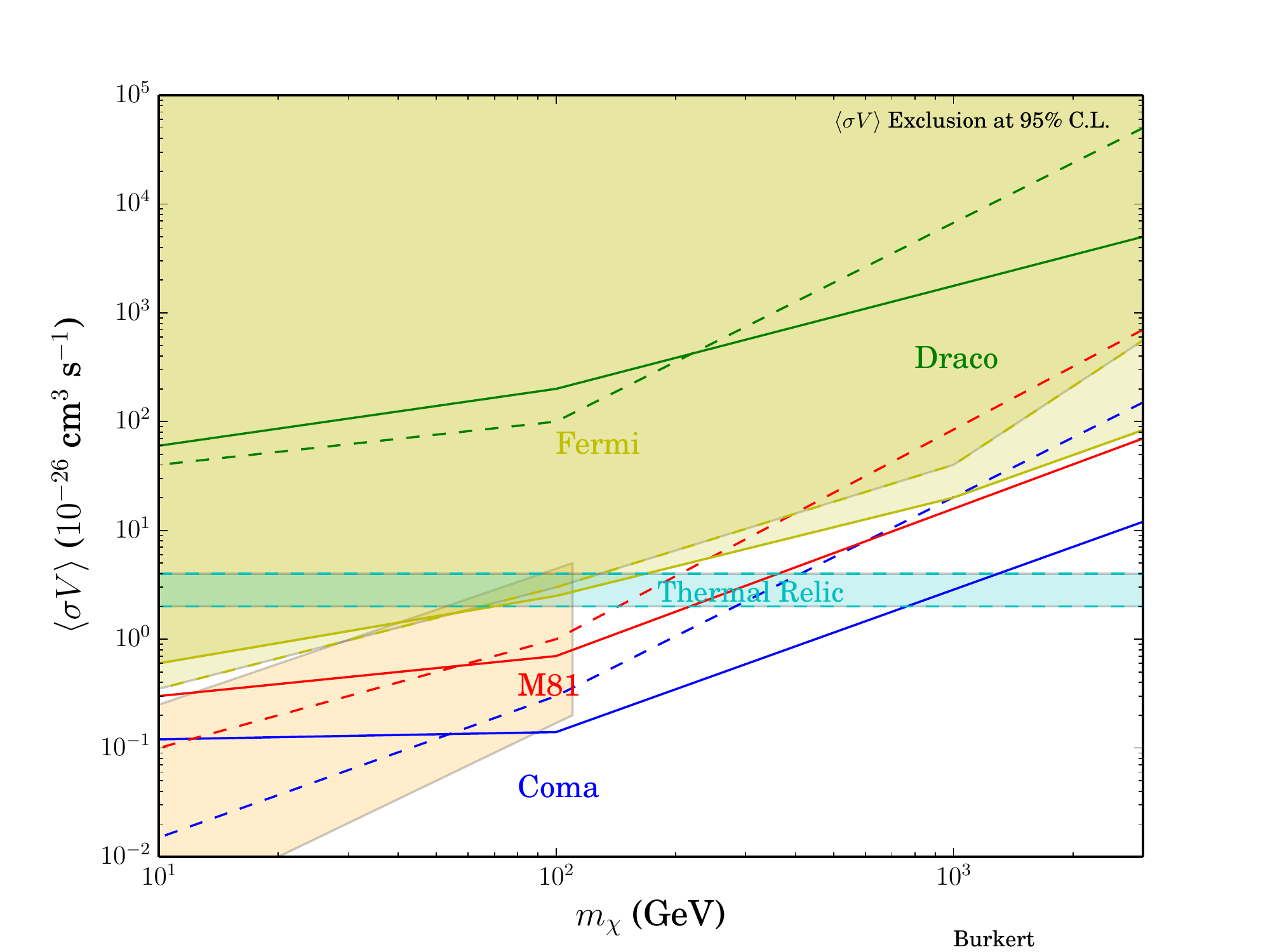}
\caption{SKA 1000 hrs sensitivity cross-section as a function of neutralino mass. The red curve corresponds to the M81 galaxy, the green curve to the draco dwarf and the blue to the Coma cluster. The yellow region is for the Fermi-LAT exclusion derived via J-factor estimation in dwarf galaxies which were assumed to be point-sources~\cite{Fermidwarves2015}, while the cyan shaded area is the thermal wimp region, and the orange region covers the Fermi-LAT GC models. Solid curves are $b\bar{b}$ and dashed are $\tau^+\tau^-$. Upper panel: NFW halo profile. Lower panel: Burkert halo profile.}
\label{fig:sigv_ska}
\end{figure}

Figure~\ref{fig:sigv} displays the constraints that can be derived from the minimal cross-section with which ASTRO-H could observe DM-induced Inverse-Compton Scattering (ICS) emission for each neutralino mass. The large background X-ray fluxes in the chosen environments mean that ASTRO-H cannot resolve DM emissions of models below the Fermi-LAT limit. However, in earlier figures it was argued that ASTRO-H is well positioned to observe features of the DM ICS, so environments with less X-ray background might be more suitable. A study of this nature will be confined to further work to determine the possible usefulness of ASTRO-H data in DM searches. The lack of higher frequency X-ray limits in the Draco dwarf will also play a role in reducing the displayed constraints.  

Given that ASTRO-H has a larger collecting area in the lower frequency part of the ICS spectrum, the $b\bar{b}$ channel has stronger fluxes and thus can provide better constraints. The use of the Burkert profile in the lower panel of Fig.~\ref{fig:sigv} weakens the constraints that can be derived by roughly an order of magnitude.


Figure~\ref{fig:sigv} should also be compared to Fig.~\ref{fig:sigv_ska} sourced from~\cite{gsp2015} but applying the above ``disentanglement test". This shows the potential cross-section constraints for the case of the SKA. We stress that the SKA constraints are several orders of magnitude better in all environments. In the case of Draco the Burkert profile constraints are substantially weakened by the fact that its flux is only integrated over a 4 arcminute squared area, as the dominant spectrum is chosen to respect the VLA limit. The limited area also greatly exaggerates the effects of spatial diffusion in reducing the severity of these constraints. The use of the Burkert halo profile in the lower panel of Fig.~\ref{fig:sigv_ska} weakens all the constraints by more than an order of magnitude. For an NFW halo profile, the SKA can probe models well below the GC favoured region in Coma and reaching $\langle \sigma V \rangle \sim 6 \times 10^{-28}$ cm$^3$ s$^{-1}$ in M81, while in Draco it can cover the region given by $\langle \sigma V\rangle \gs 5 \times 10^{-27}$ cm$^{3}$ s$^{-1}$ and $m_{\chi} \gs 20$ GeV. When the Burkert profile is used, however, Draco can no longer be used to probe the GC model region at all (for this 4 square arcminute integration area) and M81 can be used to study models with $\langle \sigma V\rangle \gs 3 \times 10^{-27}$ cm$^{3}$ s$^{-1}$ and $m_{\chi} \gs 12$ GeV. With the Burkert profile Coma can still be used to probe down to $\langle \sigma V\rangle \sim 1 \times 10^{-27}$ cm$^{3}$ s$^{-1}$ for all the GC masses.

\section{Discussion}
\label{sec:discussion}

We will proceed to discuss our results examining the consequences for each studied neutralino mass model with its best-fit cross-section. We will then examine the consequences of our results for the purported Reticulum II $\gamma$-ray excess. Finally, we will discuss general spectral features and the potential of multi-frequency searches for DM-induced signals with next coming experiments.

Before proceeding to this discussion we note that the synchrotron results in Coma are much more constraining for certain neutralino masses than others. This depends on two aspects. The first being that larger mass neutralinos produce electrons with a greater average energy, shifting the peak of the resulting synchrotron spectrum to higher frequencies (for the same choice of the magnetic field). The second aspect is the slope of the Coma diffuse radio emission spectrum, which tails off rapidly above 1 GHz. When we combine these two aspects we see that larger mass neutralino models will be more constrained by the available Coma diffuse radio emission data, as the synchrotron peak is rapidly shifted towards frequencies where the amplitude of the data is tailing off. Of course, a sufficiently large cross-section will make the spectral amplitude of any model incompatible with the Coma data at most frequencies.

For the AFP models we find that Coma predictions are incompatible with the slope (and often also with the amplitude) of its synchrotron radio spectrum, predicting unobserved flux excess and spectral flattening above 1 GHz. This conclusion is largely unmitigated by the use of the Burkert profile and, as shown in Fig.~\ref{fig:bfields}, magnetic field uncertainties cannot account for the conflict with the data.\\
The M81 spectra show no conflicts with the available SED.\\
However, the Draco spectra conflict with the VLA limit for an NFW profile and $b\bar{b}$ annihilation channel. Therefore, even though Draco does not provide a definitive dismissal of the AFP with best-fit annihilation cross-section, it does serve to suggest that there is cause for concern, and thus reinforces the Coma results. Further observations will increase the constraining power of Draco. We note that the flatness of the AFP spectra might well conflict with future radio studies of Draco, as also occurs in Coma.

Taking all these results into consideration, and as the studied representative model was the most compatible with the Planck data, we must conclude that the remaining AFP model not yet excluded by Planck must be now considered eliminated through this analysis. This is reinforced by the fact that Fig.~\ref{fig:Planck} shows that the Fermi-LAT data excludes these models in the $b\bar{b}$ channel.
 
A recent re-analysis of the AMS-2 results~\cite{mauro2015} indicates that a more sophisticated astrophysical model would allow a neutralino with mass $\sim 50$ GeV and a cross section $\sim 3 \times 10^{-26}$ cm$^{3}$ s$^{-1}$ to account for observed positron excesses. 
We note that this model will be covered by the conclusions which apply to the GC median mass model (40 GeV) with the Fermi-LAT dwarf cross-section ($\sim 1 \times 10^{-26}$ cm$^{3}$ s$^{-1}$) as their results do not differ sufficiently to alter any conclusions. Therefore, we find that this revised AFP model will be ruled out by a violation of Fermi-LAT stacked-cluster limits on Coma and by the conflict with the flux and the slope of Coma radio data (see the Fermi GC discussion below).  This is subject, however, to uncertainty over the halo density profile of Coma, as the use of the Burkert profile removes all aforementioned conflicts.

For the Fermi GC models with best-fit annihilation cross-sections we find that there are conflicts between Coma radio data and the predicted synchrotron slope of all but the minimal (10 GeV) GC model and the 40 GeV $b\bar{b}$ cases. There are, however, no significant conflicts with the $\gamma$-ray limits on Coma. Neither M81 nor Draco serve to further constrain these DM models. However, in these cases neutralino-induced emissions must be sub-dominant complicating hence attempts at robust detection in cosmic structures like dwarf galaxies.

In the case of the 3 TeV neutralino model (with AFP-consistent mass) with Reticulum II and Fermi-LAT dwarf cross-sections~\cite{Fermidwarves2015} we show that there is a conflict with Coma radio data regardless of the assumed DM halo profile. Moreover, this model also conflicts with the Fermi-LAT stacked-cluster limit as well as the direct Fermi-LAT observational limits on Coma (though for the Fermi-LAT cross-section $b\bar{b}$ 3 TeV does not conflict with $\gamma$-ray limits and the Burkert profile removes all such conflicts). This means that the Coma data is not compatible with a TeV neutralino causing Reticulum II $\gamma$-ray emission with our derived Reticulum II annihilation cross-section or that sourced from Fermi-LAT dwarf studies, which included Ret. II in their analysis~\cite{Fermidwarves2015}. Similarly, the radio spectra for M81 conflicts with the AFP mass neutralino model as a source of Reticulum II $\gamma$-ray emission for the Ret. II cross-section but no the Fermi-LAT case. Lastly, TeV neutralino models with both the Reticulum II and Fermi-LAT dwarf annihilation cross-sections are in conflict with the VLA Draco limit for both halo profiles (Ret. II) and NFW only in the of Fermi-LAT cross-sections. As argued previously, this conclusion cannot be easily mitigated by appealing to magnetic field uncertainties, as these would have to be large to account for the excesses over the data.

In the case of the neutralino models with Fermi GC masses and the Reticulum II annihilation cross-section, the Ret. II cases conflict with the Coma radio data for both NFW and Burkert halo density profiles. The Fermi-LAT dwarf cross-section cases only conflict with Coma data when the NFW profile is used, apart from the 10 GeV $b\bar{b}$ case. We note that although we do not display $W^+W^-$ spectra, their relative hardness compared to $b\bar{b}$ makes them difficult to accommodate with the slope of the Coma data~\cite{Colafrancesco2006}, which remains true in our analysis with models with similar mass to GC maximal, even with cross-sections which correspond to those used in~\cite{achterberg2015} (including comparison with appropriate $W^+W^-$ spectra). However, all of the GC masses are incompatible with the Fermi-LAT stacked cluster limit and direct $\gamma$-ray limits on Coma with both NFW and Burkert density profiles for the Ret. II cross-section. In the case of the Fermi-LAT dwarf cross-section, all neutralino masses conflict with the $\gamma$-ray limits for an NFW density profile and $b\bar{b}$ emissions, for the Burkert case there are no conflicts. For M81, all of the masses between 10 and 100 GeV conflict with the data when the Ret. II cross-section is used with an NFW profile. For the Burkert profile only 40 and 100 GeV $\tau^+\tau^-$ remain in tension. In the case of the Fermi-LAT cross-sections there are no conflicts with M81 data. For Draco, with an NFW profile and the Ret. II annihilation cross-section, only the GC minimal mass (i.e., 10 GeV) and median mass with $b\bar{b}$ are compatible with the VLA limit but all the masses/channels violate the Fermi-LAT dwarf limits from~\cite{Fermidwarves2014}. In the Burkert density profile case only the 100 GeV mass with $\tau^+\tau^-$ violates the VLA limit, and all the Fermi-LAT limit violation constraints are removed. For the Fermi-LAT dwarf annihilation cross-section, only the 100 GeV $\tau^+\tau^-$ conflicts with the VLA limit, but a Burkert profile allows all masses with no conflicts with Fermi-LAT or VLA data. This makes Draco conclusions uncertain, as there are good reasons to believe that dwarf galaxies may have cored profiles~\cite{Walker2009,Adams2014}. From the strength of the Coma data we can conclude that the violation of the Fermi-LAT stacked cluster and direct limits on Coma, as well as radio conflict, means that none of the GC masses (between 10 and 100 GeV) are compatible with being responsible for any excess Reticulum II $\gamma$-ray emission, given the current Fermi-LAT limits on dwarf galaxies. 
This is especially important as the best-fit model for the Ret. II $\gamma$-ray excess is one with mass $40$ GeV and $\langle \sigma V\rangle \sim 3 \times 10^{-26}$ cm$^3$ s$^{-1}$. We also note that the Ret. II DM explanation of its $\gamma$-ray excess is already disfavoured by dwarf galaxy observations with Fermi-LAT~\cite{Fermidwarves2015}.

As a matter of the validation of our numerical results, in Section~\ref{sec:constraints} it is shown that the most significant differences between our results for the Coma cluster radio flux and the one derived in previous work, like~\cite{Colafrancesco2006}, are due to the modelling of the magnetic field spatial profile within the inner parts of the halo. It is important to highlight the fact that we use here an updated model of the Coma cluster magnetic field derived by \cite{bonafede2010} several years after the publication of the \cite{Colafrancesco2006} results. 
	
It must be noted that, as we do not take into account the fact that sub-halos at differing radii within the parent halo would experience differing magnetic field values, our predictions of the synchrotron flux for Coma may be slightly optimistic~\cite{storm2013}. However, if sub-halos follow a similar distribution to the DM density of their parent halo, then this effect is mostly significant within cored halo profiles.

It is clear from our discussion in Section~\ref{sec:results} that the mass of the neutralino has a very prominent effect on the DM-induced SED, regardless of environment. This being that it controls the position of the $\gamma$-ray, X-ray and synchrotron peaks through the maximal energy of electrons produced in DM annihilations, and also the distinctness of the ICS and $\gamma$-ray peaks. This latter property is a result of the effect of the neutralino mass on both shifting the ICS peak towards higher energies as well as its effect on the bremsstrahlung emission, which lies between the ICS and gamma peaks. Low mass neutralinos produce electron distributions capped at lower energies and thus result in a lower energy bremsstrahlung emission, resulting in the fact that the bremsstrahlung emissions occupy a spectral region over-shadowed by ICS emissions. It is notable that the mass of the neutralino also has a suppressive effect on the intensity of the emitted radiation, as a result of suppressing collisions, which competes with the effects of the higher mass on the electron and $\gamma$-ray product distributions from annihilation. In this setting, the larger cross-section of the high mass models compensates for this. 

The effect of the dominant annihilation channel on the SED is also apparent in the results shown in Figs.~\ref{fig:coma},\ref{fig:m81}, and \ref{fig:draco}. The difference between the $b\bar{b}$ and $\tau^+\tau^-$ channels is the hardening of the spectrum induced by the latter, which produces a spectral cross-over between the two channels. The position of the cross-over is shifted by the neutralino mass as discussed above. For the GC models, these cross-over points lie within the observation ranges of the SKA (synchrotron emission) and ASTRO-H (ICS emission), opening up avenues for identifying the dominant annihilation channel of any putative neutralino DM particle observation. The DM-induced $\gamma$-ray emission exhibits the same patterns of variation due to neutralino mass and annihilation channel as the ICS and synchrotron emission processes. This means that the identified spectral characteristics are also independent of the mode of emission, making it possible to make far more robust neutralino characterizations using a multi-frequency approach than with isolated spectral region studies. 

In the case of all of the models we studied in this paper, the SKA is excellently placed to measure the slope and magnitude of the synchrotron spectrum as discussed already in~\cite{gsp2015}, as well as being able to scan the majority of the GC model parameter space in the studied environments. In contrast to this, ASTRO-H is unsuitable to extend constraints in the studied environments, due to their significant X-ray backgrounds or weak available limits. However, the ASTRO-H observation window is positioned to be sensitive to the peak of the ICS spectrum, which is determined by the mass of the neutralino. Furthermore, ASTRO-H is sensitive to a region of the spectrum that displays a large variation between neutralino annihilation channels: in fact for $10 \; \mbox{GeV} < M_{\chi} \leq 100$ GeV, this encompasses the point of crossing between $b\bar{b}$ and $\tau^+\tau^-$ ICS spectra. This means that both mass and composition can be informed by ASTRO-H observation for the whole range of masses favoured by GC observations. Therefore, bearing all of this mind, further work will be required to determine the usefulness of the ASTRO-H observations in dark matter searches. In this vein we will perform a similar study for more favourable environments in future work. This is particularly important as, combined with the analysis in~\cite{gsp2015}, X-ray results would give multi-frequency indirect observations two complimentary means of identifying the nature of the neutralino from the associated emissions. Moreover, the differing emission mechanisms mean that these two methods are not subject to the same confusion or error limits and thus can be used as independent consistency checks and to provide robustness to any putative DM detection by indirect methods.

A multi-frequency observational strategy could then combine SKA constraints and those from experiments like ASTRO-H, should favourable detection environments be determined. The importance of this is that synchrotron radiation is sensitive to the magnetic field strength, and detailed structure~\cite{Colafrancesco1998}, as well as the thermal electron density in the target environment. Thus constraints based on synchrotron radiation can be said to be degenerate with respect to the neutralino model as well as some function of the magnetic field and thermal electron density. However, inverse-Compton emissions are not sensitive to the magnetic field but still depend upon the thermal electron density. The combined constraints are then sensitive only to the neutralino as well as the magnetic field, as the thermal electron density is eliminated through recognizing that the ratio of energy densities becomes $\frac{U_B}{U_{IC}} \propto B^2$. This means that the consistency between radio and higher-energy observations purporting to identify neutralino DM is a vital piece of evidence strengthening such an identification, as these different emissions mechanisms are sensitive to differing errors and confusions and serve to eliminate common dependencies. This also emphasyses the importance of Faraday rotation and polarimetry measurements made by the SKA in order to characterize magnetic fields in the target DM environments, as this is crucial to demonstrate the robustness of any potential neutralino identification. 


\section{Conclusions}
\label{sec:conclusions}

The Coma, M81, and Draco environments were shown to be promising targets for multi-frequency analysis as well as demonstrate its power to restrict the parameter space. This is of particular significance to the further constraint of models currently favoured by the galactic centre observations. In the case of the AMS-2/Fermi/PAMELA positron excess models, with best-fit cross-sections, these environments provide evidence that the remaining models in this family are excluded by existing multi-frequency observations, although this is somewhat weakly subject to magnetic field and halo substructure uncertainties. In the case of the best-fit Fermi GC models we demonstrate similar conflicts with existing data over the whole 10-100 GeV mass range. This is already suggestive of the need of a multi-frequency approach to study these DM models. Finally, the SKA is shown to be very well placed for future study of these models, with ASTRO-H being attractive but requiring further study in more favourable detection environments. We also showed that the Coma radio data can be used to derive limits on the neutralino cross-section that are stronger than the Fermi-LAT limits, regardless of halo profile. The magnetic field uncertainties were also too small to account for this improvement.

The three environments here analyzed also demonstrate that the annihilation cross-sections which matches the reported Reticulum II $\gamma$-ray excess, under modest assumptions, are in conflict with current multi-frequency data, indicating that the DM interpretation of this excess is untenable for all the considered neutralino masses which cover a range from 10 GeV to 3 TeV. This conclusion remains relatively robust even with a far more conservative annihilation cross-section limits derived from Fermi-LAT dwarf observations. This is of particular significance as the Draco $\gamma$-ray predictions remains largely in agreement with the Fermi-LAT analysis. Moreover, this reinforces the existing conflicts between the proposed Ret. II dark matter excess and the Fermi-LAT results~\cite{Fermidwarves2015} from dwarf galaxies including Ret. II.

Given the strength of the Coma radio results reported here it is worth mentioning that the constraining power of Coma is much greater in this study than in earlier ones, like \cite{Colafrancesco2006}, as we use an updated magnetic field model for Coma that provides larger synchrotron fluxes, due to its spatial profile peaking at the centre of the cluster.
However, due to the nature of the model we employ for sub-structure flux-boosting, our synchrotron flux results for Coma may be slightly optimistic, as they do not account for sub-halos experiencing weaker magnetic fields at large distances from the cluster center. In this sense the spatial dependence of the magnetic field model used for Coma introduces some uncertainty into our results (although its magnitude was shown to provide only small uncertainties), as we have clearly discussed in this work.

Despite the fact that the chosen sources are not ideal DM detection environments, we have shown that SKA has great potential in the study of the DM parameter space. We will consider a similar analysis of more favourable detection environments in further work in order to discern the potential constraints that might be yielded by ASTRO-H. 

Multi-frequency searches for DM involving the upcoming SKA and ASTRO-H show great promise in their ability to probe the DM parameter space. Should it be possible to locate environments with weaker X-ray backgrounds, ASTRO-H will be able to provide a new window on neutralino DM while the SKA is constructed. 
We have shown, in fact, that the ASTRO-H observation window contains a spectral region that can be used to differentiate both the mass and dominant annihilation channel which characterize the neutralino. The importance of having multiple future experiments capable of furthering the neutralino search is that it opens up the potential for multi-frequency examinations of the DM parameter space. This is highly attractive given the uncertainties and errors inherent in indirect DM search methods due to the complexity of DM halo environments. Moreover, indirect multi-frequency searches benefit from the fact that the signatures of the neutralino spectrum occur for each emission mechanism, despite the fact that each has different dependencies and sources of error. This means that multi-frequency observations provide a series of consistency checks, allowing for far more robust identifications and for the elimination of common error dependencies, like the density of thermal electrons within the halo (a common dependency between synchrotron and ICS emissions). We must note, however, that the uncertainties found in substructure boosting effects, nature of the halo profile, and disentangling dark matter emissions from purely baryonic processes remain significant. Although the precision of the SKA mitigates the last point as we have shown for these unfavourable detection environments, and the angular resolution of the SKA will be able to supplement this through source-subtractions~\cite{Colafrancesco2015,Regisetal2014}.
Despite this, in the case of the SKA and ASTRO-H, a major source of error in a multi-frequency DM search, in the form of the magnetic field, can be eliminated, as this can be fully characterized by the SKA during the observations required for DM searches, as discussed in~\cite{Colafrancesco2015}. It remains to be seen whether the dominant error, in the form of resolving DM emissions from astrophysical backgrounds, can be mitigated for ASTRO-H through a choice of more favourable DM search environments.

Therefore, based upon all the preceding arguments, we conclude that multi-frequency strategies revolving around the SKA, with the possible inclusion of the ASTRO-H experiment, have considerable advantages to be leveraged in the continuing hunt for neutralino dark matter.

\section*{Acknowledgments}

S.C. acknowledges support by the South African Research Chairs Initiative of the Department of Science and Technology and National Research Foundation and by the Square Kilometre Array (SKA). G.B. acknowledges support from the DST/NRF SKA post-graduate bursary initiative.

This publication makes use of data products from the Wide-field Infrared Survey Explorer (WISE), which is a joint project of the University of California, Los Angeles, and the Jet Propulsion Laboratory/California Institute of Technology, funded by the National Aeronautics and Space Administration and  from the NASA/IPAC Extragalactic Database (NED). Part of this work is also based on archival data and online services provided by the ASI SCIENCE DATA CENTER (ASDC).

We finally thank the Referee for his/her careful review of the results, as well as for several comments and valuable suggestions that allowed us to improve the presentation of this work.

\end{document}